\documentclass[aps,showpacs,preprint,epsf,epsfig, nofootinbib]{revtex4-2}
\usepackage{mathrsfs}
\usepackage{dcolumn}
\usepackage{bm}
\usepackage{hyperref}
\usepackage{epsfig,graphicx,color,appendix}
\usepackage{multirow}
\usepackage{tabularx}
\usepackage{capt-of}
\usepackage{caption}
\usepackage[countmax]{subfloat}

\usepackage{threeparttable}
\usepackage{amssymb}
\usepackage{array}
\usepackage{amsmath}
\usepackage{rotating,tabularx}
\usepackage{float}
\usepackage{placeins}
\usepackage[english]{babel}
\usepackage{booktabs}
\usepackage{multirow}
\newcommand{\mb}[1]{\mathbf{#1}}
\newcommand{\rtbar}{\bar{\mathbf{3}}}          
\begin{document}
\title{A baryon-calibrated unified quark-diquark effective mass formalism for heavy multiquarks}
\author{\textbf{Binesh Mohan}}
\email{bineshmohan96@gmail.com}
\author{\textbf{Rohit Dhir}}
\email[Corresponding author: ]{dhir.rohit@gmail.com}
\affiliation{Department of Physics and Nanotechnology,\\SRM Institute of Science and Technology, Kattankulathur-603203, Tamil Nadu, India.}

	\long\def\symbolfootnote[#1]#2{\begingroup%
		\def\thefootnote{\fnsymbol{footnote}}\footnote[#1]{#2}\endgroup}
	\def\lsim{ {\ \lower-1.2pt\vbox{\hbox{\rlap{$<$}\lower5pt\vbox{\hbox{$\sim$}
			}}}\ } }
	\def\gsim{ {\ \lower-1.2pt\vbox{\hbox{\rlap{$>$}\lower5pt\vbox{\hbox{$\sim$}
			}}}\ } }
	
	\vskip 1.5 cm
	\small
	\vskip 1.0 cm

\date{\today}
\begin{abstract}
	We present a unified framework for heavy tetraquark and pentaquark systems within the quark–diquark effective mass formalism, extending its baryon-calibrated construction to multiquark states without introducing sector-dependent parameters. Intra-diquark color–spin correlations are encoded in effective diquark masses fixed from baryon spectroscopy, while the inter-cluster chromomagnetic scale, independently determined from vector–pseudoscalar meson splittings, is propagated unchanged to exotic configurations, ensuring residual one-gluon-exchange dynamics only between composite color sources. Within this framework, we compute the complete spectra for both $\bar{\mathbf{3}}_c\!\otimes\!\mathbf{3}_c$ and $\mathbf{6}_c\!\otimes\!\bar{\mathbf{6}}_c$ configurations in tetraquarks, whereas the pentaquark analysis focuses on the dominant $\bar{\mathbf{3}}_c\!\otimes\!\bar{\mathbf{3}}_c\!\otimes\!\bar{\mathbf{3}}_c$ clustering. Heavy-quark spin symmetry and flavor-symmetry breaking across light, charm, and bottom sectors emerge naturally through the explicit $1/(m_{D_1}m_{D_2})$ scaling of the calibrated couplings. The resulting spectra exhibit a coherent dynamical hierarchy spanning baryons and multiquark states. Established exotic candidates are reproduced within hadronic uncertainties, while the unified calibration enables quantitative predictive control across flavor sectors. The framework thus provides a parameter-economical, systematically constrained baseline with unified dynamical consistency for heavy multiquark spectroscopy.
\end{abstract}


\maketitle
\newpage

\section{Introduction}
\label{Sec1}

The past two decades have transformed exotic hadron spectroscopy from a speculative extension of the quark model into a firmly established experimental discipline. The discovery of the $X(3872)$~\cite{Belle:2003nnu} first demonstrated the existence of hadronic states beyond the conventional $q\bar{q}$ and $qqq$ configurations. Subsequent observations of charged charmonium-like structures, $Z_c(3900)$, $T_{c\bar{c}1}(4200)$, $T_{c\bar{c}}(4430)$, and their strange partners $T_{c\bar{c}\bar{s}1}(4000)$ and $T_{c\bar{c}\bar{s}1}(4220)$, together with the bottom counterparts $T_{b\bar{b}1}(10610)$ and $T_{b\bar{b}1}(10650)$, have consolidated the tetraquark sector across multiple flavor combinations~\cite{ParticleDataGroup:2024cfk,LHCb:2024vfz,LHCb:2024smc,LHCb:2022sfr,LHCb:2021uow,BESIII:2020oph,Belle:2014nuw,Belle:2011aa}. The LHCb observation of the doubly charmed tetraquark $T_{cc}^+(3875)$ established the first unambiguous open-flavor doubly heavy exotic with a near-threshold mass and confirmed $J^P=1^+$ assignment~\cite{LHCb:2021vvq,LHCb:2021auc}. In the fully heavy domain, di-$J/\psi$ resonances $X(6600)$, $X(6900)$, and $X(7100)$ reported by CMS, LHCb, and ATLAS~\cite{CMS:2023owd,CMS:2025fpt,LHCb:2020bwg,ATLAS:2023bft} provide compelling evidence for compact all-charm configurations, with the complete $J^{PC}=2^{++}$ family very recently reported~\cite{CMS:2026tiu,CMS:2025fpt}. Simultaneously, the hidden-charm pentaquark spectrum has matured from the discovery of $P_c(4380)^+$ and $P_c(4450)^+$~\cite{LHCb:2015yax} through the high-resolution measurements of $P_c(4312)^+$, $P_c(4440)^+$, $P_c(4457)^+$, and $P_c(4337)^+$~\cite{LHCb:2019kea,LHCb:2021chn} to the strange partners $P_{cs}(4338)^0$ and $P_{cs}(4459)^0$~\cite{LHCb:2022ogu,LHCb:2020jpq, Belle:2025pey}. This accumulation of states across distinct heavy-quark and light-flavor sectors has transformed multiquark spectroscopy into a systematic experimental program, with high-luminosity LHC runs and future facilities poised to enlarge it further \cite{Vagnoni:2025qfv, FCC:2018evy, CEPCStudyGroup:2018rmc, CEPCStudyGroup:2018ghi, ILC:2007bjz}. Comprehensive reviews of the experimental discoveries and their theoretical interpretations can be found in Refs.~\cite{Wang:2025sic, Johnson:2024omq, Chen:2022asf, Bicudo:2022cqi, Barabanov:2020jvn, Brambilla:2019esw}.
  
These discoveries establish exotic hadrons as a laboratory for nonperturbative Quantum Chromodynamics (QCD). While threshold effects can generate molecular components, the lowest-lying states are governed at leading order by short-distance color dynamics. Isolating this compact core is therefore essential for identifying systematic flavor patterns and the dominant binding mechanism. At short distances, the primary spin-dependent force is the color-magnetic hyperfine interaction generated by one-gluon exchange (OGE)~\cite{DeRujula:1975qlm}, which encodes the universal $1/(m_i m_j)$ scaling and controls both diquark structure and residual inter-cluster interactions in exotic multiplets~\cite{Park:2013fda}. Empirical regularities across tetraquark and pentaquark spectra, hyperfine scaling, flavor-dependent ordering, and stable color hierarchies, support OGE-driven chromomagnetic dynamics as a robust organizing principle~\cite{Dillon:1995qw,Dillon:2002ks,Durand:2001zz}.

The experimental progress has prompted diverse theoretical approaches, including color-magnetic interaction models (CMIM)~\cite{Weng:2020jao,Luo:2017eub}, Regge analyses~\cite{Song:2023izj}, heavy quark effective theory (HQET)~\cite{Braaten:2020nwp}, bag models (BM)~\cite{Yan:2023lvm, Zhang:2021yul}, relativistic quark models (RQM)~\cite{Yu:2024ljg,Lu:2020rog,Lu:2020qmp,Ebert:2007rn}, heavy-quark symmetry (HQS) constructions~\cite{Eichten:2017ffp,Karliner:2017qjm}, QCD sum rules (QCDSR)~\cite{Wang:2025sic}, nonrelativistic potential models (NRPM)~\cite{Liu:2022hbk,Liu:2024fnh,Liu:2019zuc}, diffusion Monte Carlo (DMC) calculations \cite{Gordillo:2020sgc}, and lattice QCD (LQCD) simulations of doubly heavy systems~\cite{Padmanath:2022cvl,Lyu:2023xro,Prelovsek:2025vbr,Francis:2024fwf,Bicudo:2022cqi,Francis:2016hui,Alexandrou:2024iwi,Colquhoun:2024jzh}. Despite their individual successes, these approaches often address specific sectors or require sector-dependent recalibration, and a unified framework that simultaneously spans conventional baryons, and exotics within a single calibrated dynamical scheme, without free-parameter adjustment at the multiquark level, has remained elusive. 

In this work, we develop a unified description of multiquark states within the quark–diquark effective mass formalism (QDEMF), previously applied to heavy baryons in Ref.~\cite{Mohan:2026rcg}. The guiding principle is that the same diquark correlations that organize the conventional baryon spectrum also determine the mass hierarchy of tetraquark and pentaquark multiplets without introducing new sector-dependent parameters. The calibration is sequential and unidirectional: effective scalar and axial-vector diquark masses are fixed from baryon spectroscopy, while the inter-cluster chromomagnetic scale is determined independently from vector–pseudoscalar meson splittings. These inputs are then propagated to tetraquark and pentaquark systems without readjustment. Intra-diquark color–spin correlations are absorbed into the effective diquark masses, so that residual OGE-mediated hyperfine interactions act only between composite color sources, thereby avoiding double counting of short-distance dynamics. Both $\bar{\mathbf{3}}_c\!\otimes\!\mathbf{3}_c$ and $\mathbf{6}_c\!\otimes\!\bar{\mathbf{6}}_c$ color configurations are retained through their QCD Casimir factors, without imposing off-diagonal mixing assumptions. Heavy-quark spin symmetry (HQSS) and flavor-symmetry breaking across light and heavy sectors follow from the explicit $1/(m_{D_1}m_{D_2})$ dependence of the calibrated couplings.

The observed pentaquark configurations exhibit an intrinsically asymmetric heavy-light clustering, unlike the comparatively symmetric tetraquark geometry. This structural difference necessitates a unified treatment in which short-distance color–spin dynamics are fixed consistently at the diquark level before additional dynamical effects are introduced. A compact baseline is therefore required as the leading-order approximation to both sectors. By placing conventional baryons, tetraquarks, and pentaquarks on the same calibrated dynamical footing, the QDEMF establishes diquark correlations as a parameter-economical and unified organizing principle of heavy multiquark spectroscopy.

The remainder of this paper is organized as follows. Sec.~\ref{Sec2} presents the methodology. Sec.~\ref{Sec3} discusses numerical results on tetraquark and pentaquark configurations, while Sec.~\ref{Sec4} summarizes the conclusions. The color-spin wave functions are compiled in Appendices \ref{AppA}-\ref{AppB}.

\section{Methodology}
\label{Sec2}
In the exotic sector, both the traditional constituent quark model and direct QCD-based treatments face significant challenges. The strong entanglement of color, spin, and spatial correlations in multiquark systems renders a straightforward many-body description technically demanding, while purely phenomenological approaches risk losing contact with the underlying QCD dynamics.

To bridge this gap, we formulate a unified QDEMF embedded within a QCD-motivated framework. This approach retains the transparency and calculational tractability of the constituent quark model while preserving the essential color structure dictated by QCD. The dominant short-range interaction is described by OGE, implemented with explicit color-factor resolution. In this way, the $SU(3)$ color dynamics remain consistently treated across both conventional baryons and exotic multiquark configurations.

Baryons and multiquark states are therefore described using the same OGE-based color-spin interaction. All hyperfine contributions are evaluated explicitly according to the relevant color representations, without phenomenological averaging or refitting when the number of constituents changes. This ensures dynamical consistency when transitioning from three-body to multibody systems. Correlated quark pairs are treated as effective degrees of freedom, allowing multiquark systems to be reorganized into interacting clusters while ensuring consistent implementation of color and spin couplings. The resulting cluster-based description provides a systematic reduction of the many-body problem without sacrificing the essential QCD-driven structure of the interaction. 

This formulation provides a consistent description of conventional baryon spectroscopy within the quark-diquark picture \cite{Mohan:2026rcg}, and its multiquark realization is developed in this work. A coherent extension to genuinely many-body systems requires that the underlying QCD-driven structures, including color dynamics, spin-dependent interactions, and symmetry constraints, remain preserved when transitioning from three-body to multibody configurations. In what follows, we implement these requirements through the explicit construction of multiquark color, spin, and flavor wave functions, together with the corresponding effective mass operators.

\subsection {Diquark-antidiquark model of tetraquarks}
\label{Sec2A}
The tetraquark wave function is constructed by treating the diquark and antidiquark as effective constituents with well-defined color, spin, and flavor quantum numbers. A diquark is a correlated two-quark subsystem treated as a compact effective degree of freedom. For two quarks transforming under the fundamental representation of $SU(3)_c$, the allowed color decomposition of diquark is
\[
\mb3\otimes \mb3=\rtbar \oplus \mb6.
\]

The color-antisymmetric $\overline{\mb3}$ channel is attractive under OGE and therefore forms the energetically favored diquark configuration in most phenomenological descriptions of tetraquarks. In contrast, the color-symmetric $\mb6$ channel is repulsive and corresponds to a heavier and less tightly bound diquark. While sextet diquarks are disfavored in conventional baryons due to the overall antisymmetry of the three-quark wave function, they remain viable configurations in tetraquark systems and must be retained until decisively excluded by experimental evidence. In particular, a diquark transforming in the $\mb6$ representation may combine with an antidiquark in the conjugate $\overline{\mb6}$ representation to form an overall color singlet. This establishes the sextet channel as an allowed color configuration of the tetraquark system, in addition to the conventional $\overline{\mb3}\otimes \mb3$ structure, and ensures a complete group-theoretic classification of the spectrum.

Although the $\mb6 \otimes \overline{\mb6}$ configuration is energetically disfavored due to the repulsive OGE color factor in sextet diquark, it remains a legitimate spectroscopic channel. In the present work, configuration mixing between the $\overline{\mb3}\otimes \mb3$ and $\mb6 \otimes \overline{\mb6}$ sectors is neglected, and each representation is treated independently within the QDEMF. This approximation is consistent with the hierarchy implied by the underlying color-Casimir structure and provides a controlled leading-order description of the tetraquark spectrum.\footnote{Introducing configuration mixing would require additional off-diagonal interactions and independent parameters beyond the scope of the present QDEMF. Furthermore, the distinct quadratic Casimir eigenvalues of the $\bar{\mb3}\mb3$ and $\mb6\bar{\mb6}$ color sectors imply different hyperfine scales, supporting their treatment as independent spectroscopic configurations at leading order.}

The corresponding antidiquark color structure follows directly from the conjugate representation of $SU(3)_c$, with the decomposition
\[
\bar {\mb3} \otimes \bar {\mb3} = {\mb3} \oplus \bar{\mb6}.
\]
Here, the color-triplet antidiquark provides the conjugate partner of the attractive $\overline{\mb3}$ diquark, while the $\overline{\mb6}$ antidiquark represents the conjugate sextet configuration. These structures allow the diquark and antidiquark to combine into overall color-singlet states, permitting a two-body description in terms of effective diquark and antidiquark degrees of freedom.

The full color structure of the tetraquark system follows from the decomposition
\begin{equation*}
\begin{split}
\mb3 \otimes \mb3 \otimes \bar{\mb3} \otimes \bar{\mb3}
&= (\bar{\mb3} \oplus \mb6) \otimes (\mb3 \oplus \bar{\mb6}) \\
&= (\bar{\mb3} \otimes \mb3) \oplus (\mb6 \otimes \bar{\mb6})
\oplus (\bar{\mb3} \otimes \bar{\mb6}) \oplus (\mb6 \otimes \mb3).
\end{split}
\end{equation*}

Among these combinations, only the $\bar{\mb3} \otimes \mb3$ and $\mb6 \otimes \bar{\mb6}$ channels contain a color-singlet component and therefore correspond to physical tetraquark configurations. Their subsequent decomposition is given by
\[
\bar{\mb3} \otimes \mb3 = \mb1 \oplus \mb8,
\qquad
\mb6 \otimes \bar{\mb6} = \mb1 \oplus \mb8 \oplus \mb27.
\]

Both channels yield color-singlet states, ensuring overall color neutrality as required by QCD gauge invariance. The normalized color-singlet wave functions for the $\overline{\mb3}\mb3$ and $\mb6\overline{\mb6}$ configurations are provided in Appendix~\ref{AppA}. These wave functions uniquely determine the expectation values of the color operator $\langle \vec{\lambda}_i \cdot \vec{\lambda}_j \rangle$, which in turn fix the effective color factors governing the OGE-motivated hyperfine interaction between the diquark and antidiquark constituents within the QDEMF.

Having specified the color structure of the tetraquark, we now turn to its spin degrees of freedom. For $S$-wave tetraquarks, the diquark (antidiquark) spins are constructed by first coupling the spins of two quarks (antiquarks), yielding $S_{D_1}(S_{\bar D_2})=0$ or $1$. The total tetraquark spin $J$ is then obtained through coupling of the diquark and antidiquark spins,
\[
\bm{J}=\bm{S_{D_1}}+\bm{S_{\bar D_2}}, ~\text{with}~ |S_{D_1}-S_{\bar D_2}|\leq J \leq |S_{D_1}+S_{\bar D_2}|.
\]
This generates scalar-scalar ($J^P=0^+$), scalar-axial ($J^P=1^+$), and axial-axial ($J^P=0^+,1^+,2^+$) configurations within the diquark-antidiquark framework. This hierarchical spin coupling scheme is particularly convenient for implementing hyperfine interactions and for classifying the resulting tetraquark states according to their total spin and parity. The explicit spin wave functions are given in Appendix~\ref{AppA}, together with the corresponding flavor structures.

The construction of physically admissible tetraquark wave functions must satisfy the Pauli exclusion principle for identical quarks and antiquarks within each diquark and antidiquark cluster. Consequently, the total diquark and antidiquark wave function must be antisymmetric under the exchange of identical fermions, taking into account color, spin, flavor, and spatial components. For ground-state tetraquarks, the spatial part of the wave function is always symmetric. Therefore, overall antisymmetrization for identical fermions is enforced by combining flavor with color and spin. In particular, a color-antisymmetric diquark requires a symmetric spin-flavor wave function, while a color-symmetric diquark necessitates antisymmetry in the remaining degrees of freedom. These constraints play a crucial role in determining which spin and flavor configurations are allowed or forbidden for a given color structure. The same considerations apply to the antidiquark sector, ensuring that the full tetraquark wave function is properly antisymmetrized and consistent with Fermi statistics \cite{Luo:2017eub, Lu:2020rog}. In the diquark–antidiquark picture, the resulting color–spin–flavor basis states can be written as
\begin{align}
\psi_c^1 \psi_s^1 &= 
\big| [q_1 q_2]_0^{\bar{3}} \, [\bar q_3 \bar q_4]_0^{3} \big\rangle_0,
&
\psi_c^2 \psi_s^1 &= 
\big| \{q_1 q_2\}_0^{6} \, \{\bar q_3 \bar q_4\}_0^{\bar{6}} \big\rangle_0,
\nonumber\\[1.5ex]
\psi_c^1 \psi_s^2 &= 
\big| [q_1 q_2]_0^{\bar{3}} \, \{\bar q_3 \bar q_4\}_1^{3} \big\rangle_1,
&
\psi_c^2 \psi_s^2 &= 
\big| \{q_1 q_2\}_0^{6} \, [\bar q_3 \bar q_4]_1^{\bar{6}} \big\rangle_1,
\nonumber\\
\psi_c^1 \psi_s^3 &= 
\big| \{q_1 q_2\}_1^{\bar{3}} \, [\bar q_3 \bar q_4]_0^{3} \big\rangle_1,
&
\psi_c^2 \psi_s^3 &= 
\big| [q_1 q_2]_1^{6} \, \{\bar q_3 \bar q_4\}_0^{\bar{6}} \big\rangle_1,
\nonumber\\[1.5ex]
\psi_c^1 \psi_s^4 &= 
\big| \{q_1 q_2\}_1^{\bar{3}} \, \{\bar q_3 \bar q_4\}_1^{3} \big\rangle_0,
&
\psi_c^2 \psi_s^4 &= 
\big| [q_1 q_2]_1^{6} \, [\bar q_3 \bar q_4]_1^{\bar{6}} \big\rangle_0,
\nonumber\\
\psi_c^1 \psi_s^5 &= 
\big| \{q_1 q_2\}_1^{\bar{3}} \, \{\bar q_3 \bar q_4\}_1^{3} \big\rangle_1,
&
\psi_c^2 \psi_s^5 &= 
\big| [q_1 q_2]_1^{6} \, [\bar q_3 \bar q_4]_1^{\bar{6}} \big\rangle_1,
\nonumber\\
\psi_c^1 \psi_s^6 &= 
\big| \{q_1 q_2\}_1^{\bar{3}} \, \{\bar q_3 \bar q_4\}_1^{3} \big\rangle_2,
&
\psi_c^2 \psi_s^6 &= 
\big| [q_1 q_2]_1^{6} \, [\bar q_3 \bar q_4]_1^{\bar{6}} \big\rangle_2.
\nonumber
\end{align}

Here, superscripts denote the color representations, while subscripts indicate the spin quantum numbers of the diquark, antidiquark, and total tetraquark. Permutation symmetry is enforced independently within each cluster, consistent with their treatment as correlated effective constituents in the QDEMF. These basis states uniquely determine the color–spin structure governing the hyperfine interaction. The dynamical realization of these color–spin configurations is implemented within the QDEMF, which follows the chromomagnetic constituent mass formalism \cite{DeRujula:1975qlm}. In this approach, hadron masses are expressed in terms of effective constituent masses supplemented by pairwise color-magnetic interactions arising from OGE. When extended to multiquark systems, diquarks are treated as effective constituents, and the residual inter-cluster interaction retains the same pairwise operator structure. The tetraquark masses are therefore obtained directly from the corresponding effective mass relation, ensuring continuity with the established chromomagnetic expansion.

\subsection*{Tetraquark masses in effective mass formalism}
\label{tetra_mass}

\begin{figure}[t]
    \centering
    \includegraphics[width=0.4\linewidth]{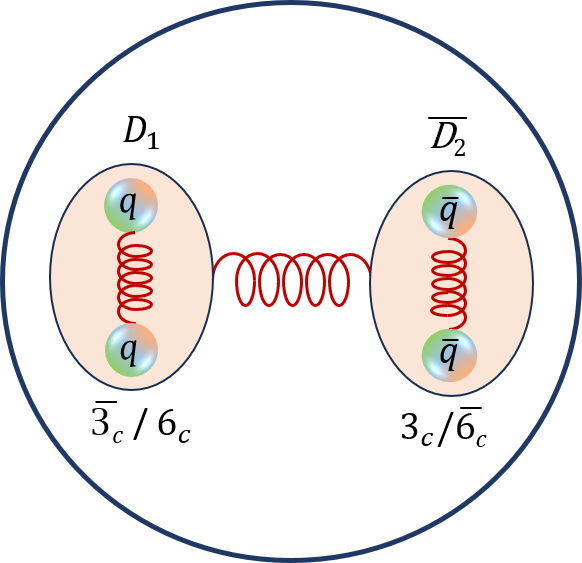}
    \caption{Schematic representation of the tetraquark as an effective diquark–antidiquark bound state.}
    \label{Fig1}
\end{figure}

The quark–diquark framework developed for baryons in Scenario II of the QDEMF \cite{Mohan:2026rcg} provides a natural extension to tetraquark systems, which are modeled here as bound states of an effective diquark and antidiquark. In this formulation, the correlated quark–quark dynamics generated by OGE are already encoded in the internal structure of the diquark degrees of freedom. The tetraquark can therefore be described, at leading order in the QDEMF, as an effective two-body system with residual interactions between the diquark and antidiquark clusters. A schematic representation of this picture is shown in Fig.~\ref{Fig1}.

In conventional four-body treatments, the chromomagnetic interaction is implemented through explicit pairwise couplings among all quarks and antiquarks, requiring a full color–spin diagonalization. While formally complete, such approaches typically introduce a large number of interaction terms whose strengths must be constrained phenomenologically. In the present framework, these short-distance correlations are incorporated hierarchically: the dominant quark–quark color–spin interactions are absorbed into the effective diquark and antidiquark masses, which are fixed independently from baryon spectroscopy, while the residual interaction between clusters is described by an effective diquark–antidiquark hyperfine term.

In this formulation, the inter-cluster hyperfine interaction is constructed to retain the same OGE color-spin operator that governs spin splittings in conventional hadrons. The corresponding color–magnetic coupling strength is fixed once from meson spectroscopy, where quark–antiquark hyperfine splittings provide a direct determination of the underlying interaction. This calibration is performed independently of the multiquark sector and is not subsequently refitted. As a result, the effective diquark–antidiquark interaction inherits its strength from the same chromomagnetic dynamics that governs mesons, ensuring universality of the interaction while avoiding the introduction of additional unconstrained parameters. This construction reduces the tetraquark system to an effective two-body problem while preserving the color–spin structure dictated by QCD at the resolved scale.

Within the diquark–antidiquark picture, the tetraquark mass separates naturally into three contributions: the effective diquark mass, the effective antidiquark mass, and the residual hyperfine interaction between the two clusters. This decomposition isolates the internal color–spin correlations encoded in the diquark substructure from the residual inter-cluster dynamics, providing a systematic and internally consistent description of multiquark states. The resulting tetraquark mass, arising from the chromomagnetic interaction induced by OGE, is therefore given by
\begin{equation}\label{Eq1}
    M_T = m_{D_1} + m_{\bar{D}_2} + b_{D_1 \bar{D}_2}\, (\bm{S}_{D_1}\!\cdot\!\bm{S}_{\bar{D}_2}),
\end{equation}
where $m_{D_1}$ and $m_{\bar{D}_2}$ denote the effective constituent masses of the diquark $D_1(ij)$ and antidiquark $\bar{D}_2(\bar k\bar l)$, respectively, and $b_{D_1 \bar{D}_2}$ characterizes the effective color-magnetic interaction between the two clusters. These masses should be understood as the effective diquark and antidiquark masses within the exotic tetraquark, encoding their internal chromomagnetic structure while serving as emergent collective parameters governing the effective mass expansion.

The effective diquark mass entering exotic hadrons incorporates the internal chromomagnetic interaction between its constituent quarks. Flavor-symmetric and antisymmetric combinations are denoted by $\{ij\}$ and $[ij]$, respectively, in the subscript, while the diquark color and spin quantum numbers $(c,S_{D_1})$ are specified explicitly in the superscript. The corresponding spin expectation values are $\langle \bm{s}_i\!\cdot\!\bm{s}_j\rangle=+\tfrac{1}{4}$ for $S_{D_1}=1$ and $-\tfrac{3}{4}$ for $S_{D_1}=0$. Therefore, for diquarks in the attractive color-antisymmetric $\overline{\mb3}$ representation, the effective diquark masses ($m_{D_i} \equiv m_{D(\bar{D})}^{(c, S_D)}$) are
\begin{equation}
\label{Eq2}
m_{\{ij\}}^{(\overline{3},\,1)}
= m_i + m_j + \frac{b^{\overline{3}}_{ij}}{4},
\end{equation}
\begin{equation}
\label{Eq3}
m_{[ij]}^{(\overline{3},\,0)}
= m_i + m_j - \frac{3\,b^{\overline{3}}_{ij}}{4},
\end{equation}
where $m_i$ and $m_j$ are the constituent quark masses. The hyperfine coefficient
\begin{equation}
\label{Eq4}
b^{\overline{3}}_{ij}
= \frac{16\pi\alpha_s}{9\,m_i m_j}
\langle \psi | \delta^{(3)}(\vec r) | \psi \rangle,
\end{equation}
encodes the short-distance chromomagnetic interaction between the quarks.

In analogy with QDEMF, these coefficients retain the same chromomagnetic operator structure as in conventional hadrons, but are reorganized in terms of effective diquark degrees of freedom appropriate for exotic systems. This reflects a reorganization of the underlying interaction rather than the introduction of new dynamics. The resulting mass expansion follows the same organizing principle as the QCD parametrization method of Morpurgo \cite{Morpurgo:1989my,Dillon:1995qw}, in which hadron masses are expressed as an operator expansion ordered by dynamical importance. In this framework, constituent mass terms and pairwise color–spin interactions provide the leading contributions, while genuine higher-body operators are subleading in the QCD operator hierarchy, consistent with the observed accuracy of baryon mass relations such as the Gell-Mann--Okubo formula \cite{Durand:2001zz,Durand:2001sz,Dillon:2002ks}.

Within QDEMF, this hierarchy is implemented by absorbing the dominant short-distance dynamics into effective constituent (di)quark masses and baryon-calibrated hyperfine couplings. No independent genuine three-body short-distance operators are therefore introduced in the multiquark sector. Any such effects, if present, are dynamically subleading at the resolved scale and are implicitly encoded in the effective parameters fixed from the conventional hadron spectrum. This controlled truncation is equivalent to a leading-order description of the underlying relativistic field theory and is sufficient for reproducing hadron masses at the level of accuracy targeted here.

The strength of the hyperfine interaction is further modulated by the color factor $\langle \vec{\lambda}_i \cdot \vec{\lambda}_j \rangle$, where $\vec{\lambda}_i$ are the Gell-Mann matrices acting in the color space of quark $i$. This color coupling can be expressed in terms of quadratic Casimir operators as
\begin{equation}
\label{Casimir}
\big\langle\vec{\lambda}_i\cdot\vec{\lambda}_j\big\rangle_{R}=\frac{1}{2}\Big(C_2(R)-C_2(R_1)-C_2(R_2)\Big).
\end{equation}
Using $C_2(3)=C_2(\overline{3})=\tfrac{4}{3}$ and $C_2(6)=\tfrac{10}{3}$ for a representation $R=R_1\otimes R_2$, one obtains the standard $SU(3)$ color factors for the $\overline{\mb3}$ and $\mb6$ color representations of diquark as follows:
\begin{equation}\label{CF}
\big\langle\vec{\lambda}_i\cdot\vec{\lambda}_j\big\rangle_{\bar 3}=-\frac{2}{3},
\qquad
\big\langle\vec{\lambda}_i\cdot\vec{\lambda}_j\big\rangle_{6}=+\frac{1}{3}.
\end{equation}
Accordingly, the color-antisymmetric $\overline{\mb3}$ diquark configuration is attractive, while the color-symmetric $\mb6$ channel is repulsive, as dictated by the sign of the associated color factors. These coefficients determine both the relative sign and magnitude of the intra-diquark hyperfine interaction and, consequently, play a central role in determining the effective diquark masses employed in the tetraquark mass construction. Aforementioned, the diquark masses in tetraquarks are effective, reflecting the internal chromomagnetic and short-distance correlations \cite{DeRujula:1975qlm}. With the color dependence made explicit, the construction of sextet diquark masses follows naturally from that of the color-antisymmetric $\overline{\mb3}$ diquarks. Since the intra-diquark hyperfine interaction (inside the tetraquark) scales linearly with the corresponding $SU(3)$ color factor, the sextet hyperfine interaction strength ($b_{ij}^{6}$) can be obtained by rescaling the antitriplet interaction ($b_{ij}^{\overline{3}}$) using the ratio of their color coefficients. Accordingly, the hyperfine coupling for a color-sextet diquark is related to that of the antitriplet by
\begin{equation}\label{Eq6}
b_{ij}^{6} = -\bigg(\frac{b_{ij}^{\bar{3}}}{2}\bigg).
\end{equation}
Consequently, the phenomenological form of $b_{ij}^{6}$ can be explicitly given as
\begin{equation}
\label{Eq7}
 b_{ij}^{6} =\frac{-8\pi\alpha_s}{9m_im_j}\,
 \langle\psi|\delta^3(\vec{r})|\psi\rangle.
\end{equation}
Using Eq. \eqref{Eq6}, the sextet diquark masses ($m_{D_1^{6}}$) can be formulated in direct analogy with the antitriplet diquark masses given by Eqs.~\eqref{Eq2} and \eqref{Eq3}, with separate expressions for axial-vector and scalar configurations. Therefore,
\begin{equation} 
\label{Eq8} 
	m_{[ij]}^{(6,\,1)} = m_i + m_j - \frac{b^{\overline{3}}_{ij}}{8},
\end{equation}
\begin{equation}
\label{Eq9} 
	m_{\{ij\}}^{(6,\,0)} = m_i + m_j + \frac{3b^{\overline{3}}_{ij}}{8}.
\end{equation}
The sign reversal and reduced magnitude of the hyperfine contribution reflect the repulsive nature of the color-sextet channel under OGE. In addition to these color-spin hyperfine contributions encoded in the effective diquark masses (both $\bar {\mb3}$ and $\mb6$), an explicit binding-energy correction is required to account for the short-range chromoelectric interaction between quarks. As established in our baryon analysis \cite{Mohan:2026rcg}, this spin-independent color-Coulomb contribution provides an additional source of attraction when more than one heavy quark is present and becomes quantitatively relevant in systems containing heavy constituents that move nonrelativistically. For diquarks in the color-antitriplet channel, the binding energy is inferred from the corresponding quark-antiquark system using standard color $SU(3)$ arguments, yielding a strength that is one-half ($\frac{1}{2}$) of that in a color-singlet $Q\bar Q^{\prime}$ configuration. This binding term is treated explicitly as a spin-independent contribution and is added along with the diquark mass in the tetraquark mass, Eq. \eqref{Eq1}. For color-sextet diquarks, the same color-factor hierarchy explained above implies a further rescaling governed by the ratio of their corresponding color coefficients \cite{Zhang:2021yul}. In close analogy with the rescaling introduced for the spin-dependent hyperfine interaction, the spin-independent color-Coulomb piece also scales with the color operator $\langle \vec{\lambda}_i \cdot \vec{\lambda}_j \rangle$, given as
\[
BE^{(6)}_{ij} = -\bigg(\frac{BE^{(\overline{3})}_{ij}}{2}\bigg),
\]
reflecting the repulsive nature of the sextet channel under OGE.\footnote{The binding energy terms are incorporated as a separate contribution to the overall tetraquark mass based on the quark content of the corresponding diquark (antidiquark).} Consequently, sextet diquarks are expected to be systematically heavier than their antitriplet counterparts for the same quark content. With this prescription, all short-range color-dependent binding effects are consistently incorporated as an explicit contribution to the effective diquark masses. These sextet diquark masses, together with the antitriplet diquark masses, provide the complete set of effective building blocks required for constructing tetraquark states within the diquark-antidiquark picture. 

As discussed above, the properties of antidiquarks follow directly from the same color–spin structure that governs diquarks. Since antiquarks transform in the conjugate representation of $SU(3)_c$, their color-couplings obey an identical Casimir hierarchy. Because the quadratic Casimir operator satisfies $C_2(R)=C_2(\overline{R})$, the chromomagnetic color factors, and consequently the effective masses and spin-dependent splittings, are numerically identical to those of the corresponding diquarks. In the QDEMF, the antidiquark therefore enters as an effective cluster whose mass and hyperfine structure are determined entirely by its internal color and spin configuration.

Once the diquark and antidiquark substructures are specified, the tetraquark is treated as an effective two-body system composed of these correlated clusters. The tetraquark mass is obtained by combining the effective cluster masses with the residual chromomagnetic interaction between their spins, as given in Eq.~\eqref{Eq1}. Since the spin-spin operator vanishes when either constituent has spin zero, chromomagnetic splittings arise only for axial-vector diquark-antidiquark configurations. The allowed tetraquark states follow directly from coupling the diquark and antidiquark spins, yielding six distinct spin-parity configurations. Their mass differences are determined entirely by the inter-cluster hyperfine interaction introduced in Eq.~\eqref{Eq1}. The explicit expressions for the corresponding tetraquark masses ($M_T$) are therefore expressed, as shown in Table~\ref{tetra_mass}, where $b_{D_1 \bar{D}_2}$ denotes the effective hyperfine coupling between the diquark and antidiquark.
\begin{table}[h!]
\centering
\setlength{\tabcolsep}{10pt}
\caption{Tetraquark mass expressions and $J^P$ quantum numbers.}
\label{tetra_mass}
\begin{tabular}{ccl}
\toprule
$S_{D_1} + S_{\bar D_2}$ & $J^P$ & Tetraquark mass ($M_T$)\footnote{The spin-spin co-efficients ($\bm{S}_{D_1}\!\cdot\!\bm{S}_{\bar{D}_2}$) for the inter-cluster hyperfine interactions are denoted by $c_J$ in our subsequent discussions.} \\
\midrule
$0 + 0$ & $0^+$ & $(m_{D_1})_{0^+} + (m_{\bar{D}_2})_{0^+}$ \\
\midrule
$0 + 1$ & $1^+$ & $(m_{D_1})_{0^+} + (m_{\bar{D}_2})_{1^+}$ \\
\midrule
$1 + 0$ & $1^+$ & $(m_{D_1})_{1^+} + (m_{\bar{D}_2})_{0^+}$ \\
\midrule
\multirow{3}{*}{$1 + 1$} & $0^+$ & $(m_{D_1})_{1^+} + (m_{\bar{D}_2})_{1^+} - 2b_{D_1 \bar{D}_2}$ \\
                         & $1^+$ & $(m_{D_1})_{1^+} + (m_{\bar{D}_2})_{1^+} - b_{D_1 \bar{D}_2}$ \\
                         & $2^+$ & $(m_{D_1})_{1^+} + (m_{\bar{D}_2})_{1^+} + b_{D_1 \bar{D}_2}$ \\
\bottomrule
\end{tabular}
\end{table}
Consequently, chromomagnetic mass splittings arise exclusively from axial-vector diquark-antidiquark configurations, as tetraquark mass differences are governed entirely by the residual chromomagnetic interaction between the effective cluster spins.

As in the diquark subsystem, the strength of the inter-cluster interaction is controlled by the corresponding color factor of the OGE operator. This factor can be evaluated directly from the quadratic Casimir relation given in Eq.~\eqref{Casimir}. For a color-singlet diquark--antidiquark system, the resulting inter-cluster color coefficients\footnote{During the finalization of this work, we noted that similar color-factor arguments were discussed in Ref.~\cite{Wang:2025sic}.} are
\begin{equation} \label{DD3}
\big\langle\vec{\lambda}_{D_1}\cdot\vec{\lambda}_{\bar{D}_2}\big\rangle_{1}
= -\frac{4}{3},
\qquad
(D_1\in\overline{\mb3},\ \bar D_2\in \mb3),
\end{equation}
\begin{equation}\label{DD6}
\big\langle\vec{\lambda}_{D_1}\cdot\vec{\lambda}_{\bar{D}_2}\big\rangle_{1}
= -\frac{10}{3},
\qquad
(D_1\in \mb6,\ \bar D_2\in \overline{\mb6}).
\end{equation}
These coefficients determine the magnitude of the residual chromomagnetic interaction between the effective clusters and thereby control the hyperfine mass splittings among tetraquark states. Therefore, the OGE-motivated diquark-antidiquark hyperfine interaction terms for $\overline{\mb3}\mb3$ and $\mb6\overline{\mb6}$ configurations are given by\footnote{Note that for brevity, we follow $m_{D_1^{\overline{3}}}\equiv m_{\{ij\}}^{(\overline{3},\,1)}$ and $m_{D_1^{6}}\equiv m_{[ij]}^{(6,\,1)}$ notation  for the remainder of the work.} 
\begin{equation}
\label{Eq10}
	b_{D_1 \bar{D}_2}^{(\overline{3}3)} =\frac{32\pi\alpha_s}{9(m_{D_1^{\overline{3}}})\,(m_{\bar {D}_2^{3}})}\langle\psi|\delta^3(\vec{r}_{ij})|\psi\rangle,
\end{equation}
\begin{equation}
\label{Eq11}
	b_{D_1 \bar{D}_2}^{(6\overline{6})} =\frac{80\pi\alpha_s}{9(m_{D_1^{6}})\,(m_{\bar {D}_2^{\overline{6}}})}\langle\psi|\delta^3(\vec{r}_{ij})|\psi\rangle.
\end{equation}
The diquark-antidiquark hyperfine interaction follows directly from the quark-quark OGE operator. At the effective level, the quark masses appearing in the quark-level hyperfine term are replaced by the corresponding effective diquark and antidiquark masses, and the quark-quark relative coordinate is replaced by the relative diquark-antidiquark coordinate entering the tetraquark wave function. The short-distance color-spin interaction therefore remains governed by the same underlying OGE dynamics. Its overall strength is fixed from baryon spectroscopy, where the color-magnetic spin-spin interaction provides a well-established description of hyperfine splittings, see Eq.~\eqref{Eq4}.

The residual diquark-antidiquark chromomagnetic scale is fixed independently from meson spectroscopy. Vector-pseudoscalar mass splittings arise from the color-magnetic contact interaction in a well-defined two-body color-singlet configuration, probing the same resolved short-distance spin-spin dynamics that determine the inter-cluster hyperfine coefficient $b_{D_1\bar{D}_2}$ in Eqs.~\eqref{Eq10} and~\eqref{Eq11}. This determination is independent of the effective diquark masses already fixed from baryon spectroscopy, which carry the internal intra-diquark chromomagnetic contributions.\footnote{Because heavy and heavy-light mesons consist of a single quark-antiquark pair in a color-singlet configuration, their vector-pseudoscalar splittings directly reflect the chromomagnetic interaction at the resolved scale. This same interaction governs the residual diquark-antidiquark hyperfine coupling, while the effective diquark masses remain those fixed from baryons and therefore encode the internal chromomagnetic contributions.} The calibration hierarchy is thus complete: baryon spectroscopy fixes the diquark masses; meson spectroscopy fixes the inter-cluster chromomagnetic scale; no free parameter enters the exotic sector.

Within the QDEMF, diquarks are treated as composite effective degrees of freedom. Their effective masses encode the dominant internal quark-level dynamics, in particular the short-distance color-spin correlations generated by OGE, and provide the leading mean-field input to the multiquark mass formula, Eq.~\eqref{Eq1}. This construction preserves the color and spin structure dictated by QCD while yielding a systematically controlled reduction of the many-body problem. All short-distance dynamics relevant for tetraquark spectroscopy are therefore encoded in the effective coefficient $b_{D_1\bar{D}_2}$ appearing in Eqs.~\eqref{Eq10} and~\eqref{Eq11}.

The replacement of the quark-quark relative coordinate $\vec{r}_{ij}$ by the inter-cluster coordinate $\vec{r}_{D\bar{D}}$ reflects a change of effective degrees of freedom rather than the introduction of a new spatial scale. The associated contact matrix element $\langle\delta^3(\vec{r}_{D\bar{D}})\rangle$ probes only the short-distance limit of the interaction and should not be interpreted as a measure of the overall spatial extent of the tetraquark system; although a diquark-antidiquark configuration may be spatially extended on average, the hyperfine interaction is sensitive only to the short-distance region resolved by the underlying quark-quark OGE dynamics. Within this hierarchical construction, the short-distance color-spin dynamics fixed in the conventional hadron sector propagate directly to the exotic sector: the residual diquark-antidiquark interaction inherits the same resolved short-distance scale without introducing new dynamical input. Accordingly, the product $\alpha_s\langle\delta^3(\vec{r})\rangle$ entering the effective coefficient $b_{D_1\bar{D}_2}$ is not treated as an independent quantity, but is consistently inherited from the same chromomagnetic interaction that governs hyperfine splittings in ordinary hadrons.

In this way, the extension from conventional hadrons to exotic multiquark systems preserves the same QCD interaction hierarchy. All short-distance quantities entering the tetraquark mass expression, Eq.~\eqref{Eq1}, are fixed \emph{a priori} from baryon and meson spectroscopy, while the tetraquark sector itself serves solely as a test of the internal consistency and predictive structure of the effective mass construction.

\subsection {Extension to Pentaquarks: Diquark-diquark-antiquark model}\label{Sec2B}
\begin{figure}[]
    \centering
    \includegraphics[width=0.4\linewidth]{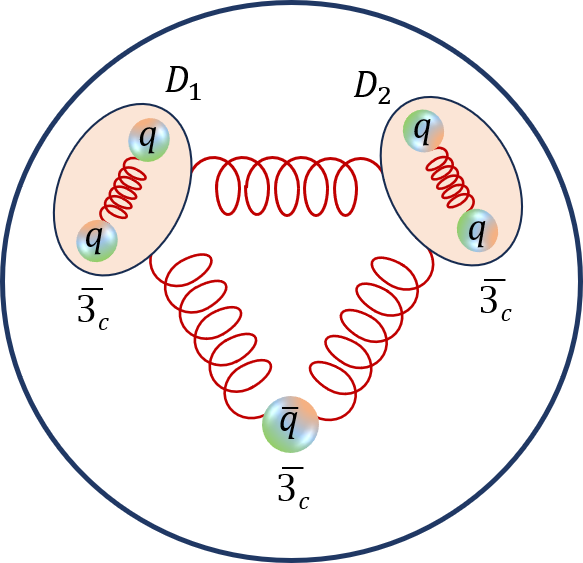}
    \caption{Diquark-diquark-antiquark interaction picture of a pentaquark.}
    \label{Fig2}
\end{figure}

The diquark-based effective framework extends systematically from tetraquarks to pentaquarks, with hidden-charm systems representing a special case motivated by experiment, treated as effective three-body bound states of two correlated diquarks and a heavy antiquark. In close analogy with the baryon case, the dominant short-distance quark-quark interactions arising from OGE are absorbed into the diquark substructures. Their effective masses and internal color-spin correlations are fixed independently from conventional baryon spectroscopy, thereby encoding the leading-order quark-level dynamics. The pentaquark is then described in terms of residual interactions among these effective degrees of freedom, yielding a transparent and systematic reduction of the underlying five-body problem. A schematic representation of the pentaquark as a three-body system, showing the diquark-diquark and diquark-antiquark interactions, is illustrated in Fig.~\ref{Fig2}.  

Within the diquark-diquark-antiquark framework for pentaquarks, the residual interactions among the two diquarks and the antiquark are implemented through cluster-level color-magnetic couplings that retain the operator structure of the underlying OGE interaction, in direct analogy with the tetraquark construction. No additional genuine three-body short-distance operators are introduced. This truncation is justified within the QDEMF by the hierarchy of short-distance contributions: pairwise color–spin interactions dominate, while higher-order multiquark operators are dynamically suppressed at the resolved scale and are effectively absorbed into the baryon-calibrated diquark masses and hyperfine coefficients. This preserves the QDEMF hierarchy and provides a systematically controlled and internally consistent description of pentaquark states without introducing additional free parameters, enabling an expansion consistent with leading-order QCD-inspired parametrizations validated in baryon and tetraquark systems.

Guided by this effective three-body picture, the admissible pentaquark configurations are determined by the requirement that the diquark-diquark-antiquark system combines into an overall color singlet, which fixes the underlying color structure of the state. Therefore, the color structure of a pentaquark in the diquark-diquark-antiquark picture can be expressed as \cite{Wang:2025sic, Chen:2022asf}
\begin{equation*}
\begin{split}
	\mb3 \otimes \mb3 \otimes \mb3 \otimes \mb3 \otimes \bar{\mb3} &= (\bar{\mb3} \oplus \mb6) \otimes (\bar{\mb3} \oplus \mb6) \otimes \bar{\mb3}\\
	&=(\bar{\mb3} \otimes \bar{\mb3} \otimes \bar{\mb3}) \oplus (\bar{\mb3} \otimes \mb6 \otimes \bar{\mb3}) \oplus (\mb6 \otimes \bar{\mb3} \otimes \bar{\mb3}).
\end{split}
\end{equation*}  
In this decomposition, we retain only those combinations that ultimately project onto a color-singlet pentaquark state. Among the allowed singlet configurations, particular emphasis is placed on the $(\bar{\mb3} \otimes \bar{\mb3} \otimes \bar{\mb3})$ channel, as it is expected to generate the energetically most favorable states. The remaining configurations involving color-sextet diquarks are systematically heavier due to the repulsive nature of their color-spin couplings and are therefore deferred to future work until further experimental data are available. To make the underlying group-theoretical structure explicit, the color wave function relevant to the $(\bar{\mb3} \otimes \bar{\mb3} \otimes \bar{\mb3})$ configuration, together with the associated spin wave functions, is presented in Appendix~\ref{AppB}. The complete color-spin-flavor bases employed in the present analysis are also included therein.

Within this construction, the pentaquark is modeled as a diquark-diquark-antiquark configuration in which the dominant color-magnetic interactions are hierarchically organized. As in the tetraquark formulation, the leading quark-quark correlations generated by OGE are absorbed into the effective diquark masses, thereby avoiding an explicit treatment of all quark-quark and quark-antiquark interactions that would otherwise require extensive color-spin diagonalization and additional phenomenological inputs. The residual dynamics is then governed by effective diquark-diquark and diquark-antiquark hyperfine interactions, consistent with the color structure dictated by QCD. Conceptually, this description parallels the three-body treatment of baryons, with the pentaquark mass decomposed into contributions from the two diquarks, the antiquark, and their residual color-magnetic couplings. The resulting pentaquark mass expression takes the general form given by
\begin{equation}\label{Eq12}
    M_P = m_{D_1} + m_{D_2} + m_{\bar q} + b_{D_1 D_2} (\bm{S_{D_1}.S_{D_2}}) + b_{D_1 \bar q} (\bm{S_{D_1}.S_{\bar q}}) + b_{D_2 \bar q} (\bm{S_{D_2}.S_{\bar q}}),
\end{equation}
where $m_{D_1}(m_{D_2})$ and $m_{\bar q}$ correspond to the effective constituent masses of the diquarks with flavor contents $ij(kl)$ and the antiquark, $\bar q$, respectively, while $b_{D_1D_2}$, $b_{D_1\bar q}$, and $b_{D_2\bar q}$ represent the corresponding effective hyperfine couplings inherited from the underlying OGE dynamics. This framework extends the diquark-based effective mass formalism used for tetraquarks to the pentaquark sector, with the essential distinction that the residual interactions now involve both diquark-diquark and diquark-antiquark pairs.

The diquark configurations entering the pentaquark system follow the same structural principles established in Sec.~\ref{Sec2A}. The effective diquark masses, including intra-diquark hyperfine effects, with the associated spin-independent color-Coulomb binding contributions, are treated consistently within the same diquark-level prescription used in the tetraquark sector. Consequently, all short-distance color dynamics are consistently absorbed into the effective diquark building blocks employed in the multiquark spectrum. Therefore, the residual spin couplings among the two diquarks and the antiquark generate ten negative-parity $S$-wave pentaquark states with total spin assignments $J^P=\frac{1}{2}^-$, $\frac{3}{2}^-$, and $\frac{5}{2}^-$. The explicit mass expressions for these configurations are summarized in Table~\ref{penta_mass}.
\begin{table}[!h]
\centering
\setlength{\tabcolsep}{10pt}
\caption{Pentaquark mass expressions and $J^P$ quantum numbers.}
\label{penta_mass}
\begin{tabular}{cccl}
\toprule
$S_{D_1}+S_{D_2}+s_{\bar{q}}$ & $S_{D_1}+S_{D_2}$ & $J^P$ & Pentaquark mass ($M_P$) \\
\midrule
$0 + 0 + \frac{1}{2}$ & $0$ & $\frac{1}{2}^-$ & $(m_{D_1})_{0^+} + (m_{D_2})_{0^+} + m_{\bar{q}}$ \\
\midrule
\multirow{2}{*}{$0 + 1 + \frac{1}{2}$} & \multirow{2}{*}{1} & $\frac{1}{2}^-$ & $(m_{D_1})_{0^+} + (m_{D_2})_{1^+} + m_{\bar{q}} - b_{D_2 \bar{q}}$ \\
 & & $\frac{3}{2}^-$ & $(m_{D_1})_{0^+} + (m_{D_2})_{1^+} + m_{\bar{q}} + \frac{1}{2} b_{D_2 \bar{q}}$ \\
\midrule
\multirow{2}{*}{$1 + 0 + \frac{1}{2}$} & \multirow{2}{*}{1} & $\frac{1}{2}^-$ & $(m_{D_1})_{1^+} + (m_{D_2})_{0^+} + m_{\bar{q}} - b_{D_1 \bar{q}}$ \\
 & & $\frac{3}{2}^-$ & $(m_{D_1})_{1^+} + (m_{D_2})_{0^+} + m_{\bar{q}} + \frac{1}{2} b_{D_1 \bar{q}}$ \\
\midrule
\multirow{5}{*}{$1 + 1 + \frac{1}{2}$} & 0 & $\frac{1}{2}^-$ & $(m_{D_1})_{1^+} + (m_{D_2})_{1^+} + m_{\bar{q}} - 2 b_{D_1 D_2}$ \\
\cmidrule{2-4}
 & \multirow{2}{*}{1} & $\frac{1}{2}^-$ & $(m_{D_1})_{1^+} + (m_{D_2})_{1^+} + m_{\bar{q}} - b_{D_1 D_2} - \frac{1}{2}(b_{D_1 \bar{q}}+b_{D_2 \bar{q}})$ \\
 & & $\frac{3}{2}^-$ & $(m_{D_1})_{1^+} + (m_{D_2})_{1^+} + m_{\bar{q}} - b_{D_1 D_2} + \frac{1}{4}(b_{D_1 \bar{q}}+b_{D_2 \bar{q}})$ \\
\cmidrule{2-4}
 & \multirow{2}{*}{2} & $\frac{3}{2}^-$ & $(m_{D_1})_{1^+} + (m_{D_2})_{1^+} + m_{\bar{q}} + b_{D_1 D_2} - \frac{3}{4}(b_{D_1 \bar{q}}+b_{D_2 \bar{q}})$ \\
 & & $\frac{5}{2}^-$ & $(m_{D_1})_{1^+} + (m_{D_2})_{1^+} + m_{\bar{q}} + b_{D_1 D_2} + \frac{1}{2}(b_{D_1 \bar{q}}+b_{D_2 \bar{q}})$ \\
\bottomrule
\end{tabular}
\end{table}
The effective hyperfine interactions between the two diquarks and between a diquark and the antiquark, parametrized by $b_{D_1D_2}$ and $b_{D_i\bar q}$, contribute only when the interacting clusters carry nonzero spin. Consequently, configurations involving scalar diquarks ($S_D=0$) do not generate inter-cluster hyperfine splittings, and the dominant mass splittings among pentaquark states arise from axial-vector diquark configurations and their couplings to the antiquark. This behavior mirrors the tetraquark case, where hyperfine structure is likewise driven by spin-carrying diquark constituents.

The color factors entering the effective diquark-diquark and diquark-antiquark OGE operators are obtained directly from quadratic Casimirs (given by Eq. \eqref{Casimir}), following the same prescription used for the diquark-antidiquark system in tetraquark. In the color-singlet projection of the full pentaquark wave function, the relevant inter-cluster color coefficients are fixed uniquely by the color representations of the interacting subsystems. These coefficients encode the net effect of the underlying quark-quark and quark-antiquark cross-couplings that would otherwise appear explicitly in a full five-body treatment. Accordingly, the effective hyperfine couplings governing the residual pentaquark dynamics can be written in the phenomenological form given by
\begin{equation}\label{penta_DD}
b_{D_1 D_2}^{(\bar 3\bar 3)} = \frac{16\pi\alpha_s}{9\,m_{D_1}\,m_{D_2}} \big\langle\psi\big|\delta^3(\vec r)\big|\psi\big\rangle,
\end{equation}
\begin{equation}\label{penta_Dq}
b_{D_i \bar q}^{(\bar 3\bar 3)} = \frac{16\pi\alpha_s}{9\,m_{D_i}\,m_{\bar q}} \big\langle\psi\big|\delta^3(\vec r)\big|\psi\big\rangle,
\end{equation}
where the appropriate color factors for both diquark-diquark and diquark-antiquark interactions determined from Casimir invariants are numerically equivalent to that of the quark-quark interaction in the antitriplet representation within a baryon (given by Eq. \eqref{CF}), \textit{i.e.}, 
\begin{equation} \label{Penta_casimir}
\vec\lambda_{D_1}.\vec\lambda_{D_2}=\vec\lambda_{D_i}.\vec\lambda_{\bar q}=-\frac{2}{3},
\end{equation}
and $\psi$ represents the spatial wave function of the effective three-body pentaquark system. 

Within the hierarchical QDEMF construction, all short-distance inputs are fixed a priori from conventional hadron spectroscopy, enabling a systematic extension from tetraquarks to pentaquarks.The diquark--antiquark hyperfine coupling $b_{D_i \bar q}$ is inherited directly from the quark-level OGE interaction calibrated in baryons, with its magnitude set by the same chromomagnetic scale that governs ordinary hyperfine splittings. The diquark-diquark coupling $b_{D_1 D_2}$ instead descends from the diquark-antidiquark interaction established in the tetraquark sector, itself anchored to vector--pseudoscalar meson splittings, where the color-singlet quark--antiquark configuration fixes the resolved contact scale.

Upon reorganizing quark coordinates into cluster degrees of freedom, the inter-cluster relative coordinate replaces the quark-level separation, reflecting a change of effective variables rather than the introduction of a new spatial scale. The contact matrix element $\langle \delta^3(\vec r) \rangle$ therefore probes only the short-distance correlations encoded in the underlying OGE dynamics and remains insensitive to the global size of the pentaquark system, preserving the internal consistency of the QDEMF hierarchy. In this sense, the short-distance contact scale is inherited rather than refitted, ensuring continuity across the baryon--tetraquark--pentaquark hierarchy.

In this framework, the numerical hierarchy follows directly from the common short-distance scale. They follow uniquely from (i) the color--Casimir algebra associated with different embeddings and (ii) the $1/(m_i m_j)$ scaling of the contact hyperfine interaction. 
Consequently,
\begin{align}
b^{(\bar{3}\bar{3})}_{D_1 D_2}
&= \tfrac{1}{2}\, b^{(\bar{3}3)}_{D_1\bar D_2}, 
\label{eq:DD_DAD_short} \\
b^{(\bar{3}\bar{3})}_{D^{\{qq\}}\bar c}
&= \tfrac{1}{2}\, b^{(\bar{3}3)}_{D^{\{qq\}}c}, 
\label{eq:DAq_Dq_short} \\
b^{(\bar{3}\bar{3})}_{D^{\{cq\}}\bar c}
&= \left(\frac{m_q}{m_{D^{\{cq\}}}}\right)
   b^{(33)}_{cq}. 
\label{eq:Dcbar_cq_short}
\end{align}

The $\tfrac{1}{2}$ factors encode the transition from $D\bar D$ or $Dq$ color embeddings to the $DD$ and $D\bar q$ configurations relevant for pentaquarks, while Eq.~\eqref{eq:Dcbar_cq_short} implements the expected inverse-mass suppression when a composite heavy--light diquark couples to a heavy antiquark. Since $b^{(\bar{3}3)}_{D^{\{qq\}}c}$ and $b^{(33)}_{cq}$ are already fixed by baryon calibration,\footnote{This refers to Scenario~II of QDEMF~\cite{Mohan:2026rcg}. Only a subset of the baryon-calibrated $\bar{\mathbf{3}}_c$ diquark mass inputs for heavy–heavy and light–light combinations are propagated here; the baryon framework is independently formulated.} the pentaquark sector introduces no additional free parameters.

This formulation makes explicit that the hyperfine splittings in pentaquarks are governed jointly by the spin content of the diquark clusters and their residual color-magnetic interactions with each other and with the antiquark. At the same time, it preserves full consistency with the tetraquark construction, differing only in the number and topology of the effective interacting subsystems. In the next section, we discuss the numerical results for the tetraquark and pentaquark states in detail.

\section{Numerical Results and Discussion}
\label{Sec3}

We present the mass spectra of heavy tetraquark and pentaquark states computed within the QDEMF. All short-distance parameters are inherited from baryon spectroscopy via the Scenario~II procedure of Ref.~\cite{Mohan:2026rcg}; no independent refitting is performed at the multiquark level. Constituent quark masses and OGE hyperfine couplings are extracted from measured baryon masses~\cite{ParticleDataGroup:2024cfk}. Scalar and axial-vector $\bar{\mathbf{3}}_c$ diquark masses follow from Eqs.~\eqref{Eq2} and~\eqref{Eq3}, with isospin breaking retained throughout; color-sextet $\mathbf{6}_c$ diquark masses are obtained by Casimir rescaling via Eqs.~\eqref{Eq8} and~\eqref{Eq9}. All diquark masses are collected in Tables~\ref{t1} and~\ref{t2}.

For the tetraquark sector, the inter-cluster hyperfine coupling is fixed from vector-pseudoscalar meson mass splittings, which probe the same resolved chromomagnetic scale that governs the residual diquark--antidiquark interaction; remaining $\bar{\mathbf{3}}\mathbf{3}$ couplings follow from flavor symmetry and $\mathbf{6}\bar{\mathbf{6}}$ couplings from the color-factor ratios of Eqs.~\eqref{DD3} and~\eqref{DD6}. The extracted inter-cluster parameters are listed in Tables~\ref{t3}--\ref{t_fullyheavy}. Subsequently, we compute tetraquark masses using Eq.~\eqref{Eq1} for all channels and both color configurations, following the mass relations in Table~\ref{tetra_mass}. In scalar–scalar and scalar–axial configurations, the masses reduce to the sum of the constituent diquark masses. For axial–axial configurations, an additional hyperfine contribution $c_J\,b_{D_1\bar{D}_2}$ appears, with $c_J = -2, -1, +1$ for $J^P = 0^+, 1^+, 2^+$, respectively. The complete $I(J^P)$-assigned spectra for singly heavy, doubly heavy, hidden-flavor, and fully heavy sectors are presented in Tables~\ref{t8}--\ref{t14_modified} and plotted against strong-decay thresholds in Figs.~\ref{Plot_SinglyCharm}--\ref{Plot_FullyHeavy}.

The pentaquark sector is treated in the diquark--diquark--antiquark picture using the same parameter set. Diquark--diquark and diquark--antiquark residual couplings are obtained from the baryon- and tetraquark-calibrated values via color-factor rescaling, introducing no additional free parameters. The resulting $P_{c\bar{c}}$ and $P_{c\bar{c}s}$ spectra are given in Tables~\ref{Penta1} and~\ref{Penta2} and displayed against open-charm thresholds in Fig.~\ref{Plot_Penta}.

\subsection{Diquark mass systematics: from baryons to exotics}
\label{Diquark_mass}

In Tables~\ref{t1} and~\ref{t2}, we list the diquark masses derived from baryon spectroscopy within the QDEMF~\cite{Mohan:2026rcg}. These inputs reproduce experimental baryon masses with a precision of $(1-3)\%$. In the heavy-flavor sectors, our previous baryon analysis in the QDEMF relied exclusively on dominant diquark configurations, specifically, light--light diquarks for singly heavy baryons and heavy--heavy configurations ($D^{cc}$, $D^{cb}$, $D^{bb}$) for doubly and triply heavy baryons. While heavy--light diquarks were omitted in that context, we consider those results a controlled baseline for our current extension to exotic states, where heavy--light diquarks emerge naturally as dictated by the specific quark content.

Rather than inferring heavy-light diquark mass parameters through symmetry (breaking) relations (Eq. \eqref{Eq4}) from heavy-heavy systems, we extract them directly from singly heavy baryons within the same QDEMF by isolating the heavy-light correlated subsystem in the baryon mass relations. We then employ these masses explicitly in our calculations for singly heavy and hidden-heavy exotic states. We adopt the heavy-heavy diquark masses from our recent analysis of heavy-flavor baryons~\cite{Mohan:2026rcg} in our doubly and fully heavy tetraquark systems. We observe that symmetry (breaking)-based estimates derived from heavy--heavy diquarks are approximately $20$~MeV lower than the corresponding values obtained from our heavy--light diquark calculations; this shift provides a useful estimate of the systematic sensitivity inherent in the heavy-flavor sector.

In the context of QDEMF baryons, light--light diquarks dominate the singly heavy baryon structure through strong scalar attraction in the $\bar{\mathbf{3}}_c$ channel. Conversely, heavy--heavy diquarks stabilize doubly and triply heavy systems via color-Coulomb binding and the approximate spin decoupling naturally favored by HQSS. In contrast to these conventional baryons, exotic multiquark configurations also admit the $\mathbf{6}_c$ diquark sector, which is otherwise absent due to color-singlet projection constraints.

The Pauli antisymmetry of the ground-state wave function restricts identical-flavor $S$-wave diquarks ($uu, dd, ss, cc, bb$) to either $\bar{\mathbf{3}}_c$ axial-vector or $\mathbf{6}_c$ scalar configurations. For non-identical flavor diquarks, both spin states are allowed; however, the mass ordering between the two color channels is reversed. This inversion stems from the opposite signs of the color-Casimir factors in the OGE hyperfine interaction, as defined in Eq.~\eqref{CF}. In the $\bar{\mathbf{3}}_c$ channel, the attractive color factor ($\langle\vec{\lambda}_i \cdot \vec{\lambda}_j\rangle = -2/3$) yields the hierarchy $m_D(0^+) < m_D(1^+)$, whereas in the $\mathbf{6}_c$ channel, the repulsive factor ($+1/3$) reverses this ordering such that $m_D(1^+) < m_D(0^+)$. We illustrate this pattern using the $ud$ system, where $m_{\bar{3}}(0^+) \simeq 625.6$~MeV~$ < m_{\bar{3}}(1^+) \simeq 732.9$~MeV, while in the sextet sector, $m_{6}(1^+) \simeq 692.7$~MeV~$ < m_{6}(0^+) \simeq 746.4$~MeV.

The mass ratio $R_{\bar{3}} = m_D(1^+)/m_D(0^+)$ provides a scale-independent measure of heavy-quark spin decoupling. In the $\bar{\mathbf{3}}_c$ channel, we find that the light sector ($ud, us, ds$) yields $R_{\bar{3}} \sim 1.17$--$1.27$ with absolute splittings $\Delta M_{\bar{3}} \sim 107$--$198$~MeV. In the charm--light sector, these values shift to
\begin{equation*}
\Delta M_{\bar{3}}(uc, dc, sc) \simeq 144\text{--}146~\text{MeV}, \qquad R_{\bar{3}} \simeq 1.07,
\end{equation*}
while in the bottom--light sector, we observe
\begin{equation*}
\Delta M_{\bar{3}}(ub, db, sb) \simeq 136\text{--}148~\text{MeV}, \qquad R_{\bar{3}} \simeq 1.03.
\end{equation*}
This trend indicates progressive spin decoupling despite only a moderate reduction in the absolute splittings. Similarly, the $\mathbf{6}_c$ channel exhibits an analogous trend with approximately half the splitting magnitudes (e.g., $\Delta M_{6} \sim 68$--$74$~MeV and $R_{6} = m_D(0^+)/m_D(1^+) \simeq 1.01$ for bottom--light), consistent with the color-Casimir ratio in Eq.~\eqref{Eq6}. 

The definitive test of this decoupling occurs in the $cb$ diquark, where Pauli constraints allow both spin states in each color representation. We find
\begin{equation*}
\Delta M_{\bar{3}}(cb) \simeq 4.5~\text{MeV}, \quad \Delta M_{6}(cb) \simeq 2.2~\text{MeV}, \quad R_{\bar{3}} \approx R_{6} \simeq 1.0,
\end{equation*}
demonstrating near-complete heavy-quark spin decoupling. For identical-flavor heavy diquarks ($cc, bb$), the near-degeneracy $m_{\bar{3}}(1^+) \approx m_{6}(0^+)$ ($\lesssim 2$~MeV for $cc$ and $\lesssim 0.2$~MeV for $bb$) reflects the $1/(m_i m_j)$ suppression governing hyperfine dynamics. Remarkably, this progression emerges naturally within the baryon-calibrated QDEMF without the external imposition of HQSS constraints.

We find that the mass differences between $\bar{\mathbf{3}}_c$ and $\mathbf{6}_c$ diquarks at fixed spin further reveal a characteristic hierarchy governed primarily by hyperfine dynamics. For scalar diquarks, the color-representation mass shift,
\[
\Delta_{\rm color}^{0^+} = m_{6}(0^+) - m_{\bar{3}}(0^+),
\]
ranges from $\sim 121$--$223$~MeV in the light sector, stabilizes at $\sim 160$~MeV in heavy--light systems, and collapses to $\sim 5$~MeV for the $cb$ case. This hierarchy reflects the significantly stronger hyperfine attraction inherent to the $\bar{\mathbf{3}}_c$ scalar channel. Conversely, for axial-vector diquarks, the ordering reverses:
\[
\Delta_{\rm color}^{1^+} = m_{\bar{3}}(1^+) - m_{6}(1^+) \approx 40\text{--}74~\text{MeV}
\]
in the light sector and $\sim 50$--$55$~MeV in heavy--light systems, eventually collapsing to $\sim 1.7$~MeV in the heavy--heavy limit. This reversal follows directly from the opposite hyperfine shifts of axial states in the two color representations. The scalar shifts remain systematically larger than the axial ones, reflecting the greater hyperfine leverage in the spin-zero channel. We propose that these diquark mass hierarchies serve as critical templates for the identification of exotic multiquark spectra, where the relative positions of $J^P$ states will provide a direct experimental probe of the underlying color-hyperfine structure.

\subsection{Inter-cluster hyperfine dynamics in exotics} \label{inter_cluster}

We employ the same OGE chromomagnetic operator that generates the intra-diquark splittings in Tables~\ref{t1} and \ref{t2} to govern the inter-cluster terms. The interaction parameters listed in Tables~\ref{t3}--\ref{t_fullyheavy} encode the residual color--spin interaction between diquark and antidiquark clusters across all tetraquark sectors, with the corresponding symmetry-breaking relations provided in the first column. We remark that within the QDEMF construction, short-distance correlations are absorbed into the baryon-calibrated diquark masses, while the remaining OGE term produces only the residual inter-cluster hyperfine structure. This separation is crucial as it prevents the double counting of quark-level correlations and ensures a consistent continuity between conventional hadrons and exotic multiquark systems.

The inter-cluster hyperfine coupling is first estimated from the vector-pseudoscalar meson splitting, yielding $b^{\rm (meson)} \simeq 36.0$~MeV from the $D^{*0}$--$D^0$ hyperfine splitting in the $D^{cu}D^{\bar{d}\bar{s}}$ channel (Sec.~\ref{Sec2A}). Interestingly, the exotic state $T^{*}_{cs0}(2900)^{++}$ with mass $(2921\pm 26)$~MeV provides an independent determination $b^{\rm (expt)} = 48.01$~MeV directly from the inter-cluster dynamics of the exotic system itself. The ${\sim}\,12$~MeV discrepancy between the two estimates is consistent with the expected correction from diquark compositeness and confirms the predictive consistency of the chromomagnetic projection. Therefore, without introducing additional uncertainties, we adopt the experimentally extracted value as the primary anchor for Tables~\ref{t3}--\ref{t_fullyheavy} in the following analysis, as it directly encodes the full inter-cluster dynamics unique to the exotic system.

We note that only axial--axial configurations carry a $b$-dependent correction; scalar--scalar and scalar--axial states are constituent sums and are therefore insensitive to the calibration choice. We observe that switching from $b^{\rm (expt)}$ to $b^{\rm (meson)}$ shifts axial--axial masses by at most $\lesssim 30$~MeV in the $\mathbf{6} \otimes \bar{\mathbf{6}}$ sector and $\lesssim 10$~MeV in the $\bar{\mathbf{3}} \otimes \mathbf{3}$ sector, both of which remain below the experimental anchor uncertainty. Notably, the scalar--axial channels, which contain the lightest $\bar{\mathbf{3}} \otimes \mathbf{3}$ candidates, remain unaffected. Consequently, the two calibrations yield effectively identical spectra at current levels of experimental precision. Propagating the anchor uncertainty $\delta b \lesssim 26$~MeV yields sub-percent shifts in total tetraquark masses and $\sim 20$--$30$~MeV uncertainties in intra-multiplet splittings, which is comparable to current experimental resolution.

In the $\bar{\mathbf{3}}\otimes\mathbf{3}$ sector, we observe that the interaction strength decreases monotonically with increasing the number of strange quarks in antidiquark. As shown in Table~\ref{t3}, the couplings are
\begin{equation}
b^{(\bar{3}3)}\simeq 61~{\rm MeV},\quad
b^{(\bar{3}3)}\simeq 48~{\rm MeV},\quad
b^{(\bar{3}3)}\simeq 42~{\rm MeV},
\end{equation}
for strangeness$=0,1$ and $2$, respectively, representing a $\sim 20\%$ reduction per strange quark. This result follows directly from the $1/(m_{D_1}m_{\bar {D_2}})$ structure of the OGE contact term, where flavor-symmetry breaking enters solely through the diquark masses without requiring additional interaction parameters. For a fixed antidiquark flavor, we find
\begin{equation}
\frac{b(D^{cu})}{b(D^{cs})}\simeq 1.09\,,
\end{equation}
demonstrating that flavor-breaking effects scale according to the inverse diquark mass product. Isospin breaking ($cu \leftrightarrow cd$) remains negligible at $< 0.1\%$, consistent with the near-degeneracy of $u$ and $d$ effective masses.

At matched flavor content, we find that the color-sector ratio
\begin{equation}
\frac{b^{(6\bar{6})}}{b^{(\bar{3}3)}} \simeq 2.75
\end{equation}
is remarkably stable across all singly charm channels. In the limit of degenerate effective diquark masses, this ratio reduces toward the pure inter-cluster Casimir value ($10/4$). The observed enhancement arises entirely from diquark mass hierarchies: since $\bar{\mathbf{3}}_c$ axial diquarks are systematically heavier than their $\mathbf{6}_c$ counterparts, the inverse-mass weighting generates a multiplicative correction:
\begin{equation} \label{Sextet_triplet}
\frac{b^{(6\bar{6})}}{b^{(\bar{3}3)}}
= \underbrace{\frac{5}{2}}_{\;\text{Casimir ratio}}
\;\times\;
\underbrace{\frac{m_{D_1^{\bar{3}}}\,m_{\bar D_2^{3}}}
{m_{D_1^{6}}\,m_{\bar D_2^{\bar 6}}}}_{\approx\,1.10\;
(\text{Mass hierarchy contribution})}
\;\simeq\; 2.75\,.
\end{equation}
The near-constancy of this ratio across flavor columns (Table \ref{t3}) indicates that the enhancement is fully determined by the internal diquark structure and introduces no additional model dependence. Furthermore, the absolute interaction scales,
\begin{equation}
b^{(\bar{3}3)} \sim 38\text{--}62~{\rm MeV},\qquad
b^{(6\bar{6})} \sim 122\text{--}167~{\rm MeV},
\end{equation}
imply a cross-sector separation $\Delta b \sim 100$~MeV at fixed flavor. This gap exceeds intra-sector flavor variations and, combined with the constituent-mass offsets discussed in Sec.~\ref{Diquark_mass}, reinforces the classification of $\bar{\mathbf{3}}\otimes\mathbf{3}$ and $\mathbf{6}\otimes\bar{\mathbf{6}}$ tetraquarks as distinct spectroscopic multiplets.

We find that the singly bottom (Table~\ref{t4}), doubly heavy (Table~\ref{t5}), hidden-heavy (Table~\ref{t6}), and fully heavy (Table~\ref{t_fullyheavy}) sectors exhibit the same qualitative patterns: strangeness suppression, near isospin degeneracy, and mass-driven flavor breaking. The extracted chromomagnetic couplings are summarized below:
\begin{center}
\begin{tabular}{lcc}
\hline\hline
Sector & ~~~$b^{(\bar{3}3)}$ (MeV)~~~& $b^{(6\bar{6})}$ (MeV) \\
\hline
Singly charm & $38$--$62$  & $122$--$167$ \\
Singly bottom & $16$--$24$  & $49$--$63$   \\
Doubly heavy ($cb$) & $13$--$19$ & $40$--$51$ \\
Hidden $c\bar c$ & $18$--$22$ & $48$--$57$ \\
Hidden $b\bar b$ & $3.0$--$3.2$ & $7.7$--$8.2$ \\
Fully heavy & $0.9$--$8.3$ & $5.2$ \\
\hline\hline
\end{tabular}
\end{center}
This uniform suppression reflects the $1/(m_{D_1}m_{\bar {D_2}})$ scaling of the chromomagnetic interaction and represents an inter-cluster manifestation of HQSS emerging dynamically within our framework. The charm-to-bottom suppression factor, $b_{\rm bottom}/b_{\rm charm} \simeq 0.4$, indicates that the dominant control parameter is the diquark mass scale rather than detailed flavor composition. Simultaneously, the color-sector ratio converges toward the Casimir limit as the mass increases:
\begin{equation}
\frac{b^{(6\bar{6})}}{b^{(\bar{3}3)}} \;:\quad
\underset{\text{singly }c}{2.75}\;\to\;
\underset{\text{singly }b}{2.70}\;\to\;
\underset{\text{doubly heavy}}{2.69}\;\to\;
\underset{\text{hidden }c\bar{c}}{2.64}\;\to\;
\underset{\text{hidden }b\bar{b}}{2.55}\;\to\;
\underset{\text{Casimir}}{2.50}\,.
\label{eq:convergence}
\end{equation}
This reflects the shrinking $\bar{\mathbf{3}}_c$--$\mathbf{6}_c$ diquark mass splitting at higher scales. In the heavy-mass limit, the hyperfine interaction becomes insensitive to the internal diquark structure, and the ratio approaches the universal color--Casimir value of $2.5$. 

We observe that the cross-sector gap $\Delta b$ systematically exceeds intra-sector flavor spreads across all configurations, maintaining $\sim 40$~MeV in the singly bottom and $\sim 5$~MeV in the hidden $b\bar{b}$ sectors. This hierarchy preserves the classification of $\bar{\mathbf{3}}\otimes\mathbf{3}$ and $\mathbf{6}\otimes\bar{\mathbf{6}}$ as distinct, non-mixing spectroscopic classes. However, the absolute $\bar{\mathbf{3}}\otimes\mathbf{3}$ splittings in $bb$ and hidden $b\bar{b}$ systems collapse to $\lesssim 13$~MeV and $\sim 3$~MeV, respectively. This extreme near-degeneracy, a rigorous consequence of $1/(m_{D_1}m_{\bar {D_2}})$ scaling, suggests that these $J^P$ multiplets will appear as overlapping resonances likely to challenge current experimental resolution.

Table~\ref{t7} summarizes the binding energies incorporated into the tetraquark mass formula. We find that mesonic ($\mathbf{1}_c$) binding is uniformly attractive and scales rapidly with the heavy-quark mass, reaching $\sim -570$~MeV for $b\bar{b}$. The $\bar{\mathbf{3}}_c$ diquark channel follows the expected Casimir scaling, yielding approximately half the mesonic attraction (e.g., $BE_{\bar{3}}(bb) \simeq -285$~MeV). Conversely, the sextet channel is universally repulsive, with $BE_{6}(bb) \simeq +143$~MeV. These opposite signs provide a significant color-separation scale that acts independently of constituent mass and chromomagnetic effects.

A characteristic feature of the QDEMF is the evolving hierarchy of residual interactions. In fully heavy systems, the diquark binding reaches $\mathcal{O}(10^2)$~MeV, thereby dominating the typical chromomagnetic splittings. In contrast, for heavy--light systems (e.g., $cs$), the binding ($\sim 40$~MeV) remains comparable to the hyperfine contributions. This behavior establishes a smooth crossover from hyperfine-dominated dynamics at lower mass scales to binding-dominated spectroscopy in the heavy-quark limit. We extract these values from meson hyperfine splittings \cite{ParticleDataGroup:2024cfk}, with the exception of $BE(\bar{b}c)$, for which we utilize the lattice input $M_{B_c^{*+}} = 6331$~MeV \cite{Mathur:2018epb}.

Since tetraquarks can occur in both $\bar{\mathbf{3}}\otimes\mathbf{3}$ and $\mathbf{6}\otimes\bar{\mathbf{6}}$ configurations, the combined effects of constituent quark masses, chromomagnetic interactions, and binding energies give rise to different spectroscopic orderings:
\begin{enumerate}
    \item In the scalar diquark sector, the sextet constituent mass shift ($\sim 120$--$220$~MeV per diquark in light/heavy--light systems) induces a baseline $\sim 250$--$400$~MeV color-sector separation. Binding reinforces this ordering through opposite signs: in the $bb$ system, the difference between attractive $\bar{\mathbf{3}}_c$ and repulsive $\mathbf{6}_c$ binding alone reaches $\sim 430$~MeV. Scalar $\mathbf{6}\otimes\bar{\mathbf{6}}$ tetraquarks are therefore shifted several hundred MeV above their $\bar{\mathbf{3}}\otimes\mathbf{3}$ counterparts, precluding significant spectral overlap.
    \item For axial diquarks, the constituent mass hierarchy is reduced ($\lesssim 50$~MeV), making inter-cluster dynamics decisive. The stronger chromomagnetic couplings of the $\mathbf{6}\otimes\bar{\mathbf{6}}$ sector compete against its repulsive binding, while $\bar{\mathbf{3}}\otimes\mathbf{3}$ states benefit from attractive binding despite weaker hyperfine effects. This reduces the color-sector separation to $\sim 100$--$250$~MeV, increasing model sensitivity and the potential for residual mixing in axial channels.
    \item With increasing heavy-quark mass, the rapid growth of $|BE_{\bar{3}}|$ drives binding-dominated spectroscopy. In the $bb$ sector, the $\sim 300$~MeV diquark binding overwhelms the $\mathcal{O}(10^1)$~MeV chromomagnetic splittings, compressing spin multiplets while preserving a sizable inter-sector offset.
\end{enumerate}

These hierarchies yield a characteristic QDEMF pattern: well-separated color multiplets in light and heavy--light sectors, persistent separations in mixed heavy-flavor systems, and compressed spin splittings in fully heavy exotics. We argue that the combined effects of Casimir scaling, mass-enhanced binding, and the repulsive nature of the $\mathbf{6}_c$ channel stabilize the interpretation of $\bar{\mathbf{3}}\otimes\mathbf{3}$ and $\mathbf{6}\otimes\bar{\mathbf{6}}$ tetraquarks as distinct spectroscopic families. The diquark mass parameters in Tables~\ref{t1} and~\ref{t2} thus provide a QCD-coherent and HQSS-consistent foundation for the exotic spectra developed in the following sections.

In the pentaquarks, the entries of Table~\ref{t_penta_params} follow directly from Eqs.~\eqref{eq:DD_DAD_short}–\eqref{eq:Dcbar_cq_short} without any retuning. The effective diquark masses absorb intra-diquark hyperfine contributions through baryon calibration; only inter-cluster terms, diquark–diquark and diquark–antiquark, are retained. With identical Casimir factors governing both diquark–diquark and diquark–antiquark interactions, the chromomagnetic couplings in the pentaquark sector are controlled entirely by inverse effective-mass scaling. The diquark–antiquark couplings $b_{D_i\bar{c}}$ are inherited from baryon calibrations,\footnote{The subsystem $D^{\{cq\}}\bar c$ has no direct analogue in conventional baryons. Its coupling is therefore fixed by matching to the heavy–light quark–quark interaction $b^{(33)}_{cq}$ extracted from baryon spectroscopy, with the appropriate inverse-mass rescaling to account for the composite diquark. The underlying chromomagnetic contact scale remains unchanged.} while the inter-diquark couplings $b_{D_1 D_2}$ are mapped from the tetraquark parameter extraction. The five effective couplings governing the spectrum are
$$
b_{\{cu\}\{ud\}} = 30.72,\quad
b_{\{cs\}\{ud\}} = 28.25,\quad
b_{\{ud\}\bar{c}} = 16.18,\quad
b_{\{cu\}\bar{c}} = 24.84,\quad
b_{\{cs\}\bar{c}} = 34.06~\text{MeV},
$$
with inter-diquark terms dominating the splittings between diquark-spin multiplets and diquark–antiquark couplings controlling the finer intra-multiplet structure. 

\subsection{Singly charm tetraquark states}
\label{Sec3A}

We summarize our predictions for the singly charm tetraquark sector in Tables~\ref{t8}--\ref{t10}. These spectra encompass states with heavy--light diquark content: $T(cn\bar{n}\bar{n})$ (Table~\ref{t8}), $T(cn\bar{n}\bar{s})$ and $T(cn\bar{s}\bar{s})$ (Table~\ref{t9}), and the $T(cs\bar{n}\bar{n})$, $T(cs\bar{n}\bar{s})$, and $T(cs\bar{s}\bar{s})$ systems (Table~\ref{t10}). We observe the following:

\begin{enumerate}
\item For a tetraquark $T(q_1 q_2\bar{q}_3\bar{q}_4)$ we define the color-sector gap at fixed $J^P$ as
\begin{equation}
\Delta(J)
= \Bigl[m^{\bar{3}}_{D}(S_D) - m^{6}_{D}(S_D^{\prime})\Bigr]
+ \Bigl[m^{3}_{\bar D}(S_{\bar D})
      - m^{\bar{6}}_{\bar D}(S_{\bar D}^{\prime})\Bigr]
+ \Bigl[c_J^{S}\,b^{(\bar{3}3)}
      - c_J^{S^{\prime}}\,b^{(6\bar{6})}\Bigr],
\label{eq:gap_general}
\end{equation}
where the first two terms encode the constituent-level color splittings of the diquark ($D$) and antidiquark ($\bar{D}$), while the last term captures the inter-cluster hyperfine difference; $c_J^{S}$ and $c_J^{S^{\prime}}$ denote the spin-coupling factors for $(S_D,S_{\bar D})\!\to\!J$ and $(S_D^{\prime},S_{\bar D}^{\prime})\!\to\!J$ in the $\bar{\mathbf{3}}\otimes\mathbf{3}$ and $\mathbf{6}\otimes\bar{\mathbf{6}}$ sectors, respectively. In the present QDEMF construction we retain distinct color sectors without mixing, so Eq.~\eqref{eq:gap_general} provides the organizing hierarchy across all flavor-asymmetric systems. For flavor-symmetric light antidiquarks ($\{\bar{u}\bar{d}\}$), the $\mathbf{6}\otimes\bar{\mathbf{6}}$ states lie consistently ${\sim}\,105$--$155$~MeV above their $\bar{\mathbf{3}}\otimes\mathbf{3}$ counterparts. In the $0^+$ channel, $m(\bar{\mathbf{3}}_1\,\mathbf{3}_1) = 2689.01$~MeV versus $m(\mathbf{6}_0\,\bar{\mathbf{6}}_0) = 2843.63$~MeV ($\Delta m \simeq +154.6$~MeV). Conversely, for flavor-antisymmetric antidiquarks ($[\bar{u}\bar{d}]$), the enhanced color factor of axial-vector sextet diquarks reverses this ordering: in the $0^+$ channel $m(\mathbf{6}_1\,\bar{\mathbf{6}}_1) = 2382.83$~MeV falls ${\sim}\,175.3$~MeV below the $\bar{\mathbf{3}}_0\,\mathbf{3}_0$ level, as the hyperfine lowering ($-2\,b^{(6\bar{6})}$) overcompensates for the sextet constituent mass. This inversion persists in the $1^+$ channel ($\Delta m \simeq -155$~MeV) before reversing again at $2^+$, as dictated by $c_J = +1$. The resulting $\bar{\mathbf{3}}$--$\mathbf{6}$ splitting at fixed $J^P$ (${\sim}\,130$--$175$~MeV) is comparatively model-independent and constitutes a robust experimental discriminator.

\item Within each color--flavor sector the axial--axial mass formula strictly implies $M(1^+) - M(0^+) = b$ and $M(2^+) - M(1^+) = 2b$. From Table~\ref{t8} we extract
\begin{alignat*}{3}
&\text{Antisymmetric }
  \mathbf{6}\otimes\bar{\mathbf{6}}:\quad
  &&b^{(6\bar{6})}
    = 2549.77 - 2382.83 &&= 166.94~\text{MeV},\\
&\text{Symmetric }
  \bar{\mathbf{3}}\otimes\mathbf{3}:\quad
  &&b^{(\bar{3}3)}
    = 2750.46 - 2689.01 &&= 61.45~\text{MeV},
\end{alignat*}
yielding $b^{(6\bar{6})}/b^{(\bar{3}3)} = 2.72$, consistent with the color-Casimir scaling of Eq.~\eqref{eq:convergence}. Isospin splittings remain a small perturbation ($\lesssim 1.5$~MeV), while each $\bar{n}\to\bar{s}$ substitution in the antidiquark raises the minimum mass by ${\sim}\,237$--$452$~MeV, reflecting the heavier strange-diquark mass.

\item All predicted states lie above their respective lowest meson--meson thresholds and are therefore subject to strong fall-apart decays. We identify three kinematic tiers based on proximity to the relevant pseudoscalar--vector and vector--vector thresholds, illustrated first for $cn\bar{n}\bar{n}$ and subsequently extended to $cn\bar{n}\bar{s}$, $cn\bar{s}\bar{s}$, $cs\bar{n}\bar{n}$, $cs\bar{n}\bar{s}$, and $cs\bar{s}\bar{s}$ (Fig.~\ref{Plot_SinglyCharm}):
\begin{enumerate}
    \item Sub-threshold tier ($m < 2638$~MeV in $cn\bar{n}\bar{n}$): The ground state at $2382.83$~MeV and the axial-vector state at $2549.77$~MeV fall below the $D\rho$ threshold, restricting their decays to $D\pi$ and $D^{*}\pi$, respectively; we expect moderate widths. In the strange-containing sectors the analogous lowest-lying states cluster between the pseudoscalar--vector and vector--vector thresholds (e.g.\ $D^{*}K$--$D^*K^{*}$ or $D_s^*K$--$D_s^*K^{*}$) rather than settling deeply below them.

    \item Near-threshold tier ($2638$--$2800$~MeV in $cn\bar{n}\bar{n}$): Several axial-vector states sit $(30$--$150)$~MeV above $D\rho$, where the simultaneous availability of $D\rho$ and $D^{*}\pi$ channels implies moderate-to-broad widths. Across all sectors, intermediate states accumulate within ${\sim}\,0$--$100$~MeV of the $DK^{*}$-, $D_sK^{*}$-, or $D^{*}K^{*}$-type thresholds. Where a state sits just below the vector-vector channel, as in the $cs\bar{n}\bar{n}$ and $cs\bar{n}\bar{s}$ spectra, kinematic suppression enhances the prospect of a relatively narrow structure sensitive to coupled-channel and threshold-cusp effects.

    \item Deeply unstable tier ($m > 2779$~MeV in $cn\bar{n}\bar{n}$): Scalar and tensor states above $D^{*}\rho$ gain access to vector--vector decay modes and correspondingly large phase space, with the notable exception of the $\bar{\mathbf{3}}_1\,\mathbf{3}_1$ scalar at $2689$~MeV, which remains below $D^{*}\rho$ and should decay dominantly through $D\pi$. In the strange-containing sectors, the majority of higher levels exceed the highest accessible thresholds by ${\sim}\,50$--$250$~MeV, opening multiple decay paths ($D^{(*)}K^{(*)}$, $D_s^{(*)}K^{(*)}$, $D_s^{(*)}\phi$) and producing broad resonance profiles, most prominently in $cs\bar{s}\bar{s}$, where every predicted state surpasses the $D_s^{(*)}\phi$ thresholds.
\end{enumerate}

\item Although the non-strange $cn\bar{n}\bar{n}$ sector lacks confirmed candidates, comparisons in the strange-partner sectors provide direct validation:
\begin{enumerate}
    \item In the $cd\bar{u}\bar{s}$ sector, the PDG scalar state at $2892 \pm 21$~MeV~\cite{ParticleDataGroup:2024cfk} is closely matched by our $(\bar{\mathbf{3}}_1\,\mathbf{3}_1,\,0^+)$ prediction at $2913.57$~MeV (${\sim}\,21$~MeV difference, well within the systematic uncertainty of our hyperfine scale).

    \item For the $T(cu\bar{d}\bar{s})$ scalar reported at $2921 \pm 26$~MeV~\cite{ParticleDataGroup:2024cfk}, our value of $2921.00$~MeV is reproduced by construction, since this state fixe the inter-cluster hyperfine interaction. Using instead the meson-calibrated coupling from Sec.~\ref{inter_cluster} ($b^{(\mathrm{meson})} \simeq 36.0$~MeV from the $D^{*0}\!-\!D^0$ splitting), the predicted mass shifts to $2946.33$~MeV, remaining within the expected $20$--$30$~MeV theoretical uncertainty. Interestingly, the framework both reproduces the observed mass and preserves stable predictions for the remaining states, indicating a substantial compact short-distance component, with threshold-driven shifts contributing only as subleading corrections.

    \item In the $cs\bar{n}\bar{n}$ sector, our closest prediction ($2841.41$~MeV) underestimates the PDG state at $2872 \pm 16$~MeV~\cite{ParticleDataGroup:2024cfk} by ${\sim}\,31$~MeV (${\sim}\,2\sigma$). While this may reflect residual coupled-channel effects, the agreement markedly improves upon NRPM~\cite{Liu:2022hbk} and RQM~\cite{Lu:2020qmp} studies, which overestimate the state by $200$--$400$~MeV.
\end{enumerate}

\item Our predicted $cn\bar{n}\bar{n}$ masses span $(2383$--$2885)$~MeV, whereas RQM calculations~\cite{Lu:2020qmp} yield $(2570$--$3327)$~MeV. The lower mean mass ($\langle m \rangle \simeq 2634$~MeV vs.\ $2949$~MeV) reflects the stronger effective color attraction inherent to the diquark-based construction, while the narrower bandwidth (${\sim}\,500$~MeV vs.~${\sim}\,750$~MeV) indicates reduced spread across $J^P$ channels. The same trend persists in the strange-containing sectors: our $cn\bar{n}\bar{s}$ masses $(2623$--$3059)$~MeV and $cn\bar{s}\bar{s}$ masses $(3072$--$3198)$~MeV remain systematically lower than the corresponding RQM~\cite{Lu:2020qmp} and NRPM~\cite{Liu:2022hbk} predictions. The $cs\bar{n}\bar{n}$, $cs\bar{n}\bar{s}$, and $cs\bar{s}\bar{s}$ sectors likewise exhibit a consistent downward mass shift and compressed spectral spacing relative to these approaches.
\end{enumerate}

The $cn\bar{n}\bar{n}$ spectrum thus organizes into three distinct bands (Fig.~\ref{Plot_SinglyCharm}): (i)~a low band ($2380$--$2560$~MeV) of maximally hyperfine-lowered states; (ii)~an intermediate region where color sectors interleave, requiring combined mass, spin--parity, and decay analyses for disentanglement; and (iii)~a high band (${\sim}\,2800$~MeV) dominated by constituent-mass penalties. These non-strange predictions, anchored by the $\mathcal{O}(10$--$30)$~MeV agreement achieved in the strange-partner sectors, remain key targets for future experimental searches.

\subsection{Singly bottom tetraquark states}
\label{Sec3B}

Tables~\ref{t11}--\ref{t13} summarize our mass predictions for singly bottom tetraquarks across both $\bar{\mathbf{3}}_c \otimes \mathbf{3}_c$ and $\mathbf{6}_c \otimes \bar{\mathbf{6}}_c$ color configurations. This subsection focuses on the quantitative shifts in the spectra induced by the $b$-quark mass scale, specifically examining how the heavy-light diquark dynamics evolve relative to the charm-sector benchmarks.

\begin{enumerate}
\item The color-sector gap defined in Eq.~\eqref{eq:gap_general} carries over to the bottom sector, but with a qualitatively different inversion pattern. For flavor-symmetric light antidiquarks ($\{\bar{u}\bar{d}\}$), the standard ordering $\bar{\mathbf{3}}\otimes\mathbf{3} < \mathbf{6}\otimes\bar{\mathbf{6}}$ persists. In the $0^+$ channel,
$$
m(\bar{\mathbf{3}}_1\,\mathbf{3}_1)=6087.59~\text{MeV},\qquad
m(\mathbf{6}_0\,\bar{\mathbf{6}}_0)=6165.30~\text{MeV},
$$
giving a modest gap of $\Delta m \simeq 78$~MeV. For flavor-antisymmetric antidiquarks ($[\bar{u}\bar{d}]$), however, the spectrum departs from the charm pattern: whereas the charm sector exhibited a full sextet inversion, the bottom ground states remain conventionally ordered. In the $0^+$ channel,
$$
m(\bar{\mathbf{3}}_0\,\mathbf{3}_0)=5891.55~\text{MeV},\qquad
m(\mathbf{6}_1\,\bar{\mathbf{6}}_1)=5917.33~\text{MeV},
$$
with a separation of only $26$~MeV, the enhanced sextet hyperfine attraction nearly compensates the constituent-mass shift. The inversion reappears only in the $1^+$ channel,
$$
m(\bar{\mathbf{3}}_0\,\mathbf{3}_0;\,1^+)=6027.55~\text{MeV}
\quad\text{vs.}\quad
m(\mathbf{6}_1\,\bar{\mathbf{6}}_1;\,1^+)=5980.48~\text{MeV},
$$
indicating that the inversion mechanism survives however is significantly weakened at the bottom scale.

\item The axial--axial splittings satisfy the QDEMF relation
$M(2^+)-M(1^+) \simeq 2\bigl(M(1^+)-M(0^+)\bigr)$, yielding:
$$
b^{(6\bar{6})} = 63.15~\text{MeV}, \qquad
b^{(\bar{3}3)} = 23.65~\text{MeV},
$$
with $b^{(6\bar{6})}/b^{(\bar{3}3)} = 2.67$, lower than the charm ($2.72$) value and confirming that this ratio is governed by color-Casimir factors rather than heavy-quark identity. The absolute hyperfine scale, by contrast, is strongly suppressed in the $c\!\to\!b$ transition:
$$
b^{(6\bar{6})}_{cn\bar{n}\bar{n}} = 166.94~\text{MeV}
\;\longrightarrow\;
b^{(6\bar{6})}_{bn\bar{n}\bar{n}} = 63.15~\text{MeV},
$$
directly reflecting the $1/(m_{D_1}m_{\bar{D}_2})$ scaling and explaining why the antisymmetric inversion becomes only partial in the bottom sector. Isospin breaking remains negligible (${\sim}\,1$--$2$~MeV). Each $\bar{n}\to\bar{s}$ substitution raises the mass by ${\sim}\,114$~MeV (first) and ${\sim}\,336$~MeV (second), while replacing $bu\to bs$ in the heavy diquark increases the ground-state mass by $126.5$~MeV, consistent with the strange-quark mass shift in the calibrated inputs and the associated binding energy $BE(bs)$ (Table~\ref{t7}).

\item All predicted $bn\bar{n}\bar{n}$ states lie significantly above the lowest open-bottom thresholds and are unstable against strong fall-apart decays. We classify the spectrum into three kinematic tiers (cf.\ Fig.~\ref{Plot_SinglyBottom}):
\begin{enumerate}
    \item Sub-$\bar{B}\rho$ tier ($m \lesssim 6050$~MeV): The ground state at $5891.55$~MeV resides ${\sim}\,470$~MeV above the $\bar{B}\pi$ threshold ($5419$~MeV), and states in this window are restricted to $\bar{B}^{(*)}\pi$ modes. The substantial phase space suggests generically broad widths.

    \item Intermediate tier ($5980$--$6112$~MeV): The $1^+$ states in this region decay predominantly through $\bar{B}^{*}\pi$. Because $S$-wave couplings are favored by parity and angular-momentum conservation, relative partial widths are governed primarily by phase space.

    \item High tier ($m \gtrsim 6107$~MeV): The $0^+, 1^+, 2^+$ states (${\sim}\,6107$--$6166$~MeV) lie near or above the $\bar{B}^{*}\rho$ threshold; and their experimental identification may be challenging due to overlapping vector–vector decay modes.
\end{enumerate}
For the strange sectors (Tables~\ref{t12} and~\ref{t13}), the relevant thresholds shift upward ($\bar{B}_sK \simeq 5863$~MeV, $\bar{B}_s^{*}K \simeq 5907$~MeV). Our lightest $bn\bar{n}\bar{s}$ configurations reside ${\sim}\,230$--$250$~MeV above $\bar{B}K$, while the $bn\bar{s}\bar{s}$ and $bs\bar{s}\bar{s}$ states ($> 6.3$~GeV) are deeply embedded in the open-bottom continuum and expected to be broad.

\item As the singly bottom sector remains entirely predictive, we advocate for these results as benchmarks for future LHCb searches, on the basis of: (i)~the $\mathcal{O}(10$--$30)$~MeV agreement achieved in the singly charm sector; (ii)~the stability of the hyperfine ratio at ${\sim}\,2.7$ across charm, bottom, and doubly heavy systems; and (iii)~consistent $c\!\to\!b$ mass scaling. Compared to RQM~\cite{Lu:2020qmp}, our $bn\bar{n}\bar{n}$ spectrum is significantly more compressed, with a mean mass lower by ${\sim}\,270$~MeV. Regarding the $T_{b\bar s}(5568)$ claim~\cite{D0:2017qqm}, we find no support for a compact $bn\bar{n}\bar{s}$ state in the $5.5$~GeV region: our lightest such configuration lies at $6005.92$~MeV, over $430$~MeV above the reported signal, and even extreme hyperfine scenarios cannot bridge this gap, in agreement with the non-confirmations by LHCb, CMS, CDF, and ATLAS \cite{LHCb:2016dxl, CMS:2017hfy, CDF:2017dwr, ATLAS:2018udc}.
\end{enumerate}

The $bn\bar{n}\bar{n}$ spectrum thus organizes into three distinct bands (cf.\ Fig.~\ref{Plot_SinglyBottom}): (i)~a low band (${\sim}\,5.9$~GeV) containing the $\bar{\mathbf{3}}_0\,\mathbf{3}_0$ scalar and the $\mathbf{6}_1\,\bar{\mathbf{6}}_1$ axial--axial $0^+$ state, separated by only ${\sim}\,26$~MeV; (ii)~an intermediate band ($5.98$--$6.11$~GeV) with interleaved color configurations, including the inverted $1^+$ state; and (iii)~a high band ($> 6.1$~GeV) dominated by the $2^+$ axial--axial states and the $\mathbf{6}_0\,\bar{\mathbf{6}}_0$ scalar.

\subsection{Doubly charm tetraquark states}
\label{Sec3B}

In this section, we present our predictions for the doubly charm tetraquark $T(cc\bar q \bar q')$ states for both $\bar{\mathbf{3}}_c\otimes \mathbf{3}_c$ and $\mathbf{6}_c\otimes\bar{\mathbf{6}}_c$ color configurations. Our analysis here focuses on the unique spectral ordering dictated by the $cc$ diquark scale.

\begin{enumerate}
\item In the $T(cc\bar{u}\bar{d})$ isoscalar channel, we identify the lightest state as the $I(J^P)=0(1^+)$ configuration at
$$
m\bigl[T(cc\bar{u}\bar{d}),\,0(1^+)\bigr]_{\bar{3}3} =
3860.35~\text{MeV},
$$
in excellent agreement with the PDG value $3874.74(10)$~MeV~\cite{LHCb:2021auc,LHCb:2021vvq}, the only experimentally established doubly heavy tetraquark to date. Our calculated mass lies $14.8$~MeV below the $D^0 D^{*+} = 3875.1$~MeV threshold and $125.9$~MeV above the $D^0 D^+\gamma = 3734.3$~MeV threshold; the ground state is therefore stable against strong decays and must proceed electromagnetically (cf.\ Fig.~\ref{Plot_cc_bb}). While most quark models predict $T(cc\bar{u}\bar{d})$ above the $D^0 D^{*+}$ threshold~\cite{Song:2023izj,Braaten:2020nwp,Zhang:2021yul,Ebert:2007rn,Lu:2020rog,Eichten:2017ffp}, our prediction falls within the narrow sub-threshold window increasingly favored by near-threshold phenomenological approaches and corroborated by recent LQCD studies~\cite{Padmanath:2022cvl,Lyu:2023xro,Prelovsek:2025vbr}, which consistently find the state a few MeV below or near threshold, with definitive conclusions awaiting future multi-channel analyses and controlled chiral extrapolations~\cite{Francis:2024fwf,Bicudo:2022cqi}.

\item A characteristic feature of the doubly heavy sector is the large separation between $\bar{\mathbf{3}}\otimes\mathbf{3}$ and $\mathbf{6}\otimes\bar{\mathbf{6}}$ configurations. Unlike hidden-charm or singly heavy systems, where inter-cluster hyperfine effects may invert the color ordering, the doubly charm spectrum exhibits a persistent hierarchy with sextet states consistently heavier. For the isoscalar $1^+$ state,
$$
m\bigl[T(cc\bar{u}\bar{d}),\,0(1^+)\bigr]_{6\bar{6}}
= 4125.86~\text{MeV},\qquad
\Delta m = 265.51~\text{MeV}.
$$
Comparable gaps persist throughout the spectrum: $\Delta m = 287.75$~MeV for $1(0^+)$, $305.81$~MeV for $\tfrac{1}{2}(1^+)_{cc\bar{u}\bar{s}}$, and $253.39$~MeV for $0(0^+)_{cc\bar{s}\bar{s}}$. This pattern reflects the suppression of heavy--heavy hyperfine effects by $1/(m_i m_j)$, which leaves the diquark binding energy, attractive in $\bar{\mathbf{3}}_c$ and repulsive in $\mathbf{6}_c$, as the dominant spectroscopic contribution.

\item The axial-axial relations $M(1^+)-M(0^+)=b$ and $M(2^+)-M(1^+)=2b$ hold with high precision across all flavor sectors. For the $\bar{\mathbf{3}}\otimes\mathbf{3}$ states we extract $b_{nn}=37.95$~MeV, $b_{ns}=29.90$~MeV, and $b_{ss}=25.85$~MeV; this monotonic decrease with strangeness confirms the expected $b \propto 1/(m_{D_1}m_{\bar{D}_2})$ scaling, consistent with Regge and HQET approaches~\cite{Song:2023izj,Braaten:2020nwp,Ebert:2007rn}. Our framework retains explicit isospin breaking through constituent masses: non-strange isovector splittings remain small ($\lesssim 3$~MeV), while strange partners exhibit a larger splitting due to the sizable $us$--$ds$ diquark mass difference,
$$
m(cc\bar{u}\bar{s}) - m(cc\bar{d}\bar{s}) = 18.56~\text{MeV}.
$$
For $T(cc\bar{u}\bar{s})$ and $T(cc\bar{d}\bar{s})$, the lowest states lie within $+18.1$~MeV and $-3.9$~MeV of the $D_s D^{*}$ thresholds, respectively, identifying them as potential narrow resonances or virtual states (cf.\ Fig.~\ref{Plot_cc_bb}).

\item In the $T(cc\bar{s}\bar{s})$ sector, the Pauli principle strictly limits the number of allowed configurations. The scalars remain the lightest and heaviest states, with masses
$$
m(0^+)_{\bar{3}3} = 4259.41~\text{MeV},\qquad
m(0^+)_{6\bar{6}} = 4512.80~\text{MeV}.
$$
All predicted $cc\bar{s}\bar{s}$ states lie well above open-charm thresholds and are therefore expected to be broad. The $cc\bar u\bar d$ isovector tensor state at $4005.63$~MeV serves as a particularly sharp experimental discriminator: while it resides above the $DD$ threshold, it lies $11.5$~MeV below $D^{*0}D^{*+}$, forbidding the $S$-wave $D^{*}D^{*}$ mode favored by other models (cf.\ Fig.~\ref{Plot_cc_bb}).

\item A comparison of the predicted masses with the corresponding open-charm two-meson thresholds, as illustrated in Fig.~\ref{Plot_cc_bb}, reveals a clear and systematic flavor dependence of the binding pattern:
\begin{enumerate}
    \item $ccu\bar d$: The $0(1^+)$ state lies slightly below the $DD^*$ threshold, with the remaining multiplet occupying the region between $DD^*$ and $D^*D^*$ thresholds. Chromomagnetic attraction produces only shallow near-threshold binding in the non-strange sector, while the color-sextet configurations remain significantly heavier and unbound.

    \item $ccu\bar s$: The ground state sits just above the $DD_s^*$ threshold, and the spectrum extends beyond $D^*D_s^*$. Strangeness suppresses the effective attraction, eliminating subthreshold stability.

    \item $ccs\bar s$: All states lie above the $D_s^{(*)}D_s^{(*)}$ thresholds; cumulative inverse-mass suppression lifts the entire multiplet into the unbound, resonant region, with the sextet configuration highest in mass and far from thresholds.
\end{enumerate}
\end{enumerate}

Our $T_{cc}$ mass resides between CMIM values~\cite{Luo:2017eub} and typical quark-model estimates \cite{Song:2023izj,Braaten:2020nwp,Zhang:2021yul,Ebert:2007rn,Lu:2020rog,Eichten:2017ffp}, while the $\bar{\mathbf{3}}$--$\mathbf{6}$ color gap remains broadly consistent across frameworks. The systematic downward shift of ${\sim}\,50$--$150$~MeV in our results relative to many quark models reflects the stronger effective color attraction inherited from our baryon-calibrated diquark masses. These features, a near-threshold $T_{cc}$ ground state, a stable color hierarchy with $\Delta m \sim 250$--$310$~MeV, and inverse-mass scaling of hyperfine coefficients, provide quantitative targets for future searches of $T_{cc}$ excitations and flavor partners near $D_s D^{(*)}$ thresholds.

\subsection{Charmed-bottom tetraquark states}
\label{Sec3C}

In Table~\ref{t15} we present the predicted masses of charmed--bottom tetraquarks $T(cb\bar{q}\bar{q}')$ for light-antidiquark flavors $(\bar{n}\bar{n},\,\bar{n}\bar{s},\,\bar{s}\bar{s})$ in both the $\bar{\mb3}_c\otimes\mb3_c$ and $\mb6_c\otimes\bar{\mb6}_c$ color sectors. The defining feature of this sector is that the heavy constituents are non-identical: the Pauli principle places no constraint on the $cb$ diquark spin, and the near-degeneracy of the scalar and axial $cb$ masses (Table~\ref{t2}), a direct consequence of the suppressed chromomagnetic interaction between unequal heavy quarks, compresses the spin-0 and spin-1 configurations to within a few MeV. This quasi-degeneracy propagates into the tetraquark spectrum and produces near-degenerate multiplets unique to the $cb$ sector. The Pauli principle does, however, restrict the light antidiquark whenever identical antiquarks are present.\footnote{For the $\bar{u}\bar{d}$ system the allowed spin depends on the interplay of color and isospin: in $I=0$ the antisymmetric flavor $[\bar{u}\bar{d}]$ requires $S_{\bar D}=0$ in $\mb3_c$ and $S_{\bar D}=1$ in $\bar{\mb6}_c$, while in $I=1$ the symmetric $\{\bar{u}\bar{d}\}$ reverses these assignments.} Combined with the two $cb$ diquark spins, the $T(cb\bar{u}\bar{d})$ system generates twelve states equally divided between isoscalar and isovector channels, as realized explicitly in Table~\ref{t15}.

\begin{enumerate}
\item For $I=0$ the $\bar{\mb3}\mb3$ ground state is a scalar--scalar configuration $(cb)^{\bar{3}}_{0}[\bar{u}\bar{d}]^{3}_{0}$ with $m\bigl[0(0^{+})\bigr]_{\bar{3}3} = 7132.56$~MeV. Pauli symmetry forces the corresponding $\mb6\bar{\mb6}$ scalar into an axial--axial configuration at $m\bigl[0(0^{+})\bigr]_{6\bar{6}} = 7368.60$~MeV, giving
$$
\Delta m(0^{+},\,I\!=\!0) = 236.04~\text{MeV}.
$$
We identify the $\bar{\mb3}\mb3$ scalar as the ground state of the $cb$ sector (cf.\ Fig.~\ref{Plot_cb}). It lies ${\sim}\,12.4$~MeV below the $\bar{B}^{0}D^{0} = 7145.0$~MeV threshold, placing it in the near-threshold regime consistent with phenomenological expectations~\cite{Karliner:2017qjm} and with recent LQCD calculations reporting an attractive $\bar{B}D$ interaction with $-39^{+12}_{-24}$~MeV binding~\cite{Radhakrishnan:2024ihu}. The quasi-degeneracy of the $cb$ diquark produces an unusually narrow low-lying band: the $\bar{\mb3}\mb3$ axial-vector state $(cb)^{\bar{3}}_{1}[\bar{u}\bar{d}]^{3}_{0}$ appears at $m\bigl[0(1^{+})\bigr]_{\bar{3}3} = 7137.02$~MeV, only $4.46$~MeV above the ground state. Because both configurations lack inter-cluster hyperfine corrections, this splitting directly measures the intrinsic scalar--axial $cb$ diquark mass difference. Collecting the relevant two-body thresholds,
$$
\bar{B}^{0}D^{0}:\,7145.0~\text{MeV};~
\bar{B}^{*0}D^{0}:\,7189.6~\text{MeV};~
\bar{B}^{0}D^{*0}:\,7286.6~\text{MeV};~
\bar{B}^{*0}D^{*0}:\,7331.6~\text{MeV},
$$
we find that both the $0(0^{+})$ and $0(1^{+})$ states lie below their respective open channels and are therefore stable against strong decay, consistent with LQCD studies indicating binding relative to $\bar BD^{(*)}$ thresholds~\cite{Padmanath:2023rdu}. The $\mb6\bar{\mb6}$ scalar and axial-vector yield $b_{6\bar{6}} = 50.53$~MeV, following the tensor relation $M(2^{+})-M(1^{+}) = 2\,b_{6\bar{6}}$. For the $cb$ sector the axial–axial $\mb6\otimes\bar{\mb6}$ ($I=0$) and $\bar{\mb3}\otimes\mb3$ ($I=1$) multiplets yield $b^{(6\bar{6})}=50.53$ MeV and $b^{(\bar{3}3)}=19.10$~MeV, giving $b^{(6\bar{6})}/b^{(\bar{3}3)}=2.65$, consistent with the ${\sim}\,2.6$--$2.7$ color-Casimir ratio observed in the singly heavy and hidden-charm sectors. In contrast, the doubly heavy $cc$ and $bb$ systems are the exceptions: identical heavy quarks force the $\bar{\mb3}_c$ diquark into a unique spin state (scalar for $cc$, axial for $bb$), eliminating one of the two color configurations needed to extract this ratio.

\item In the $I=1$ channel the Pauli constraints reverse the light-antidiquark spin assignments. The $\bar{\mb3}\mb3$ scalar now resides in an axial--axial state, $m\bigl[1(0^{+})\bigr]_{\bar{3}3} = 7206.16$~MeV, while the $\mb6\bar{\mb6}$ scalar is at $7525.54$~MeV in scalar--scalar state. The color gap increases to $\Delta m(0^{+},\,I\!=\!1) = 319.38$~MeV, larger than the isoscalar case because only the $\bar{\mb3}\mb3$ state benefits from hyperfine lowering. The ($\bar{\mb3}\mb3, 1^{+}$) state at $7225.26$~MeV gives $b_{\bar{3}3}^{(I=1)} = 19.10$~MeV, confirming $ 2\,b_{\bar{3}3}$ contribution as expected. The ordering $M(cb_{S=0}) > M(cb_{S=1})$ agrees with several independent approaches~\cite{Luo:2017eub,Song:2023izj,Zhang:2021yul,Ebert:2007rn}. The $1(0^{+})$ scalar lies above $\bar{B}D$ and is strongly unstable, whereas the $1(2^{+})$ tensor falls roughly $68$~MeV below $\bar{B}^{*}D^{*}$, suggesting it may be relatively narrow.

\item For $T(cb\bar{u}\bar{s})$ and $T(cb\bar{d}\bar{s})$ the Pauli constraints are relaxed and both antidiquark symmetries occur. The lightest scalar lie slightly above $\bar{B}_{s}D$ thresholds (cf.\ Fig.~\ref{Plot_cb}), where it can decay strongly, whereas the lowest antisymmetric axial-vector state at $7269.95$~MeV fall below the $\bar{B}_{s}^{*}D$ threshold, and we predict it as an additional stable state. The symmetric multiplets lie $\gtrsim 150$~MeV higher and are expected to be broad. The strange sector also enhances isospin breaking:
$$
m(cb\bar{u}\bar{s}) - m(cb\bar{d}\bar{s}) = 18.56~\text{MeV},
$$
substantially larger than in the nonstrange channel, owing to the $\bar{u}$--$\bar{d}$ constituent-mass difference amplified by the strange-antidiquark mass splitting. For $T(cb\bar{s}\bar{s})$ the identical antiquarks enforce $S_{\bar D}=1$ in $\mb3_c$ and $S_{\bar D}=0$ in $\bar{\mb6}_c$. The $\bar{\mb3}\mb3$ multiplet spans $7562$--$7601$~MeV with $b_{\bar{3}3} \simeq 13$~MeV, while the $\mb6\bar{\mb6}$ states cluster near $7860$~MeV. All members lie well above the $\bar{B}_{s}D_{s}$ threshold and are expected to be broad (cf.\ Fig.~\ref{Plot_cb}).

\item For the isoscalar ground state, most models predict higher masses, typically by $+100$--$160$~MeV, while the CMIM \cite{Luo:2017eub} yields a lower value. Our framework, together with the CMIM, uniquely predicts multiple subthreshold states: $I(J^{P}) = 0(0^{+})$ below $\bar{B}D$, $I(J^{P}) = 0(1^{+})$ below $\bar{B}^{(*)}D^{(*)}$, and strange antisymmetric $1^{+}$ states below $\bar{B}_{s}^{*}D$. The existence of more than one stable configuration in each flavor channel is a distinctive signature of the compact diquark--antidiquark picture.

\item A comparison of the predicted spectra with the corresponding open-charmed-bottom two-meson thresholds (Fig.~\ref{Plot_cb}) reveals a systematic flavor dependence of the binding pattern:

\begin{enumerate}
  \item \textit{$cb\bar{n}\bar{n}$:} The isoscalar $0(0^{+})$ and $0(1^{+})$ states lie below the $\bar{B}D$ and $\bar{B}^{*}D$ thresholds, indicating genuine sub-threshold binding. The isovector multiplet is pushed above $\bar{B}^{*}D$ toward the $\bar{B}^{(*)}D^{*}$ thresholds, consistent with weakly bound or resonant behavior. Color-sextet configurations are systematically heavier and remain unbound.

  \item \textit{$cb\bar{n}\bar{s}$:} The lowest states lie marginally close to the $\bar{B}_sD$ (above) and $\bar{B}_{s}^{*}D$ (below) thresholds, but the overall binding is reduced relative to the non-strange sector. Higher-spin members cluster near or above the $\bar{B}_{s}D^{*}$ and $\bar{B}_{s}^{*}D^{*}$ thresholds, signaling near-threshold resonances.

  \item \textit{$cb\bar{s}\bar{s}$:} All states lie above the $\bar B_{s}^{(*)}D_{s}^{(*)}$ thresholds; inverse-mass suppression of the hyperfine interaction lifts the entire multiplet into the unbound, resonant region. Color-sextet configurations are substantially heavier and furthest from thresholds.

\end{enumerate}
\end{enumerate}

The $T(cb\bar{q}\bar{q}')$ spectrum thus exhibits three robust features: (i)~a compressed isoscalar band near $7.13$~GeV containing two nearly degenerate stable states whose proximity is driven by $cb$ diquark quasi-degeneracy; (ii)~isospin-dependent spin inversion arising from Pauli constraints, producing markedly different hyperfine patterns in the $I=0$ and $I=1$ channels; and (iii)~an additional stable strange axial-vector state below the $\bar{B}_{s}^{*}D$ threshold. These features provide concrete experimental targets for LHCb and Belle~II and offer sharp discriminants among competing dynamical descriptions of heavy tetraquarks.

\subsection{Doubly bottom tetraquark states}
\label{Sec3D}

In Table~\ref{t16} we present the predicted masses of doubly bottom tetraquarks $T(bb\bar{q}\bar{q}')$ with light-antidiquark flavors $(\bar{n}\bar{n},\,\bar{n}\bar{s},\,\bar{s}\bar{s})$. A key structural difference from the $cb$ sector is imposed by Pauli symmetry: the identical $b$ quarks force the $\bar{\mb3}_c$ diquark into spin-1, eliminating the scalar diquark channel that appeared in lighter systems. Combined with the much stronger binding of the $bb$ diquark, this constraint produces a spectrum in which the entire $\bar{\mb3}\!\otimes\!\mb3$ band lies deeply below open-bottom thresholds.

\begin{enumerate}
\item In the isoscalar channel the antisymmetric light antidiquark $[\bar{u}\bar{d}]$ couples to the axial $bb$ diquark to form a unique $0(1^{+})$ ground state at $m\bigl[T(bb\bar{u}\bar{d}),\,0(1^{+})\bigr]_{\bar{3}3} = 10355.07$~MeV. The lowest rearrangement threshold is $\bar{B}\bar{B}^{*}$ at $10604.5$~MeV (cf.\ Fig.~\ref{Plot_cc_bb}), yielding
$$
\Delta_{\rm thr} = 249.4~\text{MeV}.
$$
Even the radiative $\bar{B}\bar{B}\gamma$ channel remains closed, so this ground state is stable against both strong and electromagnetic decays and can decay only weakly, a conclusion shared by most theoretical approaches. Quantitatively, our prediction lies closest to the CMIM~\cite{Luo:2017eub}, HQET~\cite{Braaten:2020nwp}, and HQS~\cite{Eichten:2017ffp} results, while Regge~\cite{Song:2023izj}, RQM~\cite{Ebert:2007rn,Lu:2020rog}, and BM~\cite{Zhang:2021yul} typically yield higher masses. Despite a ${\sim}\,100$--$200$~MeV spread across methods, all models except BM agree that the ground state is subthreshold. Lattice QCD independently supports a bound $bb\bar{u}\bar{d}$ tetraquark: static-potential estimates yield $E_{B} = -30(17)$ to $-90(40)$~MeV~\cite{Bicudo:2012qt,Bicudo:2015vta}, spectrum calculations find deeper binding such as $-189(10)$~MeV~\cite{Francis:2016hui}, HAL potential studies give $-83(10)$~MeV~\cite{Aoki:2023nzp}, and a recent finite-volume analysis reports $-116^{+30}_{-36}$~MeV~\cite{Tripathy:2025vao}, with modern reviews suggesting $-190 \lesssim E_{B} \lesssim -70$~MeV~\cite{Francis:2024fwf}. Our value falls at the deeper end of this range.

\item A striking difference from the charm sector is that the entire isovector $\bar{\mb3}\!\otimes\!\mb3$ multiplet:
$$
1(0^{+}):\;10436.89~\text{MeV},\quad
1(1^{+}):\;10449.65~\text{MeV},\quad
1(2^{+}):\;10475.16~\text{MeV},
$$
lies below open-bottom thresholds (cf. Fig. \ref{Plot_cc_bb}). All three states sit $120$--$175$~MeV below their respective decay channels. Most models~\cite{Song:2023izj,Braaten:2020nwp,Lu:2020rog,Zhang:2021yul,Eichten:2017ffp} place these states above threshold; only the CMIM~\cite{Luo:2017eub} shares our subthreshold prediction. The isovector triplet yields the hyperfine coefficient $b_{bb} = M(1^{+}) - M(0^{+}) = 12.76~\text{MeV}$. Compared with the doubly charm value $b_{cc} \simeq 37.95$~MeV, the suppression $b_{bb}/b_{cc} \simeq 0.34$ follows the expected HQSS scaling $b \propto 1/(m_{D_1}\,m_{D_2})$.

\item A defining feature of Table~\ref{t16} is the dramatic $\bar{\mb3}\otimes\mb3$--$\mb6\otimes\bar{\mb6}$ separation. For the isoscalar ground state,
$$
\Delta m \simeq 495~\text{MeV},
$$
substantially larger than in the $cc$ or $cb$ sectors. This growth reflects the transition from chromomagnetic dominance in charm to color-Coulomb dominance in bottom: the attractive $\bar{\mb3}_c$ channel gains binding while the repulsive $\mb6_c$ channel incurs an increasing mass shift, amplifying the color-sector gap along the sequence $cc \to cb \to bb$. Isospin breaking remains small throughout (${\sim}\,2$--$3$~MeV).

\item For $T(bb\bar{q}\bar{s})$ the antisymmetric $\tfrac{1}{2}(1^{+})$ state at $10488$~MeV lie roughly $200$~MeV below $\bar{B}_{s}\bar{B}^{*}$ and are therefore stable (cf.\ Fig.~\ref{Plot_cc_bb}). Lattice calculations support bound $bb\bar{q}\bar{s}$ states, with binding estimates ranging from $-98(10)$~MeV~\cite{Francis:2016hui} and $-87(32)$~MeV~\cite{Junnarkar:2018twb} to more recent values of $-30(3)({}^{+11}_{-31})$~\cite{Alexandrou:2024iwi} and $-47(8)$~MeV~\cite{Colquhoun:2024jzh}; our predictions again fall toward the deeper end. Symmetric light-diquark states cluster near thresholds and may appear as narrow resonances, while their sextet partners lie $\gtrsim 450$--$550$~MeV higher and are expected to be broad. For $T(bb\bar{s}\bar{s})$ the $\bar{\mb3}\mb3$ multiplet sits near the $\bar{B}_{s}^{(*)}\bar{B}_{s}^{(*)}$ thresholds (cf.\ Fig.~\ref{Plot_cc_bb}), with one tensor state slightly below $\bar{B}_{s}^{*}\bar{B}_{s}^{*}$ that is stable against $S$-wave decay and expected to be very narrow. The corresponding sextet scalar lies ${\sim}\,450$~MeV higher.

\item We examine the threshold systematics of the doubly bottom tetraquark spectra with the open-bottom thresholds shown in the Fig. \ref{Plot_cc_bb}.

\begin{enumerate}
\item $bb\bar u\bar d$ system: The $0(1^{+})$ ground state at $10355.07$~MeV lies $\sim 249$~MeV below the $\bar B \bar B^{*}$ threshold ($10604$~MeV). The isovector multiplet $(10436.89-10475.16)$~MeV also remains below the lowest open-bottom threshold $\bar B \bar B$ ($\sim10559$~MeV). Therefore, the entire spectrum is deeply sub-threshold, with $\gtrsim 120$~MeV binding for both $I=0$ and $I=1$ states.

\item $bb\bar u\bar s$ system: The spectrum spans $(10488.00-10669.85)$~MeV. The lowest state is deeply bound ($\sim 200$~MeV below $\bar B_s \bar B^{*}$ at $\sim10692$~MeV), while the remaining three states cluster between the $\bar B_s \bar B$ and $\bar B_s \bar B^{*}$ thresholds, marking a transition from strong binding to near-threshold dynamics.

\item $bb\bar s\bar s$ system: The $(10788.45-10814.51)$~MeV multiplet lies between the $\bar B_s \bar B_s^{*}$ ($\sim10733$~MeV) and $\bar B_s^{*} \bar B_s^{*}$ ($\sim10830$~MeV) thresholds, confined to a $26.06$~MeV hyperfine window. The entire sector is therefore intrinsically near-threshold rather than deeply bound.

\item Substituting $u/d \to s$ systematically raises the spectrum from the deeply bound regime ($bb\bar{u}\bar{d}$) toward and into the open-bottom thresholds ($bb\bar{s}\bar{s}$). Color-sextet states remain uniformly heavier than their antitriplet counterparts throughout and do not contribute to the low-lying bound-state spectrum.
\end{enumerate} 
\end{enumerate}

The doubly bottom spectrum is thus the most promising arena for discovering stable exotic tetraquarks: a deeply bound $0(1^{+})$ ground state ${\sim}\,250$~MeV below $\bar{B}\bar{B}^{*}$, a fully subthreshold isovector $\bar{\mb3}\!\otimes\!\mb3$ multiplet, color-sector gaps reaching ${\sim}\,450$--$535$~MeV, hyperfine suppression $b_{bb} \simeq 0.34\,b_{cc}$ consistent with HQSS, and additional strange subthreshold candidates corroborated by LQCD, all provide concrete experimental targets for LHCb and Belle~II.

\subsection{Hidden charm tetraquark states}
\label{Sec3E}

In Table~\ref{t17} we report our predicted mass spectra for hidden-charm tetraquarks composed of heavy--light diquarks. The $T(cn\bar{c}\bar{n})$ system is evaluated independently for $\bar{\mb3}_c\otimes \mb3_c$ and $\mb6_c\otimes\bar{\mb6}_c$ color configurations.

\begin{enumerate}
\item For scalar--scalar pairs the mass is a pure constituent sum (Table~\ref{tetra_mass}, first row):
$$
m(\bar{\mb3}_{0}\,\mb3_{0};\,0^{+}) \simeq 3865~\text{MeV},\qquad
m(\mb6_{0}\,\bar{\mb6}_{0};\,0^{+}) \simeq 4195~\text{MeV},\qquad
\Delta m \simeq 330~\text{MeV},
$$
where the entire gap arises from the $\mb6_c$ constituent-mass shift (${\sim}\,165$~MeV per diquark, doubled over both clusters). For hidden-flavor systems ($q_1=q_3$, $q_2=q_4$), charge-conjugation symmetry makes the diquark and antidiquark color splittings equal, reducing Eq.~\eqref{eq:gap_hidden} to
\begin{equation}
\Delta(J)
= 2\bigl[m_{D}^{\bar{3}}(S) - m_{D}^{6}(S^{\prime})\bigr]
+ c_J\bigl[b^{(\bar{3}3)} - b^{(6\bar{6})}\bigr],
\label{eq:gap_hidden}
\end{equation}
which governs the hidden-charm and hidden-bottom numerical analyses.\footnote{This hidden-flavor relation follows directly from charge-conjugation symmetry and represents a symmetry-enhanced limit of the general color-sector hierarchy.} For axial--axial pairs the ordering inverts, $m(\mb6_{1}\,\bar{\mb6}_{1}) < m(\bar{\mb3}_{1}\,\mb3_{1})$ in every $J^{P}$ channel, because the reversed constituent-level shift for axial-vector diquarks ($m_{\bar{3}}(1^{+}) > m_{6}(1^{+})$ by ${\sim}\,54$--$55$~MeV per cluster, Sec.~\ref{Diquark_mass}) is compounded over both clusters. The resulting $J^{P}$-dependent shift is:
\begin{center}
\begin{tabular}{lccc}
\hline\hline
$J^{P}$ & $c_{J}$ & Hyperfine effect & $\Delta(J)$ (MeV) \\
\hline
$0^{+}$ & $-2$ & reinforces           & $181$ \\
$1^{+}$ & $-1$ & partially reinforces & $145$ \\
$2^{+}$ & $+1$ & opposes              & $\phantom{0}74$ \\
\hline\hline
\end{tabular}
\end{center}
The step sizes $\Delta(0^{+})-\Delta(1^{+})=36$~MeV and $\Delta(1^{+})-\Delta(2^{+})=71$~MeV match $b^{(6\bar{6})}-b^{(\bar{3}3)}\simeq 35.5$~MeV and $2(b^{(6\bar{6})}-b^{(\bar{3}3)})\simeq 71$~MeV, confirming fidelity to the mass-formula structure of Table~\ref{tetra_mass}. The resulting hierarchy is distinctive: $\bar{\mb3}\!\otimes\!\mb3$ produces the overall ground state in $0^{+}$, while $\mb6\!\otimes\!\bar{\mb6}$ generates the lightest states in $1^{+}$ and $2^{+}$.

\item From Table~\ref{t17} we extract the axial--axial $M(1^{+})-M(0^{+})$ splittings,
\[
b^{(\bar{3}3)} = 4136.26 - 4114.60 = 21.66~\text{MeV};\qquad
b^{(6\bar{6})} = 57.14~\text{MeV},
\]
with $b^{(6\bar{6})}/b^{(\bar{3}3)}=2.64$, consistent with the hidden-$c\bar{c}$ color-sector ratio of Eq.~\eqref{eq:convergence}. The $T(cu\bar{c}\bar{u})$ versus $T(cd\bar{c}\bar{d})$ mass differences are uniformly $2$--$3$~MeV, consistent with $m_d - m_u$ propagating through the QDEMF. For isospin-zero $T(cu\bar{c}\bar{d})$ states, masses coincide with the arithmetic mean of the isospin-one states to sub-MeV precision. In the $1^{+}$ channel two additional states arise from diquark$\leftrightarrow$antidiquark exchange ($S_0 S_1\leftrightarrow S_1 S_0$): the resulting pairs at $(4012.47,\,4012.94)$~MeV and $(4122.50,\,4122.27)$~MeV are split by $<0.5$~MeV, reflecting approximate charge-conjugation symmetry broken only by light-quark mass differences.

\item Two PDG entries appear in the $1^{+}$ channel:
\begin{enumerate}
    \item \textit{$T_{c\bar{c}1}(3900)$}: mass $3887.1\pm 2.6$~MeV, $I^{G}(J^{PC})=1^{+}(1^{+-})$~\cite{ParticleDataGroup:2024cfk}. As an isovector charged state its exotic nature is unambiguous. Our lightest $\bar{\mb3}\!\otimes\!\mb3$ $1^{+}$ prediction (scalar--axial, $4012$~MeV) lies ${\sim}\,125$~MeV above the observed mass. Because this channel carries no hyperfine correction (Table~\ref{tetra_mass}), the deviation cannot be absorbed by tuning $b$ and instead reflects dynamics not captured within the compact tetraquark framework. The lightest $\mb6\otimes\bar{\mb6}$ $1^{+}$ state (axial–axial, $3992$ MeV) provides a closer estimate while preserving the required $C$-parity. The residual difference therefore provides an estimate of effects beyond the compact-diquark treatment; subleading hadronic dressing may be quantified as more precise data become available.

    \item \textit{$T_{c\bar{c}1}(4200)$}: mass $4196^{+35}_{-32}$~MeV~\cite{ParticleDataGroup:2024cfk}. The compact $1^{+}$ spectrum in Table~\ref{t17} organizes into two nearby color-resolved multiplets. The lower band consists of the $\mb6_{1}\bar{\mb6}_{1}$ axial--axial configuration at $3992$~MeV and the $\bar{\mb3}_{0}\mb3_{1}$ scalar--axial state at $4012$~MeV, while the upper band contains the $\mb6_{0}\bar{\mb6}_{1}$ scalar--axial state at $4122$~MeV and the $\bar{\mb3}_{1}\mb3_{1}$ axial--axial partner at $4137$~MeV. The intra--spin color splittings are $\sim110$~MeV (scalar--axial) and $\sim145$~MeV (axial--axial), reflecting the chromomagnetic hierarchy of the model. The observed mass lies only $\sim60$~MeV above the upper compact multiplet, making $T_{c\bar{c}1}(4200)$ naturally compatible with the highest $1^{+}$ color band. Within this band, mixing between the $\mb6\bar{\mb6}$ and $\bar{\mb3}\mb3$ configurations is expected, and the measured width may provide further discrimination of the dominant component. The resolved color structure implies that states with identical $J^{P}$ should occur as nearby partners rather than isolated resonances. Experimental confirmation of such same-spin structure in the $4.0$--$4.2$~GeV region would favor a compact interpretation, whereas the absence of multiplet behavior would disfavor it.
\end{enumerate}

\item The full $cn\bar{c}\bar{n}$ spectrum spans ${\sim}\,330$~MeV and naturally separates into three bands. The low band ($3865$--$3934$~MeV) contains the $\bar{\mb3}_{0}\mb3_{0}$ ground state and the lightest $\mb6_{1}\bar{\mb6}_{1}$ configuration, reflecting maximal hyperfine lowering ($-2b^{(6\bar{6})}$). The mid band ($3990$--$4137$~MeV) hosts both $\bar{\mb3}\otimes\mb3$ and $\mb6\otimes\bar{\mb6}$ states, arising from the $J^{P}$-dependent shift described by Eq.~\eqref{eq:gap_hidden}. The high band ($4180$--$4196$~MeV) is formed by the $\bar{\mb3}_{1}\mb3_{1}$ $2^{+}$ state and the $\mb6_{0}\bar{\mb6}_{0}$ scalar. Compared to the RQM~\cite{Yu:2024ljg}, which predicts higher $0^{+}$ and $1^{+}$ ground states (3970 and 4039~MeV) and a lower $2^{+}$ ($4142$~MeV), and the BM~\cite{Yan:2023lvm}, which yields a broader and generally heavier spectrum, our framework produces a more compressed level structure with lower ground-state masses.

\item We extend the analysis to the $T(cu\bar{c}\bar{s})/T(cs\bar{c}\bar{u})$ and $T(cd\bar{c}\bar{s})/T(cs\bar{c}\bar{d})$ configurations with $I=\frac{1}{2}$, summarized in Table~\ref{t18}. The qualitative band structure identified in the non-strange sector is preserved, with systematic mass shifts induced by strangeness. From the axial--axial splittings we extract $b^{(\bar{3}3)} = 19.92~\text{MeV}, \text{~and~}b^{(6\bar{6})} = 52.41~\text{MeV},$ yielding $b^{(6\bar{6})}/b^{(\bar{3}3)} = 2.63$ (Eq. \eqref{eq:convergence}), in close agreement with the non-strange ratio (2.64). The reduction relative to the non-strange $\bar{\mb3}\!\otimes\!\mb3$ value ($21.66 \to 19.92$~MeV) corresponds to an approximate $8\%$ suppression per strange quark, consistent with the $1/(m_{D_1} m_{\bar{D_2}})$ scaling of the chromomagnetic interaction. The axial-vector inversion persists across all $J^{P}$ channels. The splittings are $\Delta(0^{+}) = 115$~MeV, $\Delta(1^{+}) = 82$~MeV, and $\Delta(2^{+}) = 18$~MeV, each reduced relative to the corresponding non-strange values ($181$, $145$, and $74$~MeV), reflecting the heavier diquark mass scale. Two experimental states appear:
\begin{enumerate}
\item \textit{$T_{c\bar{c}\bar{s}1}(4000)$}: mass $3988\pm 5$~MeV, $J^{P}=1^{+}$~\cite{ParticleDataGroup:2024cfk}. Our lightest compact $1^{+}$ states lie at $4154$~MeV ($\bar{\mb3}\!\otimes\!\mb3$, scalar--axial) and $4198$~MeV ($\mb6\!\otimes\!\bar{\mb6}$, axial--axial), corresponding to upward shifts of ${\sim}166$ and ${\sim}210$~MeV, respectively. These exceed the non-strange shifts, indicating a larger residual difference between compact-diquark predictions and experiment in the strange sector. The observed mass lies close to the $D_s\bar{D}^{*}$ threshold (${\sim}3975$~MeV), indicating possible subleading contributions not captured within the compact-diquark treatment.

\item \textit{$T_{c\bar{c}\bar{s}1}(4220)$}: mass $4220^{+50}_{-40}$~MeV~\cite{ParticleDataGroup:2024cfk}. This state falls between the $\mb6_{1}\bar{\mb6}_{1}$ axial--axial $1^{+}$ at $4198$~MeV and the $\bar{\mb3}_{1}\mb3_{1}$ partner at $4281$~MeV, with the former providing the closest compact correspondence ($\Delta m \simeq 22$~MeV), well within combined uncertainties. The splitting $M(4220) - M(4000) \simeq 232~\text{MeV}$ is reduced relative to the non-strange analogue (${\sim}309$~MeV), consistent with the expected suppression from the larger strange-diquark mass scale. The $T(cd\bar{c}\bar{s})/T(cs\bar{c}\bar{d})$ systems exhibit the same pattern, with isospin-breaking shifts of only $1$--$2$~MeV.
\end{enumerate}

\item We present the $T(cs\bar{c}\bar{s})$ spectrum ($I=0$) in Table~\ref{t18}. The presence of two strange quarks further suppresses the hyperfine scales, $b^{(\bar{3}3)} = 18.32~\text{MeV}$ and $b^{(6\bar{6})} = 48.07~\text{MeV}$, while preserving the stable ratio $b^{(6\bar{6})}/b^{(\bar{3}3)}=2.62$, consistent with the hidden-charm trend ($2.64 \to 2.63 \to 2.62$ for zero, one, and two strange quarks). A qualitative new feature emerges in the $2^{+}$ channel. The axial--axial ordering, inverted for $0^{+}$ and $1^{+}$ (with $\bar{\mb3}_{1}\mb3_{1}$ heavier by $50$ and $20$~MeV), reverses here: $\mb6_{1}\bar{\mb6}_{1}$ ($4501$~MeV) lies above $\bar{\mb3}_{1}\mb3_{1}$ ($4461$~MeV) by $40$~MeV. This crossover follows directly from the mass formula: at $c_J=+1$ the $\mb6\!\otimes\!\bar{\mb6}$ state is raised by $b^{(6\bar{6})}\simeq48$~MeV, compared to only $b^{(\bar{3}3)}\simeq18$~MeV for $\bar{\mb3}\!\otimes\!\mb3$. As the constituent gap decreases with increasing strangeness, the larger hyperfine shift in the sextet sector becomes dominant,
$$
\Delta(2^{+}):\quad
+74 \;\to\; +18 \;\to\; -40~\text{MeV},
$$
for zero, one, and two strange quarks. The scalar--scalar ground state at $4155$~MeV agrees closely with the RQM~\cite{Yu:2024ljg} prediction ($4144$~MeV, $\Delta=11$~MeV), while the BM~\cite{Yan:2023lvm} yields systematically higher values. No experimental candidates are currently established; the predicted ground state near $4155$~MeV and the $2^{+}$ ordering reversal provide clear benchmarks for searches in $J/\psi\,\phi$ and $D_s\bar{D}_s^{(*)}$ channels.

\item We analyze the threshold hierarchy of the hidden-charm tetraquark spectrum relative to the relevant open-charm $S$-wave thresholds shown in Fig.~\ref{Plot_HiddenHeavy}.

\begin{enumerate}
\item $cu\bar c\bar d$ sector: A single near-threshold candidate emerges, with the $0(0^{+})$ at $3866.46$~MeV lying $\sim9$ MeV below $D\bar{D}^{*}$. States between $D\bar{D}^{*}$ and $D^{*}\bar{D}^{*}$ exhibit limited phase space and are expected to be narrow to moderate, while those above $D^{*}\bar{D}^{*}$ are fully open-channel and broad. Overall, the spectrum evolves from marginal binding to continuum-dominated dynamics.

\item $cu\bar c\bar s$ sector: The spectrum shifts upward ($\sim150$ MeV), eliminating bound configurations. The lowest $\tfrac12(0^{+})$ lies just above $D\bar{D}_s^{*}$ but well below $D^{*}\bar{D}_s^{*}$, making it the most weakly coupled state. Near-threshold clustering occurs around $D^{*}\bar{D}_s^{*}$, producing moderate widths, while higher states are fully open-channel resonances.

\item $cs\bar c\bar s$ sector: Entirely threshold-dominated. The lowest state lies $\sim75$ MeV above $D_s\bar{D}_s^{*}$, and all states reside above open S-wave thresholds. Except for one state between $D_s\bar{D}_s^{*}$ and $D_s^{*}\bar{D}_s^{*}$, the multiplet lies substantially above the highest threshold ($\gtrsim65–377$ MeV) and is expected to be broad, with no near-threshold binding.
\end{enumerate} 
\end{enumerate}

Collecting our results from Tables~\ref{t17}--\ref{t18}, the hidden-charm spectrum exhibits a coherent progression with strangeness:
\begin{center}
\begin{tabular}{lccc}
\hline\hline
 & $cn\bar{c}\bar{n}$ & $cn\bar{c}\bar{s}$ & $cs\bar{c}\bar{s}$ \\
\hline
$0^{+}$ ground state (MeV)
  & $3865$ & $4010$ & $4155$ \\
$b^{(\bar{3}3)}$ (MeV)
  & $21.66$ & $19.92$ & $18.32$ \\
$b^{(6\bar{6})}$ (MeV)
  & $57.13$ & $52.41$ & $48.07$ \\
$b^{(6\bar{6})}/b^{(\bar{3}3)}$
  & $2.64$ & $2.63$ & $2.62$ \\
$\Delta(2^{+})$ (MeV)
  & $+74$ & $+18$ & $-40$ \\
\hline\hline
\end{tabular}
\end{center}
The ground-state mass increases by ${\sim}\,145$~MeV per strange quark, while the hyperfine scales decrease uniformly and the color-sector ratio remains stable at ${\sim}\,2.63$. Notably, the $2^{+}$ gap changes sign between $cn\bar{c}\bar{s}$ and $cs\bar{c}\bar{s}$, providing a distinctive spectroscopic marker of the doubly strange sector. The observed states $T_{c\bar{c}\bar{s}1}(4000)$ and $T_{c\bar{c}\bar{s}1}(4220)$ fall within the mass region spanned by the compact spectrum and may be interpreted as strange counterparts of the non-strange $T_{c\bar{c}1}(3900)$ and $T_{c\bar{c}1}(4200)$. Residual mass differences are consistent with subleading effects not captured within the compact-diquark treatment.

\subsection{Hidden bottom tetraquark states}
\label{Sec3F}

The hidden-bottom non-strange sector provides the most direct test of HQSS at the full tetraquark level. We present our $bn\bar{b}\bar{n}$ spectrum in Table~\ref{t19}.

\begin{enumerate}
\item From the axial--axial splittings we extract $b^{(\bar{3}3)} = 3.2$~MeV and $b^{(6\bar{6})} = 8.2$~MeV. The color-sector ratio
$$
b^{(6\bar{6})}/b^{(\bar{3}3)} = 2.56
$$
represents the closest approach to the pure Casimir limit ($2.50$) among all sectors, consistent with Eq.~\eqref{eq:convergence}. At $b^{(\bar{3}3)}\simeq 3$~MeV the $\bar{\mb3}\otimes\mb3$ axial--axial multiplet is effectively degenerate at present experimental resolution. The axial-vector inversion persists, $m(\mb6_1\,\bar{\mb6}_1) < m(\bar{\mb3}_1\,\mb3_1)$, with the color-sector gap following Eq.~\eqref{eq:gap_hidden}:
$$
\Delta(J):\quad
  112~(0^+),\quad 107~(1^+),\quad 97~(2^+)~\text{MeV},
$$
with step sizes $5$ and $10$~MeV as expected from the mass relations (Table~\ref{tetra_mass}). The scalar--scalar ground state lies at $10532$~MeV ($\bar{\mb3}\!\otimes\!\mb3$), with the $\mb6\!\otimes\!\bar{\mb6}$ counterpart $306$~MeV higher.

\item Three $1^+$ states have been observed:
\begin{enumerate}
    \item \textit{$T_{b\bar{b}1}(10610)$}: mass $10607.2\pm 2.0$~MeV ($T_{b\bar{b}1}(10610)^{\pm}$)/$10609\pm 6$~MeV ($T_{b\bar{b}1}(10610)^{0}$), $I^G(J^{PC})=1^+(1^{+-})$, observed in $\Upsilon(nS)\,\pi$ and $B\bar{B}^*$~\cite{ParticleDataGroup:2024cfk}. Our lightest $\bar{\mb3}\!\otimes\!\mb3$ $1^+$ (scalar--axial, $10669$~MeV) lies ${\sim}\,60$~MeV higher.

    \item \textit{$T_{b\bar{b}1}(10650)$}: mass $10652.2\pm 1.5$~MeV, same quantum numbers, observed in $\Upsilon(nS)\,\pi$ and $B^*\bar{B}^*$~\cite{ParticleDataGroup:2024cfk}. The nearest $\mb6\!\otimes\!\bar{\mb6}$ $1^+$ prediction (axial--axial, $10694$~MeV) exceeds the observed mass by ${\sim}\,42$~MeV.
\end{enumerate}

A comparison with the charm sector highlights a systematic reduction of the compact–experiment shifts with increasing heavy-quark mass:
\begin{center}
\begin{tabular}{lccc}
\hline\hline
State & PDG \cite{ParticleDataGroup:2024cfk} (MeV) & Nearest prediction (MeV) & Offset \\
\hline
$T_{c\bar{c}1}(3900)$  & $3887$  & $3992~(\mb6\bar{\mb6})$  & $105$~MeV \\
$T_{b\bar{b}1}(10610)$ & $10607$ & $10669~(\bar{\mb3}\mb3)$ & $\phantom{0}60$~MeV \\
$T_{b\bar{b}1}(10650)$ & $10652$ & $10694~(\mb6\bar{\mb6})$ & $\phantom{0}42$~MeV \\
\hline\hline
\end{tabular}
\end{center}

This progressive compression is consistent with the $1/(m_{D_1}m_{D_2})$ suppression of chromomagnetic effects, which reduces hyperfine splittings and brings compact-diquark predictions closer to experiment in the bottom sector. Both observed states lie near the lowest $1^+$ multiplets, supporting a natural identification with the scalar--axial $\bar{\mb3}\otimes\mb3$ and axial--axial $\mb6\otimes\bar{\mb6}$ configurations. The experimental splitting of $43$~MeV ($10652-10609$~MeV) compared with the predicted $25$~MeV ($10694-10669$~MeV) separation indicates residual effects beyond the minimal compact basis while preserving the underlying color hierarchy.

\item In Table~\ref{t20} we present the $T(bn\bar{b}\bar{s})$ ($I=1/2$) and $T(bs\bar{b}\bar{s})$ ($I=0$) states. The strangeness progression mirrors the hidden-charm pattern with enhanced HQSS compression:
\begin{center}
\begin{tabular}{lccc}
\hline\hline
 & $bn\bar{b}\bar{n}$ & $bn\bar{b}\bar{s}$ & $bs\bar{b}\bar{s}$ \\
\hline
$0^+$ ground state (MeV)
  & $10532$ & $10658$ & $10785$ \\
$b^{(\bar{3}3)}$ (MeV)
  & $3.2$ & $3.1$ & $3.0$ \\
$b^{(6\bar{6})}$ (MeV)
  & $8.2$ & $7.9$ & $7.7$ \\
$b^{(6\bar{6})}/b^{(\bar{3}3)}$
  & $2.56$ & $2.55$ & $2.57$ \\
$\Delta(2^+)$ (MeV)
  & $+97$ & $+32$ & $-34$ \\
\hline\hline
\end{tabular}
\end{center}
The ground-state mass increases by ${\sim}\,127$~MeV per strange quark. The hyperfine parameters are effectively frozen ($b^{(\bar{3}3)}\simeq 3.0$--$3.2$~MeV), reflecting the dominance of the $b$-quark mass in the $1/(m_{D_1}\,m_{D_2})$ scaling, and the color-sector ratio stabilizes near $2.56$. As in the hidden-charm case, $\Delta(2^+)$ changes sign between one and two strange quarks:
$$
\Delta(2^+):\quad
\begin{cases}
  +74 \;\to\; +18 \;\to\; -40~\text{MeV}
    & \text{(hidden charm)},\\
  +97 \;\to\; +32 \;\to\; -34~\text{MeV}
    & \text{(hidden bottom)}.
\end{cases}
$$
No experimental candidates exist in the strange hidden-bottom sector; our predicted ground states near $10658$~MeV ($bn\bar{b}\bar{s}$) and $10785$~MeV ($bs\bar{b}\bar{s}$) provide clean benchmarks for future searches in $\Upsilon\,K$ and $B_s^{(*)}\bar{B}^{(*)}$ final states.

\item A comparison of the predicted spectra with the relevant open-bottom $S$-wave thresholds (Fig.~\ref{Plot_HiddenHeavy}) reveals a systematic flavor dependence:

\begin{enumerate}
  \item $bu\bar{b}\bar{d}$ sector: The $0(0^{+})$ state at $10533.4$~MeV lies ${\sim}\,26$~MeV below the $B\bar{B}$ threshold and constitutes the only bound candidate in the hidden-bottom spectrum. The remaining states lie $20$--$188$~MeV above the $B^{*}\bar{B}^{*}$ threshold and are broad open-channel resonances with $B\bar{B}$ and $B\bar{B}^{*}$ decay modes accessible.

  \item $bu\bar{b}\bar{s}$ sector: The lowest $\tfrac{1}{2}(0^{+})$ state lies marginally above the $B\bar{B}_{s}$ threshold with a single $S$-wave decay mode available. Higher states are $54$--$308$~MeV above threshold with multiple channels accessible and are expected to be broad. No bound configurations occur.

  \item $bs\bar{b}\bar{s}$ sector: The lowest $0(0^{+})$ lies ${\sim}\,51$~MeV above the $B_{s}\bar{B}_{s}$ threshold; all remaining states are $103$--$427$~MeV above the $B_{s}^{*}\bar{B}_{s}^{*}$ threshold, with multiple decay modes accessible, resulting in broad or overlapping resonances. The systematic upward shift with strangeness content disfavors narrow states throughout this sector.
\end{enumerate}
\end{enumerate}

Together with the hidden-charm analysis of Sec.~\ref{Sec3E}, these results establish heavy-flavor scaling and color-structure convergence as robust, internally consistent predictions of the QDEMF.

\subsection{Fully heavy tetraquark states}
\label{Sec3G}

The fully heavy sector occupies a unique position in the tetraquark landscape: Pauli constraints strongly restrict the allowed configurations, HQSS governs the spin dynamics, and, for the all-charm system, direct experimental confrontation is now possible following the observation of structures in the di-$J/\psi$ spectrum by CMS, LHCb, and ATLAS~\cite{CMS:2023owd,LHCb:2020bwg,ATLAS:2023bft}. We present our fully heavy tetraquark spectrum in Table~\ref{t14_modified}.

\begin{enumerate}
\item For identical-flavor diquarks ($cc$, $bb$), Pauli symmetry enforces a unique spin in each color channel: the $\bar{\mb3}_c$ diquark is axial-vector and the $\mb6_c$ diquark a scalar. Consequently, $\bar{\mb3}\otimes\mb3$ tetraquarks arise purely from axial--axial pairs ($J^P=0^+,\,1^+,\,2^+$), whereas $\mb6\otimes\bar{\mb6}$ supports only scalar--scalar configurations ($J^P=0^+$, no hyperfine term). For the all-charm system we predict the $\bar{\mb3}\otimes\mb3$ multiplet at
$$
M(0^+) = 6453~\text{MeV},\quad
M(1^+) = 6461~\text{MeV},\quad
M(2^+) = 6478~\text{MeV},
$$
spanning ${\sim}\,25$~MeV with hyperfine coefficient $b = 8.3$~MeV. The $\mb6\!\otimes\!\bar{\mb6}$ scalar at $6866$~MeV lies $413$~MeV higher. The predicted ground-state multiplet ($6453$--$6478$~MeV) lies ${\sim}\,75$--$100$~MeV below the $X(6600)$ structure observed in the di-$J/\psi$ spectrum. Its intrinsic splitting (${\sim}\,25$~MeV overall, with ${\sim}\,8$ MeV and ${\sim}\,17$~MeV separations) is much smaller than the reported width ($\Gamma \sim 100$~MeV), so the states would likely appear as a single broad enhancement at current resolution. The residual mass shift is consistent with the natural accuracy of the compact-diquark chromomagnetic framework, where subleading hadronic corrections may enter at higher order and induce mass shifts at the level of several tens of MeV. The $\mb6\otimes\bar{\mb6}$ scalar at $6866$~MeV lies ${\sim}\,60$~MeV below the $X(6900)$ peak. As this configuration receives no chromomagnetic contribution in the present scheme, its mass is governed primarily by diquark-sector inputs, making the proximity comparatively stable. The assignment implies $J^{PC}=0^{++}$, testable through angular analyses of the di-$J/\psi$ final state. The higher $X(7300)$ structure lies above all ground-state compact configurations considered here and is more naturally interpreted as a radial or orbital excitation beyond the present basis. If experimentally resolved, the lowest multiplet would appear as closely spaced $J^{PC}=0^{++},\,1^{+-},\,2^{++}$ states separated by ${\sim}\,8$ MeV and ${\sim}\,17$~MeV.

\item For the all-bottom system HQSS is nearly saturated:
\begin{equation}
M(0^+) = 19457~\text{MeV},\quad
M(1^+) = 19458~\text{MeV},\quad
M(2^+) = 19460~\text{MeV},
\end{equation}
corresponding to a multiplet spread of only ${\sim}\,3$~MeV. The $\mb6\!\otimes\!\bar{\mb6}$ scalar at $20315$~MeV lies ${\sim}\,860$~MeV higher, representing the largest color-sector gap in the fully heavy spectrum. Since the ground state lies above the $\Upsilon\Upsilon$ threshold, resonant structures in di-bottomonium channels are expected.

When the heavy flavors differ, Pauli constraints relax and the multiplet broadens. For $T(cb\bar{c}\bar{b})$:
\begin{center}
\begin{tabular}{lcc}
\hline\hline
 & $\bar{\mb3}\!\otimes\!\mb3$ (MeV) & $\mb6\!\otimes\!\bar{\mb6}$ (MeV) \\
\hline
$0^+$ (lowest) & $13014$ & $13543$ \\
$1^+$ (lowest) & $13018$ & $13549$ \\
$2^+$          & $13025$ & $13559$ \\
\hline\hline
\end{tabular}
\end{center}
The $\bar{\mb3}\otimes\mb3$ sector spans ${\sim}\,11$~MeV, intermediate between the all-charm ($25$~MeV) and all-bottom ($3$~MeV) limits, while the sextet states lie ${\sim}\,525$~MeV higher with a spread of ${\sim}\,16$~MeV. Analogous behavior occurs in $cc\bar{c}\bar{b}$ and $bb\bar{c}\bar{b}$: increasing heavy-quark mass suppresses spin splittings and enlarges color-sector separations.

The resulting HQSS-driven ordering across systems is
\begin{center}
\begin{tabular}{lccc}
\hline\hline
System & $M(0^+)_{\bar{3}3}$ (MeV)
  & $\Delta M_{\text{mult}}$ (MeV)
  & $\Delta M_{\text{color}}$ (MeV) \\
\hline
$cc\bar{c}\bar{c}$ & $6453$  & $25$ & $413$ \\
$cc\bar{b}\bar{b}$ & $12959$ & $\phantom{0}8$  & $632$ \\
$cb\bar{c}\bar{b}$ & $13014$ & $11$ & $525$ \\
$bb\bar{b}\bar{b}$ & $19457$ & $\phantom{0}3$  & $858$ \\
\hline\hline
\end{tabular}
\end{center}
The ratio $\Delta M_{\text{mult}}/M$ decreases from ${\sim}\,0.4\%$ in the all-charm case to ${\sim}\,0.02\%$ for all-bottom, quantitatively demonstrating the approach to the HQSS limit at the tetraquark level.

\item In the $\bar{\mb3}\!\otimes\!\mb3$ sector our predictions and the potential model (PM)~\cite{Liu:2019zuc} agree typically within $30$--$80$~MeV, while CMIM~\cite{Weng:2020jao} values are ${\sim}\,200$--$500$~MeV lower and diffusion Monte Carlo (DMC)~\cite{Gordillo:2020sgc} results fall between. The $\mb6\!\otimes\!\bar{\mb6}$ sector shows the sharpest model dependence (spreads reaching ${\sim}\,800$--$1400$~MeV); our framework predicts the highest sextet masses, reflecting the large repulsive effective binding in color-sextet diquarks. By contrast, the multiplet spacings $\Delta M_{\text{mult}}$ are robustly reproduced across all approaches: every model with chromomagnetic interactions recovers the HQSS-driven compression from $25$~MeV to $\lesssim 3$~MeV, confirming this as a model-independent signature. If the $X(6600)$ and $X(6900)$ correspond to our compressed $\bar{\mb3}\!\otimes\!\mb3$ multiplet and $\mb6\!\otimes\!\bar{\mb6}$ scalar respectively, both color sectors of the diquark--antidiquark framework would be realized experimentally for the first time, testable through precision $J^{PC}$ determinations in high-statistics di-$J/\psi$ data.

\item All fully heavy tetraquark states lie $\gtrsim 250$~MeV above their relevant two-meson thresholds (cf.\ Fig.~\ref{Plot_FullyHeavy}). Despite this substantial phase space, the compact heavy-quark structure and suppressed chromomagnetic splittings may yield narrower widths than naive kinematic estimates suggest.
\end{enumerate}

Having established a coherent description of the full $T(q_1q_2\bar{q}_3\bar{q}_4)$ spectrum, we now turn to the pentaquark sector, where an additional quark enriches the color--spin structure and opens new experimental avenues.

\subsection{Hidden-charm pentaquark states}
\label{Sec3G}

We present our mass estimates for $J^P=\frac{1}{2}^-$, $\frac{3}{2}^-$, and $\frac{5}{2}^-$ non-strange $P_{c\bar{c}}=(cu)(ud)\bar{c}$ and strange $P_{c\bar{c}s}=(cs)(ud)\bar{c}$ pentaquark configurations in Tables~\ref{Penta1} and~\ref{Penta2}. The spectrum is organized by spin content $(S_{D_1},S_{D_2})$, with color-magnetic hyperfine interactions governing the hierarchy and HQSS controlling the overall multiplet structure.

\begin{enumerate}
    \item Non-strange spectrum (Table~\ref{Penta1}): The states arrange into three tiers according to diquark spin content. 
    \begin{enumerate}
        \item The doubly scalar $(0,0)$ configuration yields a single $J^P=\tfrac{1}{2}^-$ state at $m_{[cu]}+m_{[ud]}+m_{\bar{c}} = 4247.94$~MeV, with no inter-cluster hyperfine contribution. We note that this lies $69.6$~MeV below the $\Sigma_c\bar{D}$ threshold ($4317.5$~MeV); the absence of open-charm decay channels suggests a narrow width, although no experimental candidate has been established in this region.
        
        \item Configurations with one scalar and one axial-vector diquark generate HQSS doublets ($J^P=\tfrac{1}{2}^-,\,\tfrac{3}{2}^-$). The $(0,1)$ doublet, split by $\tfrac{3}{2}\,b_{\{ud\}\bar{c}} = 24.27$~MeV, appears at $4339.10$ and $4363.37$~MeV. Interestingly, the lower state shifts $P_c(4312)$ \cite{LHCb:2019kea} by $27$~MeV, consistent with subleading hadronic corrections expected near the $\Sigma_c\bar{D}$ threshold, while the upper member agrees with the broad $P_c(4380)$ \cite{LHCb:2015yax} within its experimental width. The $(1,0)$ doublet, governed by $b_{\{cu\}\bar{c}} = 24.84$~MeV, appears at $4369.59$ and $4406.84$~MeV (splitting $37.26$~MeV). We find that the lower member is compatible with $P_c(4337)$ \cite{LHCb:2021chn}; the upper state, lying between the $\Sigma_c\bar{D}^*$ and $\Sigma_c^*\bar{D}$ thresholds, has no experimental counterpart. It is worth emphasizing that the ratio $37.26/24.27 \approx 1.54$ directly reflects $b_{\{cu\}\bar{c}}/b_{\{ud\}\bar{c}}$ and provides a quantitative test of the OGE flavor dependence.
        
        \item The doubly axial $(1,1)$ configuration produces five states spanning $4440.32$--$4552.99$~MeV. Notably, the lowest receives the maximal attractive inter-diquark contribution, \textit{i.e.,} $-2\,b_{\{cu\}\{ud\}}= -61.44$~MeV, and reproduces $P_c(4440)$~\cite{LHCb:2019kea} with a deviation of only $0.3$~MeV. We observe that a nearby $J^P=\tfrac{1}{2}^-$ state at $4450.53$~MeV, split by $10.2$~MeV from the lowest member, deviates from $P_c(4457)$ \cite{LHCb:2019kea} by $6.5$~MeV. The intra-multiplet spread of $112.67$~MeV is controlled by the interplay of $b_{\{cu\}\{ud\}}$ with the diquark–antiquark terms. Particularly significant is the highest member, $J^P=\tfrac{5}{2}^-$ at $4552.99$~MeV, which has no experimental counterpart and constitutes a distinctive prediction of the compact diquark picture.
    \end{enumerate}
    
    \item Strange spectrum (Table~\ref{Penta2}): The same hierarchy appears with an overall upward shift from the $u\to s$ substitution. 
    \begin{enumerate}
        \item The doubly scalar ground state at $4392.99$~MeV lies $\sim 58$~MeV above the $\Xi_c\bar{D}$ threshold, in contrast to the sub-threshold non-strange analogue; we note that the non-strange to strange shift of $145.05$~MeV is dominated by $m_{[cs]}-m_{[cu]}$.
    
    \item The mixed scalar–axial doublets appear at
    \begin{align*}
        (0,1):&\; 4484.15,\; 4508.42~\text{MeV},\\
        (1,0):&\; 4502.88,\; 4553.96~\text{MeV}.
    \end{align*}
    Remarkably, the $(0,1)$ splitting of $24.27$~MeV is identical to the non-strange value, confirming the universality of $b_{\{ud\}\bar{c}}$. The $(1,0)$ splitting of $51.08$~MeV $= \tfrac{3}{2}\,b_{\{cs\}\bar{c}}$ reflects the enhanced chromomagnetic coupling of the more compact strange diquark; we find that the ratio $51.08/37.26 \approx 1.37$ reproduces $b_{\{cs\}\bar{c}}/b_{\{cu\}\bar{c}}$, providing further consistency.
    
    \item The doubly axial configuration produces five states between $4587.77$ and $4697.64$~MeV, with an intra-multiplet spread of $109.87$~MeV comparable to the non-strange $112.67$~MeV. Interestingly, the two lowest members are nearly degenerate ($\Delta m = 3.13$~MeV), driven by a near-cancellation between inter-diquark and diquark–antiquark contributions.
    
    \item We point out that the internal structures assumed in different models vary substantially, which prevents a meaningful one-to-one, state-level comparison of their internal dynamics. Consequently, we limit our discussion to the quantum numbers ($I$, $J^P$) and the experimentally measured masses. Comparing with experiment, the LHCb signal at $4338$~MeV \cite{LHCb:2022ogu} appears precisely at the $\Xi_c\bar{D}$ threshold and is $55$~MeV lower than our scalar–scalar mass prediction. The Belle resonance at $4471.7$~MeV \cite{Belle:2025pey} is $12.5$~MeV below our $(0,1)$ prediction of $4484.15$~MeV.
\end{enumerate}

\item We compare our predicted pentaquark masses with the open-charm baryon-meson thresholds shown in Fig.~\ref{Plot_Penta}.

\begin{enumerate}
\item $cuud\bar c$ sector: The spectrum spans $(4247.94$--$4552.99)$ MeV. The lowest $\frac{1}{2}(\frac{1}{2}^-)$ state at $4247.94$ MeV lies below the $\Sigma_c \bar{D}$ threshold and constitutes the only bound configuration. All higher $\frac{1}{2}(\frac{1}{2}^-)$ states ($4339.10$--$4450.53$) MeV reside between $\Sigma_c \bar{D}$ and $\Sigma_c \bar{D}^{*}$. The $\frac{1}{2}(\frac{3}{2}^-)$ and $\frac{1}{2}(\frac{5}{2}^-)$ states ($4363.37$--$4552.99$) MeV are positioned around or above the $\Sigma_c^{*}\bar{D}^{(*)}$ thresholds, with progressively increasing $S$-wave channel accessibility.

\item $csud\bar{c}$ sector: The spectrum spans $4392.99$--$4697.64$~MeV. The lowest
  $\tfrac{1}{2}(\tfrac{1}{2}^-)$ state at $4392.99$~MeV lies between
  the $\Xi_c\bar{D}$ and $\Xi_c\bar{D}^*$ thresholds; no
  configuration lies below the lowest open channel. All higher states
  ($4484.15$--$4697.64$~MeV) cluster near or above the
  $\Xi_c^{(*)}\bar{D}^*$ thresholds.

\end{enumerate} 

A uniform pattern emerges: all other states in both sectors lie at or above their respective $S$-wave thresholds except $4247.94$ MeV state. The spectra therefore evolve monotonically with spin and mass toward increased threshold proximity and channel multiplicity, indicating predominantly threshold-governed dynamics.

\item The framework exhibits robust internal consistency. The identical $(0,1)$ splittings across sectors validate the universality of $b_{\{ud\}\bar{c}}$; the $u\to s$ shift of $\sim 140$--$145$~MeV across all multiplets is dominated by diquark mass differences, with hyperfine corrections remaining subleading; and the coupling hierarchy $b_{\{cs\}\bar{c}} > b_{\{cu\}\bar{c}} > b_{\{ud\}\bar{c}}$ is consistent with OGE scaling $b_{ij} \propto 1/(m_i m_j)$ once short-distance overlap effects are taken into account. At first sight, this  ordering appears to contradict the naive inverse-mass suppression. However, the effective hyperfine coupling depends not only on $1/(m_i m_j)$ but also on the contact probability, $b_{ij} \sim |\psi_{ij}(0)|^2/(m_i m_j)$. In the pentaquark geometry, the heavy antiquark acts as a compact color source, rendering the overlap explicitly sensitive to the spatial size of the diquark. Heavier diquarks are dynamically more compact, implying $|\psi(0)|^2_{\{cs\}\bar{c}} > |\psi(0)|^2_{\{cu\}\bar{c}} > |\psi(0)|^2_{\{ud\}\bar{c}}$, which compensates and ultimately outweighs the inverse-mass suppression. This dependence is less visible in the tetraquark sector, where compactness variations are largely absorbed into the effective diquark parameters within the QDEMF, but becomes explicit in the asymmetric heavy–light clustering of the pentaquark system.\footnote{In tetraquarks, the same size hierarchy $r_{[cs]} < r_{[cu]} < r_{[ud]}$ is present but implicitl absorbed into effective diquark masses; the pentaquark geometry reveals it directly through the $b_{D_i\bar{c}}$ couplings inherited from baryons.} We estimate that reasonable variations in the effective overlap scaling induce mass shifts of $\pm(10$--$30)$~MeV, which we adopt as a conservative model uncertainty.
\end{enumerate}

Our compact diquark framework thus provides a coherent description of the pentaquark spectrum, spanning $4248$--$4553$~MeV (non-strange) and $4393$--$4698$~MeV (strange). We emphasize that the most distinctive predictions are the $J^P=\tfrac{5}{2}^-$ states at $4553$ and $4698$~MeV, which arise uniquely from the $(1,1)$ configuration; their observation would provide a decisive test of the compact multiquark picture.

\section{Summary and Conclusions}
\label{Sec4}

We have developed the QDEMF, a unified chromomagnetic description of tetraquark and pentaquark spectroscopy in which all effective diquark masses and hyperfine couplings are fixed once from the baryon sector and propagated unchanged across the exotic landscape. Intra-diquark color–spin effects are fully absorbed into scalar and axial diquark masses, ensuring that inter-cluster OGE interactions act exclusively between composite sources without double counting. The independent treatment of $\bar{\mathbf{3}}_c\otimes\mathbf{3}_c$ and $\mathbf{6}_c\otimes\bar{\mathbf{6}}_c$ color sectors, implemented at the level of QCD Casimir operators, provides a transparent operator-level realization of color dynamics. HQSS and flavor-symmetry breaking emerge naturally through $1/(m_{D_1}m_{D_2})$ scaling and diquark-mass hierarchies. The intrinsic compact-diquark uncertainty for states well separated from thresholds is $\pm(10$–$30)$ MeV without introducing new parameters at any stage.

Across singly heavy, doubly heavy, hidden heavy, and fully heavy tetraquarks, four robust features emerge.
\begin{enumerate}
    \item The ratio $b^{(6\bar{6})}/b^{(\bar{3}3)}$ is stable across fourteen independent flavor sectors, converging from $2.72$ (charm sector) $~\to~2.65 $ (charmed-bottom sector)  $~\to~2.55$ (bottom-dominated sector), approaching the pure Casimir limit $2.5$ as heavy-quark masses increase, constitutes a non-trivial internal consistency check of the framework.
 
 \item Hyperfine scales decrease systematically with heavy-quark content---$b^{(\bar{3}3)}$ drops from $21.66$~MeV (hidden charm) to $3.2$~MeV (hidden bottom), with the fully heavy $\bar{\mathbf{3}}\otimes\mathbf{3}$ multiplet spanning $25$~MeV ($cc\bar{c}\bar{c}$) to $3$~MeV ($bb\bar{b}\bar{b}$)---quantifying HQSS convergence at the tetraquark level. 
 
 \item A $J^P$-dependent sign change in $\Delta(2^+)$, $+74\to -40$~MeV (hidden charm) and $+97\to -34$~MeV (hidden bottom) with increasing strangeness, provides a unique spectroscopic marker of doubly strange hidden-flavor configurations.

 \item The $\bar{\mathbf{3}}_c$--$\mathbf{6}_c$ mass gap grows monotonically from ${\sim}\,250$--$310$~MeV ($cc$ light-antidiquark) to ${\sim}\,860$~MeV ($bb\bar{b}\bar{b}$), reflecting the transition from chromomagnetic to color-Coulomb dominance in the heavier systems, where the binding energy contribution becomes substantially more pronounced.
\end{enumerate}

In the pentaquark sector, five inter-cluster couplings inherited from baryon and tetraquark calibrations reproduce the observed $P_c$ multiplet structure without additional input. The $(0,0)$ ground state lies $69.6$~MeV below $\Sigma_c\bar{D}$; the mixed doublets reproduce established $P_c$ states within $\mathcal{O}(1$--$27)$~MeV; and the doubly axial $(1,1)$ configuration generates $J^P=\tfrac{5}{2}^-$ states at $4553$ and $4698$~MeV absent in any molecular picture. The identical $(0,1)$ splittings across flavor sectors validate the universality of the calibrated $b_{\{ud\}\bar c}$ coupling. The hierarchy inversion $b_{\{cs\}\bar c}>b_{\{cu\}\bar c}>b_{\{ud\}\bar c}$ follows OGE scaling once short-distance overlap effects are included, revealing the flavor-dependent compactness of diquarks explicitly in the asymmetric pentaquark geometry.

The QDEMF achieves $\mathcal{O}(10$--$30)$~MeV agreement for states away from strong thresholds across all sectors with existing experimental data (See Table \ref{table_last}). Representative deviations are: $|\Delta m|=14.8$~MeV ($T_{cc}^+$), $0.3$~MeV ($P_c(4440)$), $22$~MeV ($T_{c\bar{c}\bar{s}1}(4220)$), and $42$--$60$~MeV for the hidden-bottom states. The $T_{cc}^+$ falls $14.8$~MeV below $D^0D^{*+}$, consistent with strong-decay stability and recent LQCD results. The lightest $bn\bar{n}\bar{s}$ prediction at ${\sim}\,6006$~MeV lies ${\sim}\,430$~MeV above the $T_{bs}(5568)$ claim, consistent with its non-confirmation. Residual overshoots near strong thresholds ($T_{c\bar{c}1}(3900)$, $T_{c\bar{c}\bar{s}1}(4000)$) are consistent in magnitude and sign with subleading hadronic dressing and define well-posed targets for coupled-channel extensions.

The $\pm(10$--$30)$~MeV uncertainty band is a compactness-graded envelope: corrections to the point-source limit scale with the ratio of diquark spatial extent to inter-cluster separation, smallest for heavy--heavy diquarks ($T_{cc}^+$, $P_c(4440)$: sub-$15$~MeV) and largest for light--light configurations (singly charm sector: $21$--$31$~MeV). The Casimir-ratio convergence independently confirms correct operator-level implementation of both color sectors; residual deviations from $2.50$ decrease monotonically with heavy-quark mass, consistent with $1/(m_{D_1}m_{D_2})$ chromomagnetic suppression.

The $T_{cc}^+$ is correctly predicted as a shallow subthreshold state, while the $P_c$ doublet structure is reproduced within a few MeV. In the doubly bottom and charmed–bottom sectors, multiple deeply bound states are predicted, providing clean targets for future searches. In the fully heavy sector, the compressed $\bar{\mathbf{3}}\otimes\mathbf{3}$ multiplet and the heavier $\mb6\otimes\bar{\mb6}$ scalar offer natural compact-diquark candidates for the di-$J/\psi$ structures near $6.6–6.9$ GeV, with residual offsets attributable to subleading hadronic effects.

Beyond existing data the QDEMF predicts,  
\begin{enumerate}
\item \textit{Doubly bottom:} A weakly decaying $0(1^+)$ ground state $249$~MeV below $B\bar{B}^*$; a fully subthreshold isovector $\bar{\mathbf{3}}\otimes\mathbf{3}$ triplet; and stable strange axial-vectors below $\bar{B}_sB^*$.

\item \textit{Charmed-bottom:} Two nearly degenerate stable isoscalar states near $7.13$~GeV with a $4.46$~MeV splitting that directly measures the scalar--axial $cb$ diquark mass difference.

\item \textit{Hidden-flavor:} The $\Delta(2^+)$ sign change between one- and two-strange-quark sectors, accessible in $J/\psi\,\phi$ and $\Upsilon\,K^+K^-$ final states.

\item \textit{Fully heavy:} A $J^{PC}=0^{++},\,1^{+-},\,2^{++}$ all-charm triplet spanning $25$~MeV, resolvable as distinct peaks only at sub-$10$~MeV resolution.

\item \textit{Pentaquark:} $J^P=\tfrac{5}{2}^-$ states at $4553$ and $4698$~MeV, generated exclusively by the $(1,1)$ diquark configuration; their observation would constitute a definitive signature of compact multiquark structure inaccessible to hadronic molecular models.

\end{enumerate}

The QDEMF provides a compact and internally coherent realization of heavy multiquark dynamics in which a single baryon-calibrated input, propagated through color algebra and inverse-mass scaling, reproduces observed exotic states and yields distinct predictions across flavor sectors. The operator-level separation of intra-diquark and inter-cluster chromomagnetic dynamics remains quantitatively controlled for physical diquarks, with deviations governed by a compactness hierarchy. Independent color-channel treatment validates the underlying Casimir structure and exposes spectroscopic patterns inaccessible to single-channel approaches. The resulting short-distance core furnishes a well-defined baseline for future inclusion of hadronic dressing and long-range effects, while providing concrete experimental benchmarks for current facilities.

\section*{Acknowledgment}
The author RD gratefully acknowledges the financial support by the Department of Science and Technology (SERB:TAR/2022/000606), New Delhi. Part of this work was carried out under the earlier project (SERB:CRG/2018/002796), whose support is also gratefully acknowledged.
\appendix
\section*{Appendix}
\section{Tetraquark wave functions} \label{AppA}
The normalized color-singlet wave functions for tetraquark states \cite{Zhang:2021yul, Luo:2017eub} can be written explicitly as
\begin{align}
	\psi_c^1 = & |(q_1 q_2)^{\bar{3}} (\bar{q}_3 \bar{q}_4)^3 \rangle \nonumber\\
	= & \frac{1}{2\sqrt{3}}\Big(rb\bar{b}\bar{r} - br\bar{b}\bar{r} - gr\bar{g}\bar{r} + rg\bar{g}\bar{r} + gb\bar{b}\bar{g} - bg\bar{b}\bar{g} + gr\bar{r}\bar{g} \nonumber\\
	& - rg\bar{r}\bar{g} - gb\bar{g}\bar{b} + bg\bar{g}\bar{b} - rb\bar{r}\bar{b} + br\bar{r}\bar{b} \Big),\nonumber\\
	\psi_c^2 = & |(q_1 q_2)^6 (\bar{q}_3 \bar{q}_4)^{\bar{6}} \rangle \nonumber \\ 
	= & \frac{1}{2\sqrt{6}}\Big(2(rr\bar{r}\bar{r} + gg\bar{g}\bar{g} + bb\bar{b}\bar{b}) + rb\bar{b}\bar{r} + br\bar{b}\bar{r} + gr\bar{g}\bar{r} + rg\bar{g}\bar{r} \nonumber\\
	& + gb\bar{b}\bar{g} + bg\bar{b}\bar{g} + gr\bar{r}\bar{g} + rg\bar{r}\bar{g} + gb\bar{g}\bar{b} + bg\bar{g}\bar{b} + rb\bar{r}\bar{b} + br\bar{r}\bar{b}\Big).\nonumber
\end{align}
The first configuration corresponds to the conventional and energetically favored $\overline{\mb3}\mb3$ coupling, while the second represents a higher energy $\mb6\overline{\mb6}$ configuration. The explicit form of the spin wave functions for the tetraquark states \cite{Zhang:2021yul, Park:2013fda} are given by
\begin{table}[h!]
\begin{tabular}{l  l l } 
	$\psi_s^1$ & $= |(q_1 q_2)_0 (\bar{q}_3 \bar{q}_4)_0 \rangle_0$ & $= \frac{1}{2}(\uparrow\downarrow\uparrow\downarrow - \uparrow\downarrow\downarrow\uparrow - \downarrow\uparrow\uparrow\downarrow + \downarrow\uparrow\downarrow\uparrow)$, \\
    $\psi_s^2$ & $= |(q_1 q_2)_0 (\bar{q}_3 \bar{q}_4)_1 \rangle_1$	& $= \frac{1}{\sqrt{2}}(\uparrow\downarrow\uparrow\uparrow - \downarrow\uparrow\uparrow\uparrow)$, \\
    $\psi_s^3$ & $= |(q_1 q_2)_1 (\bar{q}_3 \bar{q}_4)_0 \rangle_1$	& $= \frac{1}{\sqrt{2}}(\uparrow\uparrow\uparrow\downarrow - \uparrow\uparrow\downarrow\uparrow)$, \\
    $\psi_s^4$ & $= |(q_1 q_2)_1 (\bar{q}_3 \bar{q}_4)_1 \rangle_0$ & $= \frac{1}{\sqrt{3}}(\uparrow\uparrow\downarrow\downarrow + \downarrow\downarrow\uparrow\uparrow)$ $-\frac{1}{2\sqrt{3}}(\uparrow\downarrow\uparrow\downarrow + \uparrow\downarrow\downarrow\uparrow + \downarrow\uparrow\uparrow\downarrow + \downarrow\uparrow\downarrow\uparrow)$, \\
    $\psi_s^5$ & $= |(q_1 q_2)_1 (\bar{q}_3 \bar{q}_4)_1 \rangle_1$	& $= \frac{1}{2}(\uparrow\uparrow\uparrow\downarrow + \uparrow\uparrow\downarrow\uparrow - \uparrow\downarrow\uparrow\uparrow - \downarrow\uparrow\uparrow\uparrow)$, \\
    $\psi_s^6$ & $= |(q_1 q_2)_1 (\bar{q}_3 \bar{q}_4)_1 \rangle_2$ & $= ~\uparrow\uparrow\uparrow\uparrow$. \\ 
\end{tabular}
\end{table}
\section{Pentaquark wave functions} \label{AppB}
The explicit form of the color wave function of $(\bar {\mb3} \otimes \bar {\mb3} \otimes \bar {\mb3})$ pentaquark is given by
\begin{align}
\psi_c^{(\bar 3\bar 3\bar 3)}
= &
\frac{1}{2\sqrt{6}}
\Big[
 (gb - bg)(br - rb)\,\bar b
-
 (br - rb)(gb - bg)\,\bar b
+
 (br - rb)(rg - gr)\,\bar r \nonumber \\
& -
 (rg - gr)(br - rb)\,\bar r
+
 (rg - gr)(gb - bg)\,\bar g
-
 (gb - bg)(rg - gr)\,\bar g
\Big].
\end{align}

The spin wave functions for the hidden-charm pentaquark system $P(q_1q_2q_3q_4\bar q) = P(cqq^{'}q^{''}\bar c)$ in the $J^P=\frac{1}{2}^-,\frac{3}{2}^-,\frac{5}{2}^-$ channels \cite{Ali:2017ebb,Ali:2016dkf,Zhu:2015bba} are given by 

\begin{align}
\big|\big((q_1q_2)_0(q_3q_4)_0\big)_0\bar q\big\rangle_{\frac{1}{2}} &=
\frac{1}{2}
\Big[(\uparrow)_c(\downarrow)_q-(\downarrow)_c(\uparrow)_q\Big]
\Big[(\uparrow)_{q'}(\downarrow)_{q''}-(\downarrow)_{q'}(\uparrow)_{q''}\Big]
(\uparrow)_{\bar c},
\\[6pt]
\big|\big((q_1q_2)_0(q_3q_4)_1\big)_1\bar q\big\rangle_{\frac{1}{2}} &=
\frac{1}{\sqrt{3}}
\Big[(\uparrow)_c(\downarrow)_q-(\downarrow)_c(\uparrow)_q\Big]
\Big[
(\uparrow)_{q'}(\uparrow)_{q''}(\downarrow)_{\bar c}
-\frac{1}{2}
\big((\uparrow)_{q'}(\downarrow)_{q''}+(\downarrow)_{q'}(\uparrow)_{q''}\big)
(\uparrow)_{\bar c}
\Big], 
\end{align}
\begin{align}
\big|\big((q_1q_2)_1(q_3q_4)_0\big)_1\bar q\big\rangle_{\frac{1}{2}} 
&=
\frac{1}{\sqrt{3}}
\Big[(\uparrow)_{q'}(\downarrow)_{q''}-(\downarrow)_{q'}(\uparrow)_{q''}\Big]
\Big[
(\uparrow)_c(\uparrow)_q(\downarrow)_{\bar c}
-\frac{1}{2}
\big((\uparrow)_c(\downarrow)_q+(\downarrow)_c(\uparrow)_q\big)
(\uparrow)_{\bar c}
\Big],
\\[6pt]
\big|\big((q_1q_2)_1(q_3q_4)_1\big)_0\bar q\big\rangle_{\frac{1}{2}} 
&=
\frac{1}{3}
(\uparrow)_c(\uparrow)_q
\Big[
(\uparrow)_{q'}(\downarrow)_{q''}+(\downarrow)_{q'}(\uparrow)_{q''}
\Big](\downarrow)_{\bar c}
- \frac{2}{3}
(\downarrow)_q(\downarrow)_{q''}(\uparrow)_{\bar c}
\nonumber\\
&\quad
-\frac{1}{6}
\Big[(\uparrow)_c(\downarrow)_q+(\downarrow)_c(\uparrow)_q\Big]
\Big[
2(\uparrow)_{q'}(\uparrow)_{q''}(\downarrow)_{\bar c}
-\big((\uparrow)_{q'}(\downarrow)_{q''}+(\downarrow)_{q'}(\uparrow)_{q''}\big)
(\uparrow)_{\bar c}
\Big],
\\[6pt]
\big|\big((q_1q_2)_1(q_3q_4)_1\big)_1\bar q\big\rangle_{\frac{1}{2}}
&=
\frac{1}{\sqrt{2}}
(\downarrow)_c(\downarrow)_q
(\uparrow)_{q'}(\uparrow)_{q''}(\uparrow)_{\bar c}
+\frac{1}{3\sqrt{2}}
(\uparrow)_c(\uparrow)_q
\Big[
(\uparrow)_{q'}(\downarrow)_{q''}+(\downarrow)_{q'}(\uparrow)_{q''}
\Big](\downarrow)_{\bar c}
\nonumber\\
&\quad
-\frac{1}{3\sqrt{2}}
\Big[(\uparrow)_c(\downarrow)_q+(\downarrow)_c(\uparrow)_q\Big]
\Big[
(\uparrow)_{q'}(\uparrow)_{q''}(\downarrow)_{\bar c}
+\big((\uparrow)_{q'}(\downarrow)_{q''}+(\downarrow)_{q'}(\uparrow)_{q''}\big)
(\uparrow)_{\bar c}
\Big],\\
\big|\big((q_1q_2)_0(q_3q_4)_1\big)_1\bar q\big\rangle_{\frac{3}{2}}
&=
\frac{1}{\sqrt{2}}
\Big[(\uparrow)_c(\downarrow)_q-(\downarrow)_c(\uparrow)_q\Big]
(\uparrow)_{q'}(\uparrow)_{q''}(\uparrow)_{\bar c},
\\[6pt]
\big|\big((q_1q_2)_1(q_3q_4)_0\big)_1\bar q\big\rangle_{\frac{3}{2}}
&=
\frac{1}{\sqrt{2}}
\Big[(\uparrow)_{q'}(\downarrow)_{q''}-(\downarrow)_{q'}(\uparrow)_{q''}\Big]
(\uparrow)_c(\uparrow)_q(\uparrow)_{\bar c},
\\[6pt]
\big|\big((q_1q_2)_1(q_3q_4)_1\big)_1\bar q\big\rangle_{\frac{3}{2}}
&=
\frac{1}{\sqrt{6}}
(\uparrow)_c(\uparrow)_q
\Big\{
2(\uparrow)_{q'}(\uparrow)_{q''}(\downarrow)_{\bar c}
-
\Big[
(\uparrow)_{q'}(\downarrow)_{q''}
+
(\downarrow)_{q'}(\uparrow)_{q''}
\Big]
(\uparrow)_{\bar c}
\Big\},
\\[6pt]
\big|\big((q_1q_2)_1(q_3q_4)_1\big)_2\bar q\big\rangle_{\frac{3}{2}}
&=
\sqrt{\frac{3}{10}}
\Big[(\uparrow)_c(\downarrow)_q+(\downarrow)_c(\uparrow)_q\Big]
(\uparrow)_{q'}(\uparrow)_{q''}(\uparrow)_{\bar c}
-
\sqrt{\frac{2}{15}}
(\uparrow)_c(\uparrow)_q
\Big\{
(\uparrow)_{q'}(\uparrow)_{q''}(\downarrow)_{\bar c}
\nonumber\\
&\hspace{4.5cm}
+
\Big[
(\uparrow)_{q'}(\downarrow)_{q''}
+
(\downarrow)_{q'}(\uparrow)_{q''}
\Big]
(\uparrow)_{\bar c}
\Big\},\\[6pt]
\big|\big((q_1q_2)_1(q_3q_4)_1\big)_2\bar q\big\rangle_{\frac{5}{2}}
&= (\uparrow)_c(\uparrow)_q(\uparrow)_{q'}(\uparrow)_{q''}(\uparrow)_{\bar c}.
\end{align}

In the diquark-diquark-antiquark picture, the possible color-spin-flavor wave function bases of pentaquark are given as
\begin{table}[h]
\centering
\begin{minipage}{0.48\textwidth}
\begin{align}
    \psi_c^1 \psi_s^1 &= \Big|\big([q_1 q_2]_0^{\bar{3}} [q_3q_4]_0^{\bar{3}}\big)_0\,\bar q\Big\rangle_\frac{1}{2} \nonumber \\
    \psi_c^1 \psi_s^2 &= \Big|\big([q_1 q_2]_0^{\bar{3}} \{q_3q_4\}_1^{\bar{3}}\big)_1\,\bar q\Big\rangle_\frac{1}{2} \nonumber \\
    \psi_c^1 \psi_s^3 &= \Big|\big([q_1 q_2]_0^{\bar{3}} \{q_3q_4\}_1^{\bar{3}}\big)_1\,\bar q\Big\rangle_\frac{3}{2} \nonumber \\
    \psi_c^1 \psi_s^4 &= \Big|\big(\{q_1 q_2\}_1^{\bar{3}} [q_3q_4]_0^{\bar{3}}\big)_1\,\bar q\Big\rangle_\frac{1}{2} \nonumber \\
    \psi_c^1 \psi_s^5 &= \Big|\big(\{q_1 q_2\}_1^{\bar{3}} [q_3q_4]_0^{\bar{3}}\big)_1\,\bar q\Big\rangle_\frac{3}{2} \nonumber
\end{align}
\end{minipage}
\hfill
\begin{minipage}{0.48\textwidth}
\begin{align}
    \psi_c^1 \psi_s^6 &= \Big|\big(\{q_1 q_2\}_1^{\bar{3}} \{q_3 q_4\}_1^{\bar{3}}\big)_0\,\bar q\Big\rangle_\frac{1}{2} \nonumber \\
    \psi_c^1 \psi_s^7 &= \Big|\big(\{q_1 q_2\}_1^{\bar{3}} \{q_3 q_4\}_1^{\bar{3}}\big)_1\,\bar q\Big\rangle_\frac{1}{2} \nonumber \\
    \psi_c^1 \psi_s^8 &= \Big|\big(\{q_1 q_2\}_1^{\bar{3}} \{q_3 q_4\}_1^{\bar{3}}\big)_1\,\bar q\Big\rangle_\frac{3}{2} \nonumber \\
    \psi_c^1 \psi_s^9 &= \Big|\big(\{q_1 q_2\}_1^{\bar{3}} \{q_3 q_4\}_1^{\bar{3}}\big)_2\,\bar q\Big\rangle_\frac{3}{2} \nonumber \\
    \psi_c^1 \psi_s^{10} &= \Big|\big(\{q_1 q_2\}_1^{\bar{3}} \{q_3 q_4\}_1^{\bar{3}}\big)_2\,\bar q\Big\rangle_\frac{5}{2} 
\end{align}
\end{minipage}
\end{table}
\newpage
\bibliographystyle{apsrev4-2} 
\bibliography{Ref.bib}
\FloatBarrier
\newpage	
\begin{table}[ht]
	\centering
	\captionof{table}{Light diquark masses (in MeV).} 
	\label{t1}
	\begin{tabular}{|c|c|c|c|c|}	 \hline
		\textbf{Diquark} &\multicolumn{2}{c|}{\textbf{$\overline{\mb3}_c$ Diquarks\footnote{The $\overline{\mb3}_c$ diquark masses are adopted from our recent work \cite{Mohan:2026rcg}.}}} &\multicolumn{2}{c|}{\textbf{$\mb6_c$ Diquarks}} \\
		 \cline{2-3}  \cline{4-5}
		 \textbf{flavor content} & \textbf{Axial-vector ($1^+$)} & \textbf{Scalar ($0^+$)} & \textbf{Axial-vector ($1^+$)} & \textbf{Scalar ($0^+$)} \\ \hline \hline
	    $uu$ & $731.083$ & - & - & $744.143$ \\
		$ud$ & $732.943$ & $625.607$ & $692.692$ & $746.360$ \\
		$dd$ & $733.701$ & - & - & $746.925$ \\
		$us$ & $930.345$ & $758.537$ & $865.917$ & $951.821$ \\ 
		$ds$ & $938.057$ & $739.98$ & $863.778$ & $962.817$ \\
		$ss$ & $1076.36$ & - & - & $1079.63$ \\ \hline
	\end{tabular}
\end{table}
\begin{table}[ht]
	\centering
	\captionof{table}{Heavy diquark masses (in MeV).} 
	\label{t2}
	\begin{tabular}{|c|c|c|c|c|c|c|}	 \hline
    \textbf{Diquark} &\multicolumn{2}{c|}{\textbf{$\overline{\mb3}_c$ Diquarks}} &\multicolumn{2}{c|}{\textbf{$\mb6_c$ Diquarks}} \\
		 \cline{2-3}  \cline{4-5}
		 \textbf{flavor content} & \textbf{Axial-vector ($1^+$)} & \textbf{Scalar ($0^+$)} & \textbf{Axial-vector ($1^+$)} & \textbf{Scalar ($0^+$)} \\ \hline \hline
	    $uc$ & $2078.96$ & $1932.48$ & $2024.03$ & $2097.27$ \\
		$dc$ & $2079.99$ & $1933.98$ & $2025.23$ & $2098.24$ \\
		$sc$ & $2260.75$ & $2116.81$ & $2206.78$ & $2278.75$ \\
		$cc$\footnotemark[1] & $3365.92$ & - & - & $3367.58$ \\ \hline \hline 
		$ub$ & $5401.94$ & $5265.94$ & $5350.94$ & $5418.94$ \\
		$db$ & $5402.98$ & $5267.42$ & $5352.14$ & $5419.92$ \\
		$sb$ & $5587.42$ & $5439.22$ & $5531.85$ & $5605.95$ \\ 
		$cb$\footnotemark[1] & $6689.54$ & $6685.08$ & $6687.89$ & $6690.11$ \\
		$bb$\footnotemark[1] & $10014.6$ & - & - & $10014.8$ \\ \hline
	\end{tabular}
    \footnotetext[1]{The $\overline{\mb3}_c$ heavy-heavy diquark masses employed in this work are adopted from our recent analysis of heavy flavor baryons \cite{Mohan:2026rcg}. These masses are $\sim 20$ MeV smaller than the corresponding values obtained from our calculations involving heavy-light diquarks.} 
\end{table}
\begin{table}[ht]
\centering
\captionof{table}{Strong hyperfine interaction terms for singly charm tetraquarks (in MeV).} 
\label{t3}
\setlength{\tabcolsep}{10pt}
\renewcommand{\arraystretch}{1.5}
\begin{tabular}{|c|c|c|}	 \hline
\multirow{2}{*}{\textbf{Symmetry relations\footnotemark[1]}} & \multicolumn{2}{c|}{\textbf{Diquark-antidiquark interaction terms\footnote{As per the Eq. \eqref{Sextet_triplet} given in Sec. \ref{inter_cluster} \[b^{(6\bar{6})}_{D_1\bar{D_2}} =\frac{5}{2}\;\Bigg(\frac{m_{D_1^{\bar 3}}\,m_{\bar{D_2}^{3}}}{m_{D_1^{6}}\,m_{\bar{D_2}^{\bar 6}}}\Bigg)b^{(\bar 3 3)}_{D_1\bar{D_2}}.\]}}} \\ \cline{2-3}
& $b_{D_1\bar D_2}^{(\bar 33)}$ & $b_{D_1\bar D_2}^{(6\bar 6)}$ \\ \hline \hline
	$\displaystyle\Big(\frac{m_{D^{ds}}}{m_{D^{uu}}}\Big)b_{D^{cu}D^{\bar d\bar s}}$ & $b_{D^{cu}D^{\bar u\bar u}}(b_{D^{cd}D^{\bar u\bar u}}) = 61.60(61.57)$ & - \\ 
	$\displaystyle\Big(\frac{m_{D^{ds}}}{m_{D^{ud}}}\Big)b_{D^{cu}D^{\bar d\bar s}}$ & $b_{D^{cu}D^{\bar u\bar d}}(b_{D^{cd}D^{\bar u\bar d}}) = 61.44(61.41)$ & $b_{D^{cu}D^{\bar u\bar d}}(b_{D^{cd}D^{\bar u\bar d}}) = 166.95(166.85)$ \\ 
	$\displaystyle\Big(\frac{m_{D^{ds}}}{m_{D^{dd}}}\Big)b_{D^{cu}D^{\bar d\bar s}}$ & $b_{D^{cu}D^{\bar d\bar d}}(b_{D^{cd}D^{\bar d\bar d}}) = 61.38(61.35)$ & - \\ 
	$\displaystyle\Big(\frac{m_{D^{ds}}}{m_{D^{us}}}\Big)b_{D^{cu}D^{\bar d\bar s}}$ & $b_{D^{cu}D^{\bar u\bar s}}(b_{D^{cd}D^{\bar u\bar s}}) = 48.41(48.38)$ & $b_{D^{cu}D^{\bar u\bar s}}(b_{D^{cd}D^{\bar u\bar s}}) = 133.55(133.47)$ \\ 
	- & $b_{D^{cu}D^{\bar d\bar s}}(b_{D^{cd}D^{\bar d\bar s}}) = 48.01(47.99)$ & $b_{D^{cu}D^{\bar d\bar s}}(b_{D^{cd}D^{\bar d\bar s}}) = 133.88(133.80)$ \\ 
	$\displaystyle\Big(\frac{m_{D^{ds}}}{m_{D^{ss}}}\Big)b_{D^{cu}D^{\bar d\bar s}}$ & $b_{D^{cu}D^{\bar s\bar s}}(b_{D^{cd}D^{\bar s\bar s}}) = 41.84(41.82)$ & - \\  \hline \hline 
	$\displaystyle\Big(\frac{m_{D^{cu}}m_{D^{ds}}}{m_{D^{cs}}m_{D^{uu}}}\Big)b_{D^{cu}D^{\bar d\bar s}}$ & $b_{D^{cs}D^{\bar u\bar u}} = 56.65$ & - \\ 
	$\displaystyle\Big(\frac{m_{D^{cu}}m_{D^{ds}}}{m_{D^{cs}}m_{D^{ud}}}\Big)b_{D^{cu}D^{\bar d\bar s}}$ & $b_{D^{cs}D^{\bar u\bar d}} = 56.50$ & $b_{D^{cs}D^{\bar u\bar d}} = 153.12$ \\ 
	$\displaystyle\Big(\frac{m_{D^{cu}}m_{D^{ds}}}{m_{D^{cs}}m_{D^{dd}}}\Big)b_{D^{cu}D^{\bar d\bar s}}$ & $b_{D^{cs}D^{\bar d\bar d}} = 56.44$ & - \\ 
	$\displaystyle\Big(\frac{m_{D^{cu}}m_{D^{ds}}}{m_{D^{cs}}m_{D^{us}}}\Big)b_{D^{cu}D^{\bar d\bar s}}$ & $b_{D^{cs}D^{\bar u\bar s}} = 44.51$ & $b_{D^{cs}D^{\bar u\bar s}} = 122.49$ \\ 
	$\displaystyle\Big(\frac{m_{D^{cu}}}{m_{D^{cs}}}\Big)b_{D^{cu}D^{\bar d\bar s}}$ & $b_{D^{cs}D^{\bar d\bar s}} = 44.15$ & $b_{D^{cs}D^{\bar d\bar s}} = 122.79$ \\ 
	$\displaystyle\Big(\frac{m_{D^{cu}}m_{D^{ds}}}{m_{D^{cs}}m_{D^{ss}}}\Big)b_{D^{cu}D^{\bar d\bar s}}$ & $b_{D^{cs}D^{\bar s\bar s}} = 38.48$ & - \\ \hline
\end{tabular}
\footnotetext{Symmetry relations are used to calculate the diquark-antidiquark interaction terms utilizing diquark masses from Table \ref{t1} and \ref{t2}, and $b_{D^{cu}D^{\bar d\bar s}} = 48.01$ MeV extracted from the experimental mass $T_{c\bar s0}^{*}(2900)^{++} = (2921\pm 26)$ MeV \cite{LHCb:2022sfr}.} 
\end{table}
\begin{table}[ht]
\centering
\captionof{table}{Strong hyperfine interaction terms for singly bottom tetraquarks (in MeV).} 
\label{t4}
\setlength{\tabcolsep}{10pt}
\renewcommand{\arraystretch}{1.5}
\begin{tabular}{|c|c|c|}	 \hline
\multirow{2}{*}{\textbf{Symmetry relations\footnotemark[1]}} & \multicolumn{2}{c|}{\textbf{Diquark-antidiquark interaction terms}} \\ \cline{2-3}
& $b_{D_1\bar D_2}^{(\bar 33)}$ & $b_{D_1\bar D_2}^{(6\bar 6)}$ \\ \hline \hline
	$\displaystyle\Big(\frac{m_{D^{cu}}m_{D^{ds}}}{m_{D^{bu}}m_{D^{uu}}}\Big)b_{D^{cu}D^{\bar d\bar s}}$ & $b_{D^{bu}D^{\bar u\bar u}}(b_{D^{bd}D^{\bar u\bar u}}) = 23.71(23.70)$ & - \\ 
	$\displaystyle\Big(\frac{m_{D^{cu}}m_{D^{ds}}}{m_{D^{bu}}m_{D^{ud}}}\Big)b_{D^{cu}D^{\bar d\bar s}}$ & $b_{D^{bu}D^{\bar u\bar d}}(b_{D^{bd}D^{\bar u\bar d}}) = 23.65(23.64)$ & $b_{D^{bu}D^{\bar u\bar d}}(b_{D^{bd}D^{\bar u\bar d}}) = 63.15(63.14)$ \\ 
	$\displaystyle\Big(\frac{m_{D^{cu}}m_{D^{ds}}}{m_{D^{bu}}m_{D^{dd}}}\Big)b_{D^{cu}D^{\bar d\bar s}}$ & $b_{D^{bu}D^{\bar d\bar d}}(b_{D^{bd}D^{\bar d\bar d}}) = 23.62(23.62)$ & - \\ 
	$\displaystyle\Big(\frac{m_{D^{cu}}m_{D^{ds}}}{m_{D^{bu}}m_{D^{us}}}\Big)b_{D^{cu}D^{\bar d\bar s}}$ & $b_{D^{bu}D^{\bar u\bar s}}(b_{D^{bd}D^{\bar u\bar s}}) = 18.63(18.63)$ & $b_{D^{bu}D^{\bar u\bar s}}(b_{D^{bd}D^{\bar u\bar s}}) = 50.52(50.51)$ \\ 
	$\displaystyle\Big(\frac{m_{D^{cu}}}{m_{D^{bu}}}\Big)b_{D^{cu}D^{\bar d\bar s}}$ & $b_{D^{bu}D^{\bar d\bar s}}(b_{D^{bd}D^{\bar d\bar s}}) = 18.48(18.47)$ & $b_{D^{bu}D^{\bar d\bar s}}(b_{D^{bd}D^{\bar d\bar s}}) = 50.64(50.63)$ \\ 
	$\displaystyle\Big(\frac{m_{D^{cu}}m_{D^{ds}}}{m_{D^{bu}}m_{D^{ss}}}\Big)b_{D^{cu}D^{\bar d\bar s}}$ & $b_{D^{bu}D^{\bar s\bar s}}(b_{D^{bd}D^{\bar s\bar s}}) = 16.10(16.10)$ & - \\  \hline \hline 
	$\displaystyle\Big(\frac{m_{D^{cu}}m_{D^{ds}}}{m_{D^{bs}}m_{D^{uu}}}\Big)b_{D^{cu}D^{\bar d\bar s}}$ & $b_{D^{bs}D^{\bar u\bar u}} = 22.92$ & - \\ 
	$\displaystyle\Big(\frac{m_{D^{cu}}m_{D^{ds}}}{m_{D^{bs}}m_{D^{ud}}}\Big)b_{D^{cu}D^{\bar d\bar s}}$ & $b_{D^{bs}D^{\bar u\bar d}} = 22.86$ & $b_{D^{bs}D^{\bar u\bar d}} = 61.08$ \\ 
	$\displaystyle\Big(\frac{m_{D^{cu}}m_{D^{ds}}}{m_{D^{bs}}m_{D^{dd}}}\Big)b_{D^{cu}D^{\bar d\bar s}}$ & $b_{D^{bs}D^{\bar d\bar d}} = 22.84$ & - \\ 
	$\displaystyle\Big(\frac{m_{D^{cu}}m_{D^{ds}}}{m_{D^{bs}}m_{D^{us}}}\Big)b_{D^{cu}D^{\bar d\bar s}}$ & $b_{D^{bs}D^{\bar u\bar s}} = 18.01$ & $b_{D^{bs}D^{\bar u\bar s}} = 48.86$ \\ 
	$\displaystyle\Big(\frac{m_{D^{cu}}}{m_{D^{bs}}}\Big)b_{D^{cu}D^{\bar d\bar s}}$ & $b_{D^{bs}D^{\bar d\bar s}} = 17.86$ & $b_{D^{bs}D^{\bar d\bar s}} = 48.99$ \\ 
	$\displaystyle\Big(\frac{m_{D^{cu}}m_{D^{ds}}}{m_{D^{bs}}m_{D^{ss}}}\Big)b_{D^{cu}D^{\bar d\bar s}}$ & $b_{D^{bs}D^{\bar s\bar s}} = 15.57$ & - \\ \hline  
\end{tabular}
\footnotetext{Symmetry relations are used to calculate the diquark-antidiquark interaction terms utilizing diquark masses from Table \ref{t1} and \ref{t2}, and $b_{D^{cu}D^{\bar d\bar s}} = 48.01$ MeV extracted from the experimental mass $T_{c\bar s0}^{*}(2900)^{++} = (2921\pm 26)$ MeV \cite{LHCb:2022sfr}.} 
\end{table}
\begin{table}[ht]
\centering
\captionof{table}{Strong hyperfine interaction terms for doubly heavy tetraquarks (in MeV).} 
\label{t5}
\setlength{\tabcolsep}{10pt}
\renewcommand{\arraystretch}{1.5}
\begin{tabular}{|c|c|c|}	 \hline
\multirow{2}{*}{\textbf{Symmetry relations\footnotemark[1]}} & \multicolumn{2}{c|}{\textbf{Diquark-antidiquark interaction terms}} \\ \cline{2-3}
& $b_{D_1\bar D_2}^{(\bar 33)}$ & $b_{D_1\bar D_2}^{(6\bar 6)}$ \\ \hline \hline
	$\displaystyle\Big(\frac{m_{D^{cu}}m_{D^{ds}}}{m_{D^{cc}}m_{D^{uu}}}\Big)b_{D^{cu}D^{\bar d\bar s}}$ & $b_{D^{cc}D^{\bar u\bar u}} = 38.05$ & - \\ 
	$\displaystyle\Big(\frac{m_{D^{cu}}m_{D^{ds}}}{m_{D^{cc}}m_{D^{ud}}}\Big)b_{D^{cu}D^{\bar d\bar s}}$ & $b_{D^{cc}D^{\bar u\bar d}} = 37.95$ & - \\ 
	$\displaystyle\Big(\frac{m_{D^{cu}}m_{D^{ds}}}{m_{D^{cc}}m_{D^{dd}}}\Big)b_{D^{cu}D^{\bar d\bar s}}$ & $b_{D^{cc}D^{\bar d\bar d}} = 37.91$ & - \\ 
	$\displaystyle\Big(\frac{m_{D^{cu}}m_{D^{ds}}}{m_{D^{cc}}m_{D^{us}}}\Big)b_{D^{cu}D^{\bar d\bar s}}$ & $b_{D^{cc}D^{\bar u\bar s}} = 29.90$ & - \\ 
	$\displaystyle\Big(\frac{m_{D^{cu}}}{m_{D^{cc}}}\Big)b_{D^{cu}D^{\bar d\bar s}}$ & $b_{D^{cc}D^{\bar d\bar s}} = 29.65$ & - \\ 
	$\displaystyle\Big(\frac{m_{D^{cu}}m_{D^{ds}}}{m_{D^{cc}}m_{D^{ss}}}\Big)b_{D^{cu}D^{\bar d\bar s}}$ & $b_{D^{cc}D^{\bar s\bar s}} = 25.84$ & - \\  \hline \hline 
	$\displaystyle\Big(\frac{m_{D^{cu}}m_{D^{ds}}}{m_{D^{cb}}m_{D^{uu}}}\Big)b_{D^{cu}D^{\bar d\bar s}}$ & $b_{D^{cb}D^{\bar u\bar u}} = 19.14$ & - \\ 
	$\displaystyle\Big(\frac{m_{D^{cu}}m_{D^{ds}}}{m_{D^{cb}}m_{D^{ud}}}\Big)b_{D^{cu}D^{\bar d\bar s}}$ & $b_{D^{cb}D^{\bar u\bar d}} = 19.10$ & $b_{D^{cb}D^{\bar u\bar d}} = 50.53$ \\ 
	$\displaystyle\Big(\frac{m_{D^{cu}}m_{D^{ds}}}{m_{D^{cb}}m_{D^{dd}}}\Big)b_{D^{cu}D^{\bar d\bar s}}$ & $b_{D^{cb}D^{\bar d\bar d}} = 19.08$ & - \\ 
	$\displaystyle\Big(\frac{m_{D^{cu}}m_{D^{ds}}}{m_{D^{cb}}m_{D^{us}}}\Big)b_{D^{cu}D^{\bar d\bar s}}$ & $b_{D^{cb}D^{\bar u\bar s}} = 15.04$ & $b_{D^{cb}D^{\bar u\bar s}} = 40.42$ \\ 
	$\displaystyle\Big(\frac{m_{D^{cu}}}{m_{D^{cb}}}\Big)b_{D^{cu}D^{\bar d\bar s}}$ & $b_{D^{cb}D^{\bar d\bar s}} = 14.92$ & $b_{D^{cb}D^{\bar d\bar s}} = 40.52$ \\ 
	$\displaystyle\Big(\frac{m_{D^{cu}}m_{D^{ds}}}{m_{D^{cb}}m_{D^{ss}}}\Big)b_{D^{cu}D^{\bar d\bar s}}$ & $b_{D^{cb}D^{\bar s\bar s}} = 13.00$ & - \\ \hline \hline 
	$\displaystyle\Big(\frac{m_{D^{cu}}m_{D^{ds}}}{m_{D^{bb}}m_{D^{uu}}}\Big)b_{D^{cu}D^{\bar d\bar s}}$ & $b_{D^{bb}D^{\bar u\bar u}} = 12.79$ & - \\ 
	$\displaystyle\Big(\frac{m_{D^{cu}}m_{D^{ds}}}{m_{D^{bb}}m_{D^{ud}}}\Big)b_{D^{cu}D^{\bar d\bar s}}$ & $b_{D^{bb}D^{\bar u\bar d}} = 12.76$ & - \\ 
	$\displaystyle\Big(\frac{m_{D^{cu}}m_{D^{ds}}}{m_{D^{bb}}m_{D^{dd}}}\Big)b_{D^{cu}D^{\bar d\bar s}}$ & $b_{D^{bb}D^{\bar d\bar d}} = 12.74$ & - \\ 
	$\displaystyle\Big(\frac{m_{D^{cu}}m_{D^{ds}}}{m_{D^{bb}}m_{D^{us}}}\Big)b_{D^{cu}D^{\bar d\bar s}}$ & $b_{D^{bb}D^{\bar u\bar s}} = 10.05$ & - \\ 
	$\displaystyle\Big(\frac{m_{D^{cu}}}{m_{D^{bb}}}\Big)b_{D^{cu}D^{\bar d\bar s}}$ & $b_{D^{bb}D^{\bar d\bar s}} = 9.97$ & - \\ 
	$\displaystyle\Big(\frac{m_{D^{cu}}m_{D^{ds}}}{m_{D^{bb}}m_{D^{ss}}}\Big)b_{D^{cu}D^{\bar d\bar s}}$ & $b_{D^{bb}D^{\bar s\bar s}} = 8.69$ & - \\ \hline
\end{tabular}
\footnotetext{Symmetry relations are used to calculate the diquark-antidiquark interaction terms utilizing diquark masses from Table \ref{t1} and \ref{t2}, and $b_{D^{cu}D^{\bar d\bar s}} = 48.01$ MeV extracted from the experimental mass $T_{c\bar s0}^{*}(2900)^{++} = (2921\pm 26)$ MeV \cite{LHCb:2022sfr}.} 
\end{table}
\begin{table}[ht]
\centering
\captionof{table}{Strong hyperfine interaction terms for hidden heavy flavor tetraquarks (in MeV).} 
\label{t6}
\setlength{\tabcolsep}{10pt}
\renewcommand{\arraystretch}{1.5}
\begin{tabular}{|c|c|c|}	 \hline
\multirow{2}{*}{\textbf{Symmetry relations\footnotemark[1]}} & \multicolumn{2}{c|}{\textbf{Diquark-antidiquark interaction terms}} \\ \cline{2-3}
& $b_{D_1\bar D_2}^{(\bar 33)}$ & $b_{D_1\bar D_2}^{(6\bar 6)}$ \\ \hline \hline
	$\displaystyle\Big(\frac{m_{D^{ds}}}{m_{D^{cu}}}\Big)b_{D^{cu}D^{\bar d\bar s}}$ & $b_{D^{cu}D^{\bar c\bar u}}(b_{D^{cd}D^{\bar c\bar d}}) = 21.66(21.64)$ & $b_{D^{cu}D^{\bar c\bar u}}(b_{D^{cd}D^{\bar c\bar d}}) = 57.14(57.07)$ \\ 
	$\displaystyle\Big(\frac{m_{D^{ds}}}{m_{D^{cd}}}\Big)b_{D^{cu}D^{\bar d\bar s}}$ & $b_{D^{cu}D^{\bar c\bar d}}=b_{D^{cd}D^{\bar c\bar u}} = 21.65$ & $b_{D^{cu}D^{\bar c\bar d}}=b_{D^{cd}D^{\bar c\bar u}} = 57.10$ \\ 
	$\displaystyle\Big(\frac{m_{D^{ds}}}{m_{D^{cs}}}\Big)b_{D^{cu}D^{\bar d\bar s}}$ & $b_{D^{cu}D^{\bar c\bar s}}=b_{D^{cs}D^{\bar c\bar u}} = 19.92$ & $b_{D^{cu}D^{\bar c\bar s}}=b_{D^{cs}D^{\bar c\bar u}} = 52.40$ \\ 
	$\displaystyle\Big(\frac{m_{D^{cu}}m_{D^{ds}}}{m_{D^{cd}}m_{D^{cs}}}\Big)b_{D^{cu}D^{\bar d\bar s}}$ & $b_{D^{cd}D^{\bar c\bar s}}=b_{D^{cs}D^{\bar c\bar d}} = 19.91$ & $b_{D^{cd}D^{\bar c\bar s}}=b_{D^{cs}D^{\bar c\bar d}} = 52.37$ \\ 
	$\displaystyle\Big(\frac{m_{D^{cu}}m_{D^{ds}}}{m_{D^{cs}}m_{D^{cs}}}\Big)b_{D^{cu}D^{\bar d\bar s}}$ & $b_{D^{cs}D^{\bar c\bar s}} = 18.32$ & $b_{D^{cs}D^{\bar c\bar s}} = 48.06$ \\  \hline \hline 
	$\displaystyle\Big(\frac{m_{D^{cu}}m_{D^{ds}}}{m_{D^{bu}}m_{D^{bu}}}\Big)b_{D^{cu}D^{\bar d\bar s}}$ & $b_{D^{bu}D^{\bar b\bar u}}(b_{D^{bd}D^{\bar b\bar d}}) = 3.21(3.21)$ & $b_{D^{bu}D^{\bar b\bar u}}(b_{D^{bd}D^{\bar b\bar d}}) = 8.17(8.17)$ \\ 
	$\displaystyle\Big(\frac{m_{D^{cu}}m_{D^{ds}}}{m_{D^{bu}}m_{D^{bd}}}\Big)b_{D^{cu}D^{\bar d\bar s}}$ & $b_{D^{bu}D^{\bar b\bar d}}=b_{D^{bd}D^{\bar b\bar u}} = 3.21$ & $b_{D^{bu}D^{\bar b\bar d}}=b_{D^{bd}D^{\bar b\bar u}} = 8.17$ \\ 
	$\displaystyle\Big(\frac{m_{D^{cu}}m_{D^{ds}}}{m_{D^{bu}}m_{D^{bs}}}\Big)b_{D^{cu}D^{\bar d\bar s}}$ & $b_{D^{bu}D^{\bar b\bar s}}=b_{D^{bs}D^{\bar b\bar u}} = 3.10$ & $b_{D^{bu}D^{\bar b\bar s}}=b_{D^{bs}D^{\bar b\bar u}} = 7.91$ \\ 
	$\displaystyle\Big(\frac{m_{D^{cu}}m_{D^{ds}}}{m_{D^{bd}}m_{D^{bs}}}\Big)b_{D^{cu}D^{\bar d\bar s}}$ & $b_{D^{bd}D^{\bar b\bar s}}=b_{D^{bs}D^{\bar b\bar d}} = 3.10$ & $b_{D^{bd}D^{\bar b\bar s}}=b_{D^{bs}D^{\bar b\bar d}} = 7.91$ \\ 
	$\displaystyle\Big(\frac{m_{D^{cu}}m_{D^{ds}}}{m_{D^{bs}}m_{D^{bs}}}\Big)b_{D^{cu}D^{\bar d\bar s}}$ & $b_{D^{bs}D^{\bar b\bar s}} = 3.00$ & $b_{D^{bs}D^{\bar b\bar s}} = 7.65$ \\  \hline  
\end{tabular}
\footnotetext{Symmetry relations are used to calculate the diquark-antidiquark interaction terms utilizing diquark masses from Table \ref{t1} and \ref{t2}, and $b_{D^{cu}D^{\bar d\bar s}} = 48.01$ MeV extracted from the experimental mass $T_{c\bar s0}^{*}(2900)^{++} = (2921\pm 26)$ MeV \cite{LHCb:2022sfr}.} 
\end{table}
\begin{table}[ht]
\centering
\captionof{table}{Strong hyperfine interaction terms for fully heavy tetraquarks (in MeV).} 
\label{t_fullyheavy}
\setlength{\tabcolsep}{10pt}
\renewcommand{\arraystretch}{1.5}
\begin{tabular}{|c|c|c|} \hline
\multirow{2}{*}{\textbf{Symmetry relations\footnotemark[1]}} 
& \multicolumn{2}{c|}{\textbf{Diquark-antidiquark interaction terms}} \\ \cline{2-3}
& $b_{D_1\bar D_2}^{(\bar 33)}$ & $b_{D_1\bar D_2}^{(6\bar 6)}$ \\ 
\hline \hline

$\displaystyle \Big(\frac{m_{D^{cu}}m_{D^{ds}}}{(m_{D^{cc}})^2}\Big)b_{D^{cu}D^{\bar d\bar s}}$ 
& $b_{D^{cc}D^{\bar c\bar c}}=8.26$ 
& - \\ 

$\displaystyle \Big(\frac{m_{D^{cu}}m_{D^{ds}}}{m_{D^{cb}}m_{D^{cc}}}\Big)b_{D^{cu}D^{\bar d\bar s}}$
& $b_{D^{cb}D^{\bar c\bar c}}=4.16$ 
& - \\ 

$\displaystyle \Big(\frac{m_{D^{cu}}m_{D^{ds}}}{(m_{D^{cb}})^2}\Big)b_{D^{cu}D^{\bar d\bar s}}$
& $b_{D^{cb}D^{\bar c\bar b}}=2.09$ 
& $b_{D^{cb}D^{\bar c\bar b}}=5.23$ \\ 

$\displaystyle \Big(\frac{m_{D^{cu}}m_{D^{ds}}}{m_{D^{bb}}m_{D^{cc}}}\Big)b_{D^{cu}D^{\bar d\bar s}}$
& $b_{D^{bb}D^{\bar c\bar c}}=2.78$ 
& - \\ 

$\displaystyle \Big(\frac{m_{D^{cu}}m_{D^{ds}}}{m_{D^{bb}}m_{D^{cb}}}\Big)b_{D^{cu}D^{\bar d\bar s}}$
& $b_{D^{bb}D^{\bar c\bar b}}=1.40$ 
& - \\ 

$\displaystyle \Big(\frac{m_{D^{cu}}m_{D^{ds}}}{(m_{D^{bb}})^2}\Big)b_{D^{cu}D^{\bar d\bar s}}$
& $b_{D^{bb}D^{\bar b\bar b}}=0.93$ 
& - \\ \hline
\end{tabular}

\footnotetext{Symmetry relations are used to calculate the diquark-antidiquark interaction terms utilizing diquark masses from Table \ref{t1} and \ref{t2}, and $b_{D^{cu}D^{\bar d\bar s}} = 48.01$ MeV extracted from the experimental mass $T_{c\bar s0}^{*}(2900)^{++} = (2921\pm 26)$ MeV \cite{LHCb:2022sfr}.}
\end{table}
\begin{table}[ht]
	\centering
	\captionof{table}{Binding energy (in MeV).}
	\label{t7}
	\begin{tabular}{|c|c|c|c|c|c|}\hline
		Experimental & \multicolumn{2}{c|}{$BE(Q\bar{Q}^{\prime})$} & \multicolumn{3}{c|}{$BE(QQ^{\prime})$}  \\ \cline{2-6}
        inputs \cite{ParticleDataGroup:2024cfk} & Flavor content & $1_c$ & Flavor content & $\bar {\mb3}_c$ & $\mb6_c$ \\ \hline \hline
		$D_s^{*+}, D_s^{+}$ & $c\bar{s}$ & $-78.55$ & $cs$ & $-39.28$ & $19.64$ \\
		$J/\Psi, \eta_c$ & $c\bar{c}$ & $-262.36$ & $cc$ & $-131.18$ & $65.59$ \\ \hline
		$B_s^{*0}, B_s^{0}$ & $\bar{b}s$ & $-93.57$ & $bs$ & $-46.79$ & $23.40$ \\
		$B_c^{*+}$\footnote{The $B_c^{*+}$ mass of $6331$ MeV from LQCD~\cite{Mathur:2018epb} is used to compute the binding energy term $BE(\bar{b}c)$.}, $B_c^{+}$ & $\bar{b}c$ & $-356.26$ & $bc$ & $-178.13$ & $89.07$ \\
		$\Upsilon, \eta_b$ & $\bar{b}b$ & $-570.29$ & $bb$ & $-285.14$ & $142.57$ \\ \hline
	\end{tabular}
\end{table}	
\begin{table}[ht]
\centering
\caption{Effective inter-cluster hyperfine interaction terms entering the pentaquark mass (in MeV).}
\label{t_penta_params}
\setlength{\tabcolsep}{10pt}
\renewcommand{\arraystretch}{1.5}
\begin{tabular}{|c|c|} \hline
\textbf{Symmetry relations\footnotemark[1]} 
& \textbf{Interaction terms} \\ 
\hline \hline

$\displaystyle 
\Big(\frac{1}{2}\Big)
b_{D^{cu}D^{\bar u\bar d}}^{(\bar 33)}
$
& $b_{D^{cu}D^{ud}}^{(\bar 3\bar 3)} = 30.72$ \\ 

$\displaystyle 
\Big(\frac{1}{2}\Big)
b_{D^{cs}D^{\bar u\bar d}}^{(\bar 33)}
$
& $b_{D^{cs}D^{ud}}^{(\bar 3\bar 3)} = 28.25$ \\  \hline

$\displaystyle 
\Big(\frac{1}{2}\Big)
b_{D^{\{ud\}}c}^{(\bar 33)}
$
& $b_{D^{\{ud\}}\bar c}^{(\bar 3\bar 3)} = 16.18$ \\ \hline

$\displaystyle 
\Big(\frac{m_u}{m_{D^{cu}}}\Big)
b_{cu}^{(33)}
$
& $b_{D^{\{cu\}}\bar c}^{(\bar 3\bar 3)} = 24.84$ \\  

$\displaystyle 
\Big(\frac{m_s}{m_{D^{cs}}}\Big)
b_{cs}^{(33)}
$
& $b_{D^{\{cs\}}\bar c}^{(\bar 3\bar 3)} = 34.06$ \\  

\hline
\end{tabular}

\footnotetext{
The inter-diquark coupling $b_{D^{cu}D^{ud}}$ is obtained from the corresponding 
diquark–antidiquark interaction calibrated in the tetraquark sector, while the diquark–antiquark couplings 
$b_{D^{\{cu\}}\bar c}$ and $b_{D^{\{ud\}}\bar c}$ are inherited from baryon-calibrated 
quark-diquark interactions.}
\end{table}
\begin{table}[ht]
\centering
\captionof{table}{Masses of $(\overline{\mb3}_c\mb3_c)$ and $(\mb6_c\overline{\mb6}_c)$ open charm $cn\bar n\bar n$ tetraquark states using heavy-light diquarks.}
\label{t8}
\begin{tabular}{|c|c|c|c|c|}\hline
\multirow{2}{*}{\textbf{States}} & \multirow{2}{*}{$I(J^{P})$\footnotemark[1]} & \multirow{2}{*}{\textbf{Configuration\footnotemark[2]}} & \textbf{This} & \bf RQM  \\
& & & \textbf{work} & \bf \cite{Lu:2020qmp} \\ \hline

$T(cu\bar u\bar d)/$ & $\frac{1}{2}(0^+)$ & $\big|(cu)_0^{\bar 3}[\bar u\bar d]_0^3 \big\rangle$ & $2558.09/2559.59$ & $2570$ \\
$T(cd\bar u\bar d)$ & & $\big|(cu)_1^{6}[\bar u\bar d]_1^{\bar 6} \big\rangle$ & $2382.83/2384.22$ & $3064$ \\
& $\frac{1}{2}(1^+)$ & $\big|(cu)_1^{\bar 3}[\bar u\bar d]_0^3 \big\rangle$ & $2704.57/2705.60$ & $2802$ \\
& & $\big|(cu)_0^{6}[\bar u\bar d]_1^{\bar 6} \big\rangle$ & $2789.96/2790.93$ & $3019$ \\
& & $\big|(cu)_1^{6}[\bar u\bar d]_1^{\bar 6} \big\rangle$ & $2549.77/2551.07$ & $3190$ \\
& $\frac{1}{2}(2^+)$ & $\big|(cu)_1^{6}[\bar u\bar d]_1^{\bar 6} \big\rangle$ & $2883.67/2884.77$ & $3240$ \\
& $\frac{1}{2}/\frac{3}{2}(0^+)$ & $\big|(cu)_1^{\bar 3}\{\bar u\bar d\}_1^3 \big\rangle$ & $2689.01/2690.11$ & $2915$ \\
& & $\big|(cu)_0^{6}\{\bar u\bar d\}_0^{\bar 6} \big\rangle$ & $2843.63/2844.60$ & $3327$ \\
& $\frac{1}{2}/\frac{3}{2}(1^+)$ & $\big|(cu)_0^{\bar 3}\{\bar u\bar d\}_1^3 \big\rangle$ & $2665.42/2666.92$ & $2980$ \\
& & $\big|(cu)_1^{\bar 3}\{\bar u\bar d\}_1^3 \big\rangle$ & $2750.46/2751.52$ & $3140$ \\
& & $\big|(cu)_1^{6}\{\bar u\bar d\}_0^{\bar 6} \big\rangle$ & $2770.39/2771.59$ & $3260$ \\
& $\frac{1}{2}/\frac{3}{2}(2^+)$ & $\big|(cu)_1^{\bar 3}\{\bar u\bar d\}_1^3 \big\rangle$ & $2873.35/2874.35$ & $3210$ \\ \hline

$T(cu\bar u\bar u)/$ & $\frac{1}{2}/\frac{3}{2}(0^+)$ & $\big|(cu)_1^{\bar 3}\{\bar u\bar u\}_1^3 \big\rangle$ & $2686.84/2687.93$ & - \\
$T(cd\bar u\bar u)$ & & $\big|(cu)_0^{6}\{\bar u\bar u\}_0^{\bar 6} \big\rangle$ & $2841.41/2842.38$ & - \\
& $\frac{1}{2}/\frac{3}{2}(1^+)$ & $\big|(cu)_0^{\bar 3}\{\bar u\bar u\}_1^3 \big\rangle$ & $2663.56/2665.06$ & - \\
& & $\big|(cu)_1^{\bar 3}\{\bar u\bar u\}_1^3 \big\rangle$ & $2748.44/2749.50$ & - \\
& & $\big|(cu)_1^{6}\{\bar u\bar u\}_0^{\bar 6} \big\rangle$ & $2768.17/2769.37$ & - \\
& $\frac{1}{2}/\frac{3}{2}(2^+)$ & $\big|(cu)_1^{\bar 3}\{\bar u\bar u\}_1^3 \big\rangle$ & $2871.64/2872.64$ & - \\ \hline

$T(cu\bar d\bar d)/$ & $\frac{1}{2}/\frac{3}{2}(0^+)$ & $\big|(cu)_1^{\bar 3}\{\bar d\bar d\}_1^3 \big\rangle$ & $2689.90/2690.99$ & - \\
$T(cd\bar d\bar d)$ & & $\big|(cu)_0^{6}\{\bar d\bar d\}_0^{\bar 6} \big\rangle$ & $2844.20/2845.17$ & - \\
& $\frac{1}{2}/\frac{3}{2}(1^+)$ & $\big|(cu)_0^{\bar 3}\{\bar d\bar d\}_1^3 \big\rangle$ & $2666.18/2667.68$ & - \\
& & $\big|(cu)_1^{\bar 3}\{\bar d\bar d\}_1^3 \big\rangle$ & $2751.28/2752.34$ & - \\
& & $\big|(cu)_1^{6}\{\bar d\bar d\}_0^{\bar 6} \big\rangle$ & $2770.96/2772.16$ & - \\
& $\frac{1}{2}/\frac{3}{2}(2^+)$ & $\big|(cu)_1^{\bar 3}\{\bar d\bar d\}_1^3 \big\rangle$ & $2874.04/2875.04$ & - \\ \hline
\end{tabular}
\footnotetext[1]{The isospin labels denote the allowed flavor symmetry classification of the light-quark sector. For symmetric light-quark configurations, both $I=\frac{1}{2}$ and $\frac{3}{2}$ are allowed, whereas antisymmetric configurations correspond to $I=\frac{1}{2}$ only. The quoted masses separated by slash ($/$) refer to explicit flavor states $cu\bar u\bar d$ and $cd\bar u\bar d$, respectively, and are therefore not associated with particular isospin eigenstates.}
\footnotetext[2]{Square brackets $[\,]$ and curly braces $\{\,\}$ denote antisymmetric and symmetric flavor wave functions, respectively, while parentheses $(\,)$ indicate subsystems for which no permutation symmetry is imposed. The same notation is used consistently throughout this work to represent the state configuration of exotic states.}
\end{table}
\begin{table}[ht]
\centering
\captionof{table}{Masses of $(\overline{\mb3}_c\mb3_c)$ and $(\mb6_c\overline{\mb6}_c)$ open charm ($cn\bar n\bar s, cn\bar s\bar s$) tetraquark states using heavy-light diquarks.}
\label{t9}
\begin{tabular}{|c|c|c|c|c|c|c|}\hline
\multirow{2}{*}{\textbf{States}} & \multirow{2}{*}{$I(J^{P})$} & \multirow{2}{*}{\textbf{Configuration}} & \textbf{This} & \bf NRPM & \bf RQM & \bf PDG \\
& & & \textbf{work} & \bf \cite{Liu:2022hbk} & \bf \cite{Lu:2020qmp} & \bf \cite{ParticleDataGroup:2024cfk}  \\ \hline

$T(cu\bar u\bar s)/$ & $0/1(0^+)$ & $\big|(cu)_0^{\bar 3}[\bar u\bar s]_0^3 \big\rangle$ & $2691.02/2692.52$ & $2828$ & $2807$ & - \\
$T(cd\bar u\bar s)$ & & $\big|(cu)_1^{6}[\bar u\bar s]_1^{\bar 6} \big\rangle$ & $2622.85/2624.21$ & $3150$ & $3224$ & - \\
& $0/1(1^+)$ & $\big|(cu)_1^{\bar 3}[\bar u\bar s]_0^3 \big\rangle$ & $2837.50/2838.53$ & $2949$ & $3013$ & - \\
& & $\big|(cu)_0^{6}[\bar u\bar s]_1^{\bar 6} \big\rangle$ & $2963.19/2964.16$ & $3119$ & $3181$ & - \\
& & $\big|(cu)_1^{6}[\bar u\bar s]_1^{\bar 6} \big\rangle$ & $2756.40/2757.68$ & $3199$ & $3325$ & - \\
& $0/1(2^+)$ & $\big|(cu)_1^{6}[\bar u\bar s]_1^{\bar 6} \big\rangle$ & $3023.50/3024.62$ & $3214$ & $3364$ & - \\
& $0/1(0^+)$ & $\big|(cu)_1^{\bar 3}\{\bar u\bar s\}_1^3 \big\rangle$ & $2912.49/2913.57$ & $3279$ & $3087$ & $2892\pm 21$ \\
& & $\big|(cu)_0^{6}\{\bar u\bar s\}_0^{\bar 6} \big\rangle$ & $3049.09/3050.06$ & $3046$ & $3435$ & - \\
& $0/1(1^+)$ & $\big|(cu)_0^{\bar 3}\{\bar u\bar s\}_1^3 \big\rangle$ & $2862.83/2864.33$ & $3201$ & $3139$ & - \\
& & $\big|(cu)_1^{\bar 3}\{\bar u\bar s\}_1^3 \big\rangle$ & $2960.90/2961.95$ & $3245$ & $3275$ & - \\
& & $\big|(cu)_1^{6}\{\bar u\bar s\}_0^{\bar 6} \big\rangle$ & $2975.85/2977.05$ & $3067$ & $3372$ & - \\
& $0/1(2^+)$ & $\big|(cu)_1^{\bar 3}\{\bar u\bar s\}_1^3 \big\rangle$ & $3057.71/3058.72$ & $3239$ & $3339$ & - \\ \hline

$T(cu\bar d\bar s)/$ & $0/1(0^+)$ & $\big|(cu)_0^{\bar 3}[\bar d\bar s]_0^3 \big\rangle$ & $2672.46/2673.96$ & - & - & - \\
$T(cd\bar d\bar s)$ & & $\big|(cu)_1^{6}[\bar d\bar s]_1^{\bar 6} \big\rangle$ & $2620.05/2621.40$ & - & - & - \\
& $0/1(1^+)$ & $\big|(cu)_1^{\bar 3}[\bar d\bar s]_0^3 \big\rangle$ & $2818.94/2819.97$ & - & - & - \\
& & $\big|(cu)_0^{6}[\bar d\bar s]_1^{\bar 6} \big\rangle$ & $2961.05/2962.02$ & - & - & - \\
& & $\big|(cu)_1^{6}[\bar d\bar s]_1^{\bar 6} \big\rangle$ & $2753.93/2755.21$ & - & - & - \\
& $0/1(2^+)$ & $\big|(cu)_1^{6}[\bar d\bar s]_1^{\bar 6} \big\rangle$ & $3021.69/3022.81$ & - & - & - \\
& $0/1(0^+)$ & $\big|(cu)_1^{\bar 3}\{\bar d\bar s\}_1^3 \big\rangle$ & $2921.00/2922.08$ & - & - & $2921\pm 26$ \\
& & $\big|(cu)_0^{6}\{\bar d\bar s\}_0^{\bar 6} \big\rangle$ & $3060.09/3061.06$ & - & - & - \\
& $0/1(1^+)$ & $\big|(cu)_0^{\bar 3}\{\bar d\bar s\}_1^3 \big\rangle$ & $2870.54/2872.04$ & - & - & - \\
& & $\big|(cu)_1^{\bar 3}\{\bar d\bar s\}_1^3 \big\rangle$ & $2969.01/2970.07$ & - & - & - \\
& & $\big|(cu)_1^{6}\{\bar d\bar s\}_0^{\bar 6} \big\rangle$ & $2986.85/2988.05$ & - & - & - \\
& $0/1(2^+)$ & $\big|(cu)_1^{\bar 3}\{\bar d\bar s\}_1^3 \big\rangle$ & $3065.03/3066.03$ & - & - & - \\ \hline

$T(cu\bar s\bar s)/$ & $\frac{1}{2}(0^+)$ & $\big|(cu)_1^{\bar 3}\{\bar s\bar s\}_1^3 \big\rangle$ & $3071.64/3072.71$ & - & $3219$ & - \\
$T(cd\bar s\bar s)$ & & $\big|(cu)_0^{6}\{\bar s\bar s\}_0^{\bar 6} \big\rangle$ & $3176.90/3177.87$ & - & $3520$ & - \\
& $\frac{1}{2}(1^+)$ & $\big|(cu)_0^{\bar 3}\{\bar s\bar s\}_1^3 \big\rangle$ & $3008.84/3010.34$ & - & $3263$ & - \\
& & $\big|(cu)_1^{\bar 3}\{\bar s\bar s\}_1^3 \big\rangle$ & $3113.48/3114.53$ & - & $3382$ & - \\
& & $\big|(cu)_1^{6}\{\bar s\bar s\}_0^{\bar 6} \big\rangle$ & $3103.66/3104.86$ & - & $3458$ & - \\
& $\frac{1}{2}(2^+)$ & $\big|(cu)_1^{\bar 3}\{\bar s\bar s\}_1^3 \big\rangle$ & $3197.16/3198.17$ & - & $3443$ & - \\ \hline
\end{tabular}
\end{table}
\begin{table}[ht]
\centering
\captionof{table}{Masses of $(\overline{\mb3}_c\mb3_c)$ and $(\mb6_c\overline{\mb6}_c)$ open charm ($cs\bar n\bar n, cs\bar n\bar s, cs\bar s\bar s$) tetraquark states using heavy-light diquarks.}
\label{t10}
\begin{tabular}{|c|c|c|c|c|c|c|}\hline
\multirow{2}{*}{\textbf{States}} & \multirow{2}{*}{$I(J^{P})$} & \multirow{2}{*}{\textbf{Configuration}} & \textbf{This} & \bf NRPM & \bf RQM & \bf PDG \\
& & & \textbf{work} & \bf \cite{Liu:2022hbk} & \bf \cite{Lu:2020qmp} & \bf \cite{ParticleDataGroup:2024cfk}  \\ \hline

$T(cs\bar u\bar d)$ & $0(0^+)$ & $\big|(cs)_0^{\bar 3}[\bar u\bar d]_0^3 \big\rangle$ & $2703.14$ & $2818$ & $2765$ & - \\
& & $\big|(cs)_1^{6}[\bar u\bar d]_1^{\bar 6} \big\rangle$ & $2612.87$ & $3095$ & $3152$ & - \\
& $0(1^+)$ & $\big|(cs)_1^{\bar 3}[\bar u\bar d]_0^3 \big\rangle$ & $2847.08$ & $2946$ & $2964$ & - \\
& & $\big|(cs)_0^{6}[\bar u\bar d]_1^{\bar 6} \big\rangle$ & $2991.08$ & $3067$ & $3108$ & - \\
& & $\big|(cs)_1^{6}[\bar u\bar d]_1^{\bar 6} \big\rangle$ & $2765.99$ & $3161$ & $3263$ & - \\
& $0(2^+)$ & $\big|(cs)_1^{6}[\bar u\bar d]_1^{\bar 6} \big\rangle$ & $3072.23$ & $3188$ & $3316$ & - \\
& $1(0^+)$ & $\big|(cs)_1^{\bar 3}\{\bar u\bar d\}_1^3 \big\rangle$ & $2841.41$ & $3260$ & $3065$ & $2872\pm 16$ \\
& & $\big|(cs)_0^{6}\{\bar u\bar d\}_0^{\bar 6} \big\rangle$ & $3044.75$ & $3046$ & $3396$ & - \\
& $1(1^+)$ & $\big|(cs)_0^{\bar 3}\{\bar u\bar d\}_1^3 \big\rangle$ & $2810.47$ & $3185$ & $3130$ & - \\
& & $\big|(cs)_1^{\bar 3}\{\bar u\bar d\}_1^3 \big\rangle$ & $2897.91$ & $3227$ & $3235$ & - \\
& & $\big|(cs)_1^{6}\{\bar u\bar d\}_0^{\bar 6} \big\rangle$ & $2972.78$ & $3079$ & $3339$ & - \\
& $1(2^+)$ & $\big|(cs)_1^{\bar 3}\{\bar u\bar d\}_1^3 \big\rangle$ & $3010.92$ & $3226$ & $3302$ & - \\ \hline

$T(cs\bar u\bar u)/$ & $1(0^+)$ & $\big|(cs)_1^{\bar 3}\{\bar u\bar u\}_1^3 \big\rangle$ & $2839.26/2842.28$ & - & - & - \\
$T(cs\bar d\bar d)$ & & $\big|(cs)_0^{6}\{\bar u\bar u\}_0^{\bar 6} \big\rangle$ & $3042.53/3045.32$ & - & - & - \\
& $1(1^+)$ & $\big|(cs)_0^{\bar 3}\{\bar u\bar u\}_1^3 \big\rangle$ & $2808.61/2811.23$ & - & - & - \\
& & $\big|(cs)_1^{\bar 3}\{\bar u\bar u\}_1^3 \big\rangle$ & $2895.91/2898.73$ & - & - & - \\
& & $\big|(cs)_1^{6}\{\bar u\bar u\}_0^{\bar 6} \big\rangle$ & $2970.56/2973.35$ & - & - & - \\
& $1(2^+)$ & $\big|(cs)_1^{\bar 3}\{\bar u\bar u\}_1^3 \big\rangle$ & $3009.20/3011.62$ & - & - & - \\ \hline

$T(cs\bar u\bar s)/$ & $\frac{1}{2}(0^+)$ & $\big|(cs)_0^{\bar 3}[\bar u\bar s]_0^3 \big\rangle$ & $2836.07/2817.51$ & - & $2967$ & - \\
$T(cs\bar d\bar s)$ & & $\big|(cs)_1^{6}[\bar u\bar s]_1^{\bar 6} \big\rangle$ & $2847.36/2844.61$ & - & $3312$ & - \\
& $\frac{1}{2}(1^+)$ & $\big|(cs)_1^{\bar 3}[\bar u\bar s]_0^3 \big\rangle$ & $2980.01/2961.45$ & - & $3156$ & - \\
& & $\big|(cs)_0^{6}[\bar u\bar s]_1^{\bar 6} \big\rangle$ & $3164.31/3162.17$ & - & $3261$ & - \\
& & $\big|(cs)_1^{6}[\bar u\bar s]_1^{\bar 6} \big\rangle$ & $2969.85/2967.40$ & - & $3397$ & - \\
& $\frac{1}{2}(2^+)$ & $\big|(cs)_1^{6}[\bar u\bar s]_1^{\bar 6} \big\rangle$ & $3214.83/3212.99$ & - & $3438$ & - \\
& $\frac{1}{2}(0^+)$ & $\big|(cs)_1^{\bar 3}\{\bar u\bar s\}_1^3 \big\rangle$ & $3062.79/3071.23$ & - & $3217$ & - \\
& & $\big|(cs)_0^{6}\{\bar u\bar s\}_0^{\bar 6} \big\rangle$ & $3250.21/3261.21$ & - & $3506$ & - \\
& $\frac{1}{2}(1^+)$ & $\big|(cs)_0^{\bar 3}\{\bar u\bar s\}_1^3 \big\rangle$ & $3007.88/3015.59$ & - & $3271$ & - \\
& & $\big|(cs)_1^{\bar 3}\{\bar u\bar s\}_1^3 \big\rangle$ & $3107.30/3115.38$ & - & $3370$ & - \\
& & $\big|(cs)_1^{6}\{\bar u\bar s\}_0^{\bar 6} \big\rangle$ & $3178.24/3189.24$ & - & $3451$ & - \\
& $\frac{1}{2}(2^+)$ & $\big|(cs)_1^{\bar 3}\{\bar u\bar s\}_1^3 \big\rangle$ & $3196.33/3203.68$ & - & $3431$ & - \\ \hline

$T(cs\bar s\bar s)$ & $0(0^+)$ & $\big|(cs)_1^{\bar 3}\{\bar s\bar s\}_1^3 \big\rangle$ & $3220.88$ & - & $3331$ & - \\
& & $\big|(cs)_0^{6}\{\bar s\bar s\}_0^{\bar 6} \big\rangle$ & $3378.02$ & - & $3592$ & - \\
& $0(1^+)$ & $\big|(cs)_0^{\bar 3}\{\bar s\bar s\}_1^3 \big\rangle$ & $3153.89$ & - & $3379$ & - \\
& & $\big|(cs)_1^{\bar 3}\{\bar s\bar s\}_1^3 \big\rangle$ & $3259.35$ & - & $3475$ & - \\
& & $\big|(cs)_1^{6}\{\bar s\bar s\}_0^{\bar 6} \big\rangle$ & $3306.05$ & - & $3539$ & - \\
& $0(2^+)$ & $\big|(cs)_1^{\bar 3}\{\bar s\bar s\}_1^3 \big\rangle$ & $3336.31$ & - & $3533$ & - \\ \hline
\end{tabular}
\end{table}
\begin{figure}[t]
    \centering
    \includegraphics[width=1.0\linewidth]{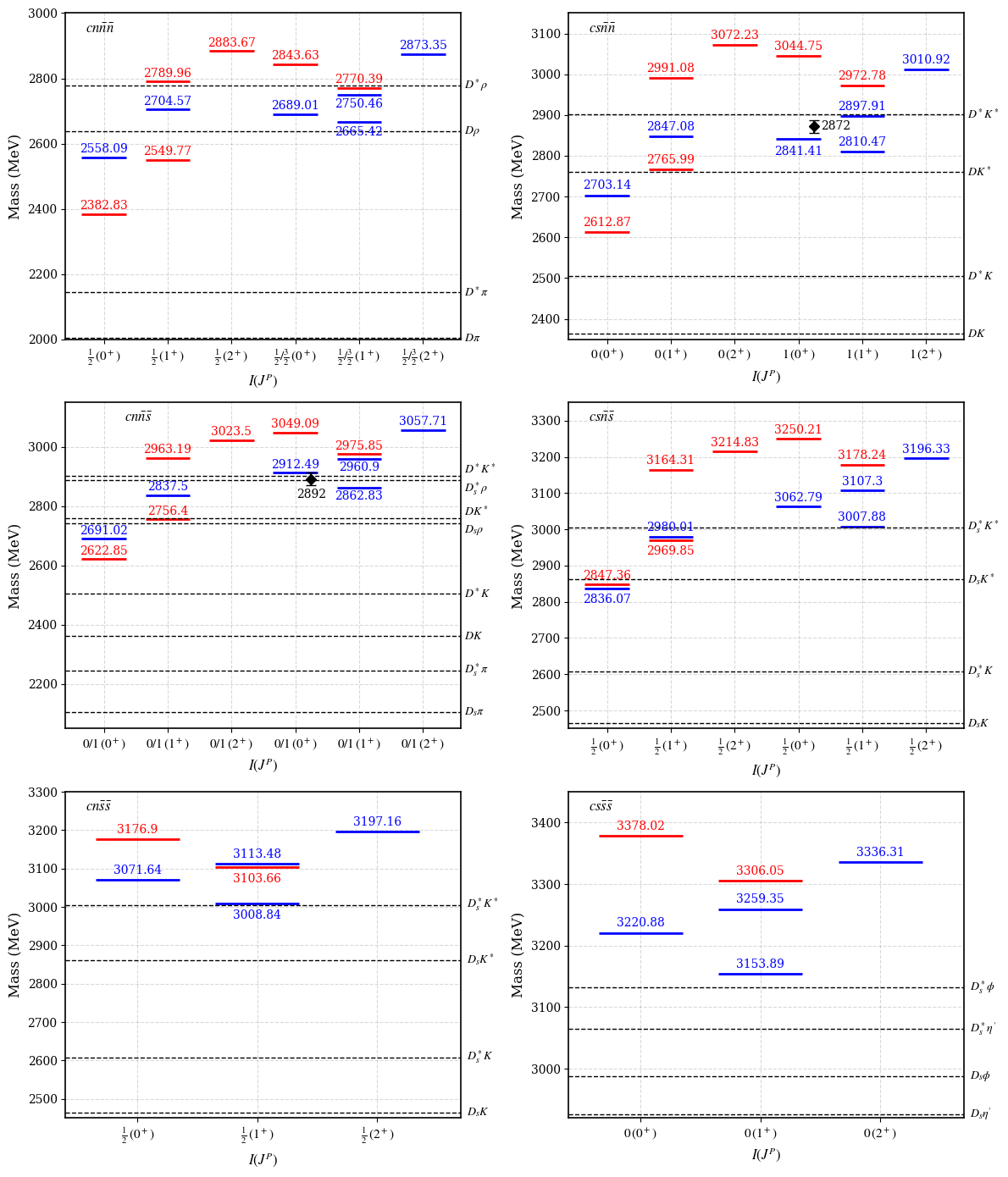}
    \caption{Mass spectra of singly-charm tetraquark states: $cn\bar n\bar n, cn\bar n\bar s, cn\bar s\bar s$ (left panel) and $cs\bar n\bar n, cs\bar n\bar s, cs\bar s\bar s$ (right panel). The $\bar {\mb3}\otimes\mb3~(\mb6\otimes\bar{\mb6})$ configurations are shown in blue (red), while black diamonds correspond to available experimental states. The relevant thresholds are indicated by dashed lines. The same legend is followed throughout.}
    \label{Plot_SinglyCharm}
\end{figure}
\begin{table}[ht]
\centering
\captionof{table}{Masses of $(\overline{\mb3}_c\mb3_c)$ and $(\mb6_c\overline{\mb6}_c)$ open bottom $bn\bar n\bar n$ tetraquark states using heavy-light diquarks.}
\label{t11}
\begin{tabular}{|c|c|c|c|c|}\hline
\multirow{2}{*}{\textbf{States}} & \multirow{2}{*}{$I(J^{P})$} & \multirow{2}{*}{\textbf{Configuration}} & \textbf{This} & \bf RQM \\
& & & \textbf{work} & \bf \cite{Lu:2020qmp}  \\ \hline

$T(bu\bar u\bar d)/$ & $\frac{1}{2}(0^+)$ & $\big|(bu)_0^{\bar 3}[\bar u\bar d]_0^3 \big\rangle$ & $5891.55/5893.03$ & $5977$ \\
$T(bd\bar u\bar d)$ & & $\big|(bu)_1^{6}[\bar u\bar d]_1^{\bar 6} \big\rangle$ & $5917.33/5918.56$ & $6382$ \\
& $\frac{1}{2}(1^+)$ & $\big|(bu)_1^{\bar 3}[\bar u\bar d]_0^3 \big\rangle$ & $6027.55/6028.59$ & $6080$ \\
& & $\big|(bu)_0^{6}[\bar u\bar d]_1^{\bar 6} \big\rangle$ & $6111.63/6112.61$ & $6373$ \\
& & $\big|(bu)_1^{6}[\bar u\bar d]_1^{\bar 6} \big\rangle$ & $5980.48/5981.70$ & $6534$ \\
& $\frac{1}{2}(2^+)$ & $\big|(bu)_1^{6}[\bar u\bar d]_1^{\bar 6} \big\rangle$ & $6106.78/6107.97$ & $6552$ \\
& $\frac{1}{2}/\frac{3}{2}(0^+)$ & $\big|(bu)_1^{\bar 3}\{\bar u\bar d\}_1^3 \big\rangle$ & $6087.59/6088.64$ & $6256$ \\
& & $\big|(bu)_0^{6}\{\bar u\bar d\}_0^{\bar 6} \big\rangle$ & $6165.30/6166.28$ & $6621$ \\
& $\frac{1}{2}/\frac{3}{2}(1^+)$ & $\big|(bu)_0^{\bar 3}\{\bar u\bar d\}_1^3 \big\rangle$ & $5998.88/6000.36$ & $6286$ \\
& & $\big|(bu)_1^{\bar 3}\{\bar u\bar d\}_1^3 \big\rangle$ & $6111.24/6112.28$ & $6478$ \\
& & $\big|(bu)_1^{6}\{\bar u\bar d\}_0^{\bar 6} \big\rangle$ & $6097.30/6098.50$ & $6591$ \\
& $\frac{1}{2}/\frac{3}{2}(2^+)$ & $\big|(bu)_1^{\bar 3}\{\bar u\bar d\}_1^3 \big\rangle$ & $6158.53/6159.57$ & $6503$ \\ \hline

$T(bu\bar u\bar u)/$ & $\frac{1}{2}/\frac{3}{2}(0^+)$ & $\big|(bu)_1^{\bar 3}\{\bar u\bar u\}_1^3 \big\rangle$ & $6085.61/6086.66$ &  - \\
$T(bd\bar u\bar u)$ & & $\big|(bu)_0^{6}\{\bar u\bar u\}_0^{\bar 6} \big\rangle$ & $6163.08/6164.06$ & - \\
& $\frac{1}{2}/\frac{3}{2}(1^+)$ & $\big|(bu)_0^{\bar 3}\{\bar u\bar u\}_1^3 \big\rangle$ & $5997.02/5998.50$ & - \\
& & $\big|(bu)_1^{\bar 3}\{\bar u\bar u\}_1^3 \big\rangle$ & $6109.32/6110.36$ & - \\
& & $\big|(bu)_1^{6}\{\bar u\bar u\}_0^{\bar 6} \big\rangle$ & $6095.08/6096.28$ & - \\
& $\frac{1}{2}/\frac{3}{2}(2^+)$ & $\big|(bu)_1^{\bar 3}\{\bar u\bar u\}_1^3 \big\rangle$ & $6156.73/6157.77$ & - \\ \hline

$T(bu\bar d\bar d)/$ & $\frac{1}{2}/\frac{3}{2}(0^+)$ & $\big|(bu)_1^{\bar 3}\{\bar d\bar d\}_1^3 \big\rangle$ & $6088.40/6089.44$ & - \\
$T(bd\bar d\bar d)$ & & $\big|(bu)_0^{6}\{\bar d\bar d\}_0^{\bar 6} \big\rangle$ & $6165.87/6166.85$ & - \\
& $\frac{1}{2}/\frac{3}{2}(1^+)$ & $\big|(bu)_0^{\bar 3}\{\bar d\bar d\}_1^3 \big\rangle$ & $5999.64/6001.12$ & - \\
& & $\big|(bu)_1^{\bar 3}\{\bar d\bar d\}_1^3 \big\rangle$ & $6112.02/6113.06$ & - \\
& & $\big|(bu)_1^{6}\{\bar d\bar d\}_0^{\bar 6} \big\rangle$ & $6097.87/6099.07$ & - \\
& $\frac{1}{2}/\frac{3}{2}(2^+)$ & $\big|(bu)_1^{\bar 3}\{\bar d\bar d\}_1^3 \big\rangle$ & $6159.26/6160.30$ & - \\ \hline
\end{tabular}
\end{table}
\begin{table}[ht]
\centering
\captionof{table}{Masses of $(\overline{\mb3}_c\mb3_c)$ and $(\mb6_c\overline{\mb6}_c)$ open bottom ($bn\bar n\bar s, bn\bar s\bar s$) tetraquark states using heavy-light diquarks.}
\label{t12}
\begin{tabular}{|c|c|c|c|c|}\hline
\multirow{2}{*}{\textbf{States}} & \multirow{2}{*}{$I(J^{P})$} & \multirow{2}{*}{\textbf{Configuration}} & \textbf{This} & \bf RQM \\
& & & \textbf{work} & \bf \cite{Lu:2020qmp} \\ \hline

$T(bu\bar u\bar s)/$ & $0/1(0^+)$ & $\big|(bu)_0^{\bar 3}[\bar u\bar s]_0^3 \big\rangle$ & $6024.48/6025.96$ & $6203$ \\
$T(bd\bar u\bar s)$ & & $\big|(bu)_1^{6}[\bar u\bar s]_1^{\bar 6} \big\rangle$ & $6115.82/6117.05$ & $6538$ \\
& $0/1(1^+)$ & $\big|(bu)_1^{\bar 3}[\bar u\bar s]_0^3 \big\rangle$ & $6160.48/6161.52$ & $6292$ \\
& & $\big|(bu)_0^{6}[\bar u\bar s]_1^{\bar 6} \big\rangle$ & $6284.86/6285.84$ & $6531$ \\
& & $\big|(bu)_1^{6}[\bar u\bar s]_1^{\bar 6} \big\rangle$ & $6166.34/6167.55$ & $6650$ \\
& $0/1(2^+)$ & $\big|(bu)_1^{6}[\bar u\bar s]_1^{\bar 6} \big\rangle$ & $6267.37/6268.56$ & $6668$ \\
& $0/1(0^+)$ & $\big|(bu)_1^{\bar 3}\{\bar u\bar s\}_1^3 \big\rangle$ & $6295.03/6296.07$ & $6421$ \\
& & $\big|(bu)_0^{6}\{\bar u\bar s\}_0^{\bar 6} \big\rangle$ & $6370.76/6371.74$ & $6723$ \\
& $0/1(1^+)$ & $\big|(bu)_0^{\bar 3}\{\bar u\bar s\}_1^3 \big\rangle$ & $6196.29/6197.77$ & $6446$ \\
& & $\big|(bu)_1^{\bar 3}\{\bar u\bar s\}_1^3 \big\rangle$ & $6313.66/6314.70$ & $6605$ \\
& & $\big|(bu)_1^{6}\{\bar u\bar s\}_0^{\bar 6} \big\rangle$ & $6302.76/6303.96$ & $6694$ \\
& $0/1(2^+)$ & $\big|(bu)_1^{\bar 3}\{\bar u\bar s\}_1^3 \big\rangle$ & $6350.91/6351.95$ & $6630$ \\ \hline

$T(bu\bar d\bar s)/$ & $0/1(0^+)$ & $\big|(bu)_0^{\bar 3}[\bar d\bar s]_0^3 \big\rangle$ & $6005.92/6007.40$ & - \\
$T(bd\bar d\bar s)$ & & $\big|(bu)_1^{6}[\bar d\bar s]_1^{\bar 6} \big\rangle$ & $6113.44/6114.66$ & - \\
& $0/1(1^+)$ & $\big|(bu)_1^{\bar 3}[\bar d\bar s]_0^3 \big\rangle$ & $6141.92/6142.96$ & - \\
& & $\big|(bu)_0^{6}[\bar d\bar s]_1^{\bar 6} \big\rangle$ & $6282.72/6283.70$ & - \\
& & $\big|(bu)_1^{6}[\bar d\bar s]_1^{\bar 6} \big\rangle$ & $6164.08/6165.29$ & - \\
& $0/1(2^+)$ & $\big|(bu)_1^{6}[\bar d\bar s]_1^{\bar 6} \big\rangle$ & $6265.36/6266.55$ & - \\
& $0/1(0^+)$ & $\big|(bu)_1^{\bar 3}\{\bar d\bar s\}_1^3 \big\rangle$ & $6303.04/6304.09$ & - \\
& & $\big|(bu)_0^{6}\{\bar d\bar s\}_0^{\bar 6} \big\rangle$ & $6381.76/6382.74$ & - \\
& $0/1(1^+)$ & $\big|(bu)_0^{\bar 3}\{\bar d\bar s\}_1^3 \big\rangle$ & $6204.00/6205.48$ & - \\
& & $\big|(bu)_1^{\bar 3}\{\bar d\bar s\}_1^3 \big\rangle$ & $6321.52/6322.56$ & - \\
& & $\big|(bu)_1^{6}\{\bar d\bar s\}_0^{\bar 6} \big\rangle$ & $6313.76/6314.96$ & - \\
& $0/1(2^+)$ & $\big|(bu)_1^{\bar 3}\{\bar d\bar s\}_1^3 \big\rangle$ & $6358.47/6359.51$ & - \\ \hline

$T(bu\bar s\bar s)/$ & $\frac{1}{2}(0^+)$ & $\big|(bu)_1^{\bar 3}\{\bar s\bar s\}_1^3 \big\rangle$ & $6446.10/6447.14$ & $6547$ \\
$T(bd\bar s\bar s)$ & & $\big|(bu)_0^{6}\{\bar s\bar s\}_0^{\bar 6} \big\rangle$ & $6498.57/6499.55$ & $6800$ \\
& $\frac{1}{2}(1^+)$ & $\big|(bu)_0^{\bar 3}\{\bar s\bar s\}_1^3 \big\rangle$ & $6342.30/6343.78$ & $6569$ \\
& & $\big|(bu)_1^{\bar 3}\{\bar s\bar s\}_1^3 \big\rangle$ & $6462.20/6463.24$ & $6704$ \\
& & $\big|(bu)_1^{6}\{\bar s\bar s\}_0^{\bar 6} \big\rangle$ & $6430.57/6431.77$ & $6773$ \\
& $\frac{1}{2}(2^+)$ & $\big|(bu)_1^{\bar 3}\{\bar s\bar s\}_1^3 \big\rangle$ & $6494.40/6495.44$ & $6729$ \\ \hline
\end{tabular}
\end{table}
\begin{table}[ht]
\centering
\captionof{table}{Masses of $(\overline{\mb3}_c\mb3_c)$ and $(\mb6_c\overline{\mb6}_c)$ open bottom ($bs\bar n\bar n, bs\bar n\bar s, bs\bar s\bar s$) tetraquark states using heavy-light diquarks.}
\label{t13}
\begin{tabular}{|c|c|c|c|c|}\hline
\multirow{2}{*}{\textbf{States}} & \multirow{2}{*}{$I(J^{P})$} & \multirow{2}{*}{\textbf{Configuration}} & \textbf{This} & \bf RQM \\
& & & \textbf{work} & \bf \cite{Lu:2020qmp} \\ \hline

$T(bs\bar u\bar d)$ & $0(0^+)$ & $\big|(bs)_0^{\bar 3}[\bar u\bar d]_0^3 \big\rangle$ & $6018.04$ & $6194$ \\
& & $\big|(bs)_1^{6}[\bar u\bar d]_1^{\bar 6} \big\rangle$ & $6125.77$ & $6493$ \\
& $0(1^+)$ & $\big|(bs)_1^{\bar 3}[\bar u\bar d]_0^3 \big\rangle$ & $6166.24$ & $6272$ \\
& & $\big|(bs)_0^{6}[\bar u\bar d]_1^{\bar 6} \big\rangle$ & $6322.04$ & $6492$ \\
& & $\big|(bs)_1^{6}[\bar u\bar d]_1^{\bar 6} \big\rangle$ & $6186.86$ & $6630$ \\
& $0(2^+)$ & $\big|(bs)_1^{6}[\bar u\bar d]_1^{\bar 6} \big\rangle$ & $6309.03$ & $6656$ \\
& $1(0^+)$ & $\big|(bs)_1^{\bar 3}\{\bar u\bar d\}_1^3 \big\rangle$ & $6227.85$ & $6430$ \\
& & $\big|(bs)_0^{6}\{\bar u\bar d\}_0^{\bar 6} \big\rangle$ & $6375.71$ & $6714$ \\
& $1(1^+)$ & $\big|(bs)_0^{\bar 3}\{\bar u\bar d\}_1^3 \big\rangle$ & $6125.37$ & $6457$ \\
& & $\big|(bs)_1^{\bar 3}\{\bar u\bar d\}_1^3 \big\rangle$ & $6250.71$ & $6577$ \\
& & $\big|(bs)_1^{6}\{\bar u\bar d\}_0^{\bar 6} \big\rangle$ & $6301.61$ & $6689$ \\
& $1(2^+)$ & $\big|(bs)_1^{\bar 3}\{\bar u\bar d\}_1^3 \big\rangle$ & $6296.44$ & $6602$ \\ \hline

$T(bs\bar u\bar u)/$ & $1(0^+)$ & $\big|(bs)_1^{\bar 3}\{\bar u\bar u\}_1^3 \big\rangle$ & $6225.87/6228.65$ & - \\
$T(bs\bar d\bar d)$ & & $\big|(bs)_0^{6}\{\bar u\bar u\}_0^{\bar 6} \big\rangle$ & $6373.49/6376.28$ & - \\
& $1(1^+)$ & $\big|(bs)_0^{\bar 3}\{\bar u\bar u\}_1^3 \big\rangle$ & $6123.51/6126.13$ & - \\
& & $\big|(bs)_1^{\bar 3}\{\bar u\bar u\}_1^3 \big\rangle$ & $6248.79/6251.49$ & - \\
& & $\big|(bs)_1^{6}\{\bar u\bar u\}_0^{\bar 6} \big\rangle$ & $6299.39/6302.18$ & - \\
& $1(2^+)$ & $\big|(bs)_1^{\bar 3}\{\bar u\bar u\}_1^3 \big\rangle$ & $6294.63/6297.17$ & - \\ \hline

$T(bs\bar u\bar s)/$ & $\frac{1}{2}(0^+)$ & $\big|(bs)_0^{\bar 3}[\bar u\bar s]_0^3 \big\rangle$ & $6150.97/6132.41$ & $6388$ \\
$T(bs\bar d\bar s)$ & & $\big|(bs)_1^{6}[\bar u\bar s]_1^{\bar 6} \big\rangle$ & $6323.44/6321.06$ & $6643$ \\
& $\frac{1}{2}(1^+)$ & $\big|(bs)_1^{\bar 3}[\bar u\bar s]_0^3 \big\rangle$ & $6299.17/6280.61$ & $6464$ \\
& & $\big|(bs)_0^{6}[\bar u\bar s]_1^{\bar 6} \big\rangle$ & $6495.27/6493.13$ & $6639$ \\
& & $\big|(bs)_1^{6}[\bar u\bar s]_1^{\bar 6} \big\rangle$ & $6372.30/6370.04$ & $6748$ \\
& $\frac{1}{2}(2^+)$ & $\big|(bs)_1^{6}[\bar u\bar s]_1^{\bar 6} \big\rangle$ & $6470.03/6468.01$ & $6772$ \\
& $\frac{1}{2}(0^+)$ & $\big|(bs)_1^{\bar 3}\{\bar u\bar s\}_1^3 \big\rangle$ & $6434.95/6442.96$ & $6573$ \\
& & $\big|(bs)_0^{6}\{\bar u\bar s\}_0^{\bar 6} \big\rangle$ & $6581.17/6592.17$ & $6818$ \\
& $\frac{1}{2}(1^+)$ & $\big|(bs)_0^{\bar 3}\{\bar u\bar s\}_1^3 \big\rangle$ & $6322.78/6330.49$ & $6598$ \\
& & $\big|(bs)_1^{\bar 3}\{\bar u\bar s\}_1^3 \big\rangle$ & $6452.96/6460.82$ & $6704$ \\
& & $\big|(bs)_1^{6}\{\bar u\bar s\}_0^{\bar 6} \big\rangle$ & $6507.07/6518.07$ & $6793$ \\
& $\frac{1}{2}(2^+)$ & $\big|(bs)_1^{\bar 3}\{\bar u\bar s\}_1^3 \big\rangle$ & $6488.98/6496.55$ & $6728$ \\ \hline

$T(bs\bar s\bar s)$ & $0(0^+)$ & $\big|(bs)_1^{\bar 3}\{\bar s\bar s\}_1^3 \big\rangle$ & $6585.85$ & $6682$ \\
& & $\big|(bs)_0^{6}\{\bar s\bar s\}_0^{\bar 6} \big\rangle$ & $6708.98$ & $6898$ \\
& $0(1^+)$ & $\big|(bs)_0^{\bar 3}\{\bar s\bar s\}_1^3 \big\rangle$ & $6468.79$ & $6705$ \\
& & $\big|(bs)_1^{\bar 3}\{\bar s\bar s\}_1^3 \big\rangle$ & $6601.42$ & $6802$ \\
& & $\big|(bs)_1^{6}\{\bar s\bar s\}_0^{\bar 6} \big\rangle$ & $6634.88$ & $6874$ \\
& $0(2^+)$ & $\big|(bs)_1^{\bar 3}\{\bar s\bar s\}_1^3 \big\rangle$ & $6632.56$ & $6826$ \\ \hline
\end{tabular}
\end{table}
\begin{figure}[t]
    \centering
    \includegraphics[width=1.0\linewidth]{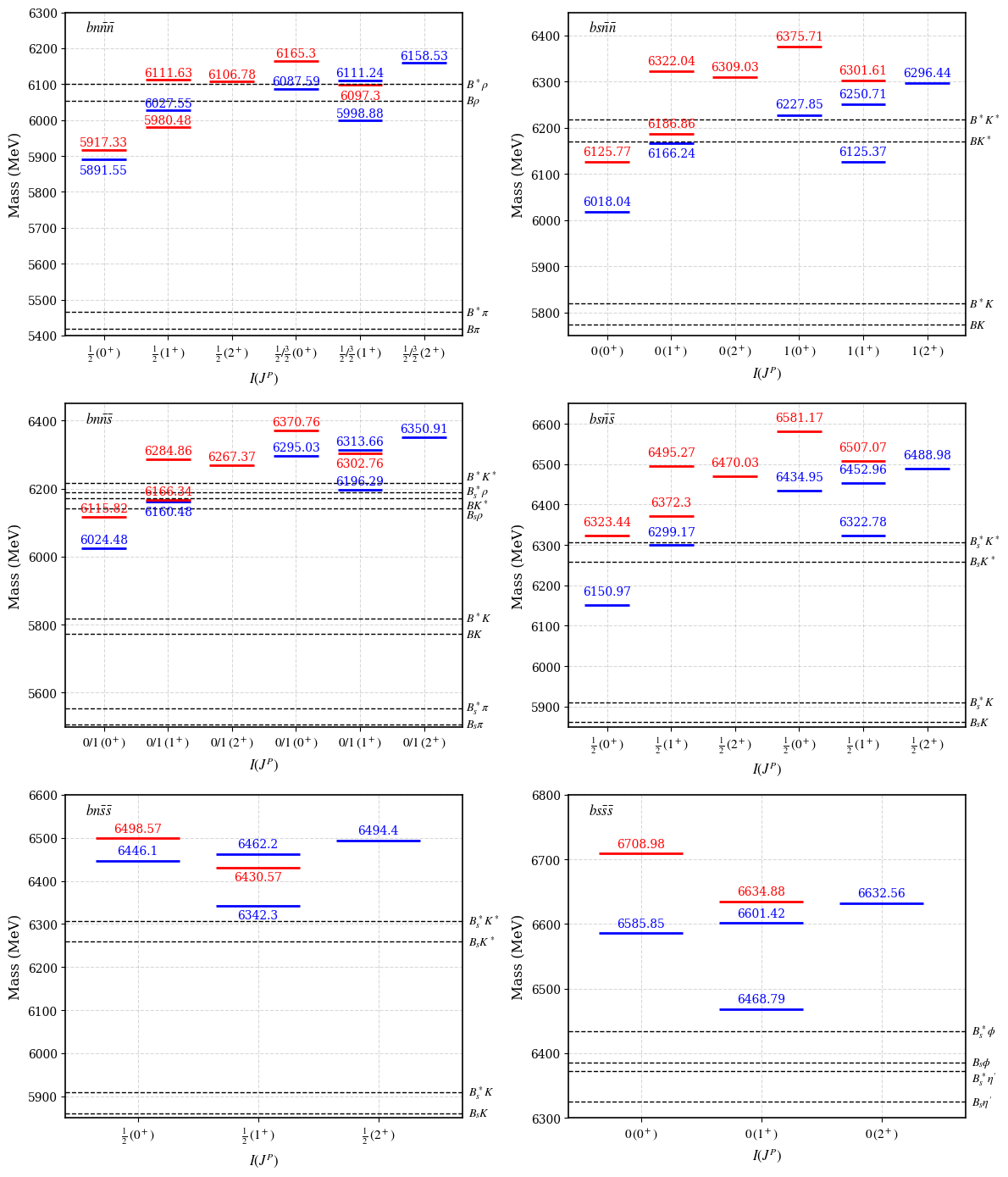}
    \caption{Mass spectra of singly-bottom tetraquark states: $bn\bar n\bar n, bn\bar n\bar s, bn\bar s\bar s$ (left panel) and $bs\bar n\bar n, bs\bar n\bar s, bs\bar s\bar s$ (right panel).}
    \label{Plot_SinglyBottom}
\end{figure}
\begin{table}[ht]
\centering
\captionof{table}{Masses of $(\overline{\mb3}_c\mb3_c)$ and $(\mb6_c\overline{\mb6}_c)$ doubly charm tetraquark states (in MeV) using heavy-heavy diquarks.}
\label{t14}
\begin{tabular}{|c|c|c|c|c|c|c|c|c|c|c|c|}\hline
\multirow{2}{*}{\textbf{States}} & \multirow{2}{*}{$I(J^{P})$} & \multirow{2}{*}{\textbf{Configuration}} & \multirow{2}{*}{\textbf{This work}} & \textbf{CMIM} & \textbf{Regge} & \textbf{HQET}\footnotemark[2] & \textbf{BM} & \textbf{RQM} & \textbf{RQM} & \textbf{HQS}\\
&  &  & & \bf \cite{Luo:2017eub} & \bf \cite{Song:2023izj} & \bf \cite{Braaten:2020nwp} & \bf \cite{Zhang:2021yul} & \bf \cite{Ebert:2007rn} & \bf \cite{Lu:2020rog} & \bf \cite{Eichten:2017ffp} \\ \hline
		
$T(cc\bar u\bar d)$\footnotemark[1] & $0(1^+)$ & $\big|\{cc\}_1^{\bar 3}[\bar u\bar d]_0^3 \big\rangle$ & $3860.35$ & $3779$ & $3997$ & $3947(11)$ & $3925$ &  $3935$ & $4041$ & $3978$ \\
& & $\big|\{cc\}_0^{6}[\bar u\bar d]_1^{\bar 6} \big\rangle$ & $4125.86$ & $3977$ & - & - & $4205$ &  - & $4313$ & - \\ \cline{2-11}
& $1(0^+)$ & $\big|\{cc\}_1^{\bar 3}\{\bar u\bar d\}_1^3 \big\rangle$ & $3891.78$ & $4128$ & $4163$ & $4111(11)$ & $4342$ & $4056$ & $4195$ & $4146$ \\
& & $\big|\{cc\}_0^{6}\{\bar u\bar d\}_0^{\bar 6} \big\rangle$ & $4179.53$ & $3850$ & - & - & $4032$ & - & $4414$ & - \\ \cline{2-11}
& $1(1^+)$ & $\big|\{cc\}_1^{\bar 3}\{\bar u\bar d\}_1^3 \big\rangle$ & $3929.73$ & $3973$ & $4185$ & $4133(11)$ & $4117$ &  $4079$ & $4268$ & $4167$ \\ \cline{2-11}
& $1(2^+)$ & $\big|\{cc\}_1^{\bar 3}\{\bar u\bar d\}_1^3 \big\rangle$ & $4005.63$ & $4044$ & $4229$ & $4177(11)$ & $4179$ &  $4118$ & $4318$ & $4210$ \\ \hline
    
$T(cc\bar u\bar u)/$ & $1(0^+)$ & $\big|\{cc\}_1^{\bar 3}\{\bar u\bar u\}_1^3 \big\rangle$ & $3889.73/3892.62$ & - & - & - & - & - & - & - \\
$T(cc\bar d\bar d)$ & & $\big|\{cc\}_0^{6}\{\bar u\bar u\}_0^{\bar 6} \big\rangle$ & $4177.31/4180.10$ & - & - & - & - & - & - & - \\ \cline{2-11}
& $1(1^+)$ & $\big|\{cc\}_1^{\bar 3}\{\bar u\bar u\}_1^3 \big\rangle$ & $3927.78/3930.53$ & - & - & - & - & - & - & - \\ \cline{2-11}
& $1(2^+)$ & $\big|\{cc\}_1^{\bar 3}\{\bar u\bar u\}_1^3 \big\rangle$ & $4003.87/4006.35$ & - & - & - & - & - & - & - \\ \hline
    
$T(cc\bar u\bar s)/$ & $\frac{1}{2}(1^+)$ & $\big|\{cc\}_1^{\bar 3}[\bar u\bar s]_0^3 \big\rangle$ & $3993.28/3974.72$ & $3921$ & $4180$ & $4124(12)$ & $4091$ &  $4143$ & $4232$ & $4156$ \\
$T(cc\bar d\bar s)$ & & $\big|\{cc\}_0^{6}[\bar u\bar s]_1^{\bar 6} \big\rangle$ & $4299.09/4296.95$ & $4096$ & - & - & $4314$ & - & $4427$ & - \\ \cline{2-11}
& $\frac{1}{2}(0^+)$ & $\big|\{cc\}_1^{\bar 3}\{\bar u\bar s\}_1^3 \big\rangle$ & $4105.29/4113.49$ & $4210$ & $4297$ & $4232(11)$ & $4429$ &  $4221$ & $4323$ & - \\
& & $\big|\{cc\}_0^{6}\{\bar u\bar s\}_0^{\bar 6} \big\rangle$ & $4384.99/4395.99$ & $3933$ & - & - & $4165$ & - & $4512$ & - \\ \cline{2-11}
& $\frac{1}{2}(1^+)$ & $\big|\{cc\}_1^{\bar 3}\{\bar u\bar s\}_1^3 \big\rangle$ & $4135.19/4143.14$ & $4060$ & $4315$ & $4254(11)$ & $4247$ &  $4239$ & $4394$ & - \\ \cline{2-11}
& $\frac{1}{2}(2^+)$ & $\big|\{cc\}_1^{\bar 3}\{\bar u\bar s\}_1^3 \big\rangle$ & $4194.98/4202.45$ & $4131$ & $4352$ & $4298(11)$ & $4305$ &  $4271$ & $4440$ & - \\ \hline
    
$T(cc\bar s\bar s)$ & $0(0^+)$ & $\big|\{cc\}_1^{\bar 3}\{\bar s\bar s\}_1^3 \big\rangle$ & $4259.41$ & $4293$ & $4420$ & $4352(12)$ & $4521$ &  $4359$ & $4417$ & - \\
& & $\big|\{cc\}_0^{6}\{\bar s\bar s\}_0^{\bar 6} \big\rangle$ & $4512.80$ & $4016$ & - & - & $4300$ &  - & $4587$ & - \\ \cline{2-11}
& $0(1^+)$ & $\big|\{cc\}_1^{\bar 3}\{\bar s\bar s\}_1^3 \big\rangle$ & $4285.26$ & $4146$ & $4436$ & $4374(12)$ & $4382$ &  $4375$ & $4493$ & - \\ \cline{2-11}
& $0(2^+)$ & $\big|\{cc\}_1^{\bar 3}\{\bar s\bar s\}_1^3 \big\rangle$ & $4336.94$ & $4218$ & $4470$ & $4418(12)$ & $4433$ &  $4402$ & $4536$ & - \\ \hline 
\end{tabular}
\footnotetext[1]{PDG mass of $T(cc\bar u\bar d)$ state is given as $T_{cc}(3875)^+ = 3874.74(10)$ MeV \cite{ParticleDataGroup:2024cfk}, identified with $J^P=1^+$ as per~\cite{LHCb:2021vvq}.}
\footnotetext[2]{Uncertainties in masses are given within parentheses.}
\end{table}
\begin{table}[ht]
\centering
\captionof{table}{Masses of $(\overline{\mb3}_c\mb3_c)$ and $(\mb6_c\overline{\mb6}_c)$ charmed bottom tetraquark states (in MeV)  using heavy-heavy diquarks.}
\label{t15}
\begin{tabular}{|c|c|c|c|c|c|c|c|c|c|c|c|}\hline
\multirow{2}{*}{\textbf{States}} & \multirow{2}{*}{$I(J^{P})$} & \multirow{2}{*}{\textbf{Configuration}} & \multirow{2}{*}{\textbf{This work}} & \textbf{CMIM} & \textbf{Regge} & \textbf{HQET} & \textbf{BM} & \textbf{RQM} & \textbf{RQM} & \textbf{HQS} \\
&  &  & & \bf \cite{Luo:2017eub} & \bf \cite{Song:2023izj} & \bf \cite{Braaten:2020nwp} & \bf \cite{Zhang:2021yul} & \bf \cite{Ebert:2007rn} & \bf \cite{Lu:2020rog} & \bf \cite{Eichten:2017ffp} \\\hline
		
$T(cb\bar u\bar d)$ & $0(0^+)$ & $\big|(cb)_0^{\bar 3}[\bar u\bar d]_0^3 \big\rangle$ & $7132.56$ & $7041$ & $7268$ & $7248(37)$ & $7260$ &  $7239$ & $7297$ & $7229$ \\
& & $\big|(cb)_1^{6}[\bar u\bar d]_1^{\bar 6} \big\rangle$ & $7368.60$ & $7213$ & - & - & $7502$ & - & $7580$ & - \\ \cline{2-11}
& $0(1^+)$ & $\big|(cb)_1^{\bar 3}[\bar u\bar d]_0^3 \big\rangle$ & $7137.02$ & $7106$ & $7269$ & $7282(28)$ & $7288$ & $7246$ & $7325$ & $7272$ \\ \cline{3-11}
& & $\big|(cb)_0^{6}[\bar u\bar d]_1^{\bar 6} \big\rangle$ & $7471.87$ & $7301$ & - & - & $7605$ & - & $7607$ & - \\
& & $\big|(cb)_1^{6}[\bar u\bar d]_1^{\bar 6} \big\rangle$ & $7419.13$ & $7215$ & - & - & $7518$ & - & $7666$ & - \\ \cline{2-11}
& $0(2^+)$ & $\big|(cb)_1^{6}[\bar u\bar d]_1^{\bar 6} \big\rangle$ & $7520.18$ & $7315$ & - & - & $7483$ & - & $7697$ & - \\ \cline{2-11}
& $1(0^+)$ & $\big|(cb)_1^{\bar 3}\{\bar u\bar d\}_1^3 \big\rangle$ & $7206.16$ & $7428$ & $7458$ & $7460(27)$ & $7714$ &  $7383$ & $7519$ & $7461$ \\
& & $\big|(cb)_0^{6}\{\bar u\bar d\}_0^{\bar 6} \big\rangle$ & $7525.54$ & $7241$ & - & - & $7438$ &  - & $7740$ & - \\ \cline{2-11}
& $1(1^+)$ & $\big|(cb)_0^{\bar 3}\{\bar u\bar d\}_1^3 \big\rangle$ & $7239.89$ & $7332$ & $7478$ & $7455(37)$ & $7509$ & $7403$ & $7537$ & $7439$ \\
& & $\big|(cb)_1^{\bar 3}\{\bar u\bar d\}_1^3 \big\rangle$ & $7225.26$ & $7258$ & $7469$ & $7474(27)$ & $7465$ &  $7396$ & $7561$ & $7472$ \\ \cline{3-11}
& & $\big|(cb)_1^{6}\{\bar u\bar d\}_0^{\bar 6} \big\rangle$ & $7523.32$ & $7393$ & - & - & $7699$ & - & $7729$ & - \\ \cline{2-11}
& $1(2^+)$ & $\big|(cb)_1^{\bar 3}\{\bar u\bar d\}_1^3 \big\rangle$ & $7263.45$ & $7367$ & $7490$ & $7503(27)$ & $7531$ &  $7422$ & $7586$ & $7493$ \\ \hline
    
$T(cb\bar u\bar u)/$ & $1(0^+)$ & $\big|(cb)_1^{\bar 3}\{\bar u\bar u\}_1^3 \big\rangle$ & $7204.20/7206.96$ & - & - & - & - & - & - & - \\
$T(cb\bar d\bar d)$ & & $\big|(cb)_0^{6}\{\bar u\bar u\}_0^{\bar 6} \big\rangle$ & $7523.32/7526.11$ & - & - & - & - & - & - & - \\ \cline{2-11}
& $1(1^+)$ & $\big|(cb)_0^{\bar 3}\{\bar u\bar u\}_1^3 \big\rangle$ & $7238.03/7240.65$ & - & - & - & - & - & - & - \\
& & $\big|(cb)_1^{\bar 3}\{\bar u\bar u\}_1^3 \big\rangle$ & $7223.35/7226.04$ & - & - & - & - & - & - & - \\ \cline{3-11}
& & $\big|(cb)_1^{6}\{\bar u\bar u\}_0^{\bar 6} \big\rangle$ & $7521.10/7523.89$ & - & - & - & - & - & - & - \\ \cline{2-11}
& $1(2^+)$ & $\big|(cb)_1^{\bar 3}\{\bar u\bar u\}_1^3 \big\rangle$ & $7261.64/7264.19$ & - & - & - & - & - & - & - \\ \hline
    
$T(cb\bar u\bar s)/$ & $\frac{1}{2}(0^+)$ & $\big|(cb)_0^{\bar 3}[\bar u\bar s]_0^3 \big\rangle$ & $7265.49/7246.93$ & $7158$ & $7451$ & $7422(38)$ & - &  $7444$ & $7483$ & $7406$ \\
$T(cb\bar d\bar s)$ & & $\big|(cb)_1^{6}[\bar u\bar s]_1^{\bar 6} \big\rangle$ & $7562.04/7559.70$ & $7498$ & - & - & - &  - & $7693$ & - \\ \cline{2-11}
& $\frac{1}{2}(1^+)$ & $\big|(cb)_1^{\bar 3}[\bar u\bar s]_0^3 \big\rangle$ & $7269.95/7251.39$ & $7227$ & $7452$ & $7456(28)$ & - &  $7451$ & $7514$ & $7445$ \\ \cline{3-11}
& & $\big|(cb)_0^{6}[\bar u\bar s]_1^{\bar 6} \big\rangle$ & $7645.10/7642.96$ & $7462$ & - & - & - & - & $7714$ & - \\
& & $\big|(cb)_1^{6}[\bar u\bar s]_1^{\bar 6} \big\rangle$ & $7602.46/7600.22$ & $7327$ & - & - & - &  - & $7769$ & - \\ \cline{2-11}
& $\frac{1}{2}(2^+)$ & $\big|(cb)_1^{6}[\bar u\bar s]_1^{\bar 6} \big\rangle$ & $7683.29/7681.26$ & $7442$ & - & - & - & - & $7796$ & \\ \cline{2-11}
& $\frac{1}{2}(0^+)$ & $\big|(cb)_1^{\bar 3}\{\bar u\bar s\}_1^3 \big\rangle$ & $7411.67/7419.63$ & $7312$ & $7558$ & $7578(28)$ & - &  $7540$ & $7643$ & - \\
& & $\big|(cb)_0^{6}\{\bar u\bar s\}_0^{\bar 6} \big\rangle$ & $7731.00/7742.00$ & $7332$ & - & - & - & - & $7827$ & - \\ \cline{2-11}
& $\frac{1}{2}(1^+)$ & $\big|(cb)_0^{\bar 3}\{\bar u\bar s\}_1^3 \big\rangle$ & $7437.30/7445.01$ & $7330$ & $7605$ & $7573(37)$ & - &  $7555$ & $7659$ & - \\
& & $\big|(cb)_1^{\bar 3}\{\bar u\bar s\}_1^3 \big\rangle$ & $7426.71/7434.55$ & $7402$ & $7579$ & $7592(28)$ & - &  $7552$ & $7682$ & - \\ \cline{3-11}
& & $\big|(cb)_1^{6}\{\bar u\bar s\}_0^{\bar 6} \big\rangle$ & $7728.78/7739.78$ & $7407$ & - & - & - & - & $7816$ & - \\ \cline{2-11}
& $\frac{1}{2}(2^+)$ & $\big|(cb)_1^{\bar 3}\{\bar u\bar s\}_1^3 \big\rangle$ & $7456.80/7464.39$ & $7415$ & $7616$ & $7621(28)$ & - &  $7572$ & $7705$ & - \\ \hline
    
$T(cb\bar s\bar s)$ & $0(0^+)$ & $\big|(cb)_1^{\bar 3}\{\bar s\bar s\}_1^3 \big\rangle$ & $7561.76$ & $7581$ & $7717$ & $7696(28)$ & $7875$ &  $7673$ & $7735$ & - \\
& & $\big|(cb)_0^{6}\{\bar s\bar s\}_0^{\bar 6} \big\rangle$ & $7858.81$ & $7394$ & - & - & $7693$ & - & $7894$ & - \\ \cline{2-11}
& $0(1^+)$ & $\big|(cb)_0^{\bar 3}\{\bar s\bar s\}_1^3 \big\rangle$ & $7583.31$ & $7414$ & $7710$ & $7691(38)$ & $7716$ &  $7684$ & $7752$ & - \\
& & $\big|(cb)_1^{\bar 3}\{\bar s\bar s\}_1^3 \big\rangle$ & $7574.77$ & $7545$ & $7733$ & $7710(28)$ & $7858$ &  $7683$ & $7775$ & - \\ \cline{3-11}
& & $\big|(cb)_1^{6}\{\bar s\bar s\}_0^{\bar 6} \big\rangle$ & $7856.59$ & $7493$ & - & - & $7757$ & - & $7881$ & - \\ \cline{2-11}
& $0(2^+)$ & $\big|(cb)_1^{\bar 3}\{\bar s\bar s\}_1^3 \big\rangle$ & $7600.77$ & $7529$ & - & $7739(28)$ & $7779$ &  $7701$ & $7798$ & - \\  \hline 
\end{tabular}
\end{table}
\begin{table}[ht]
\centering
\captionof{table}{Masses of $(\overline{\mb3}_c\mb3_c)$ and $(\mb6_c\overline{\mb6}_c)$ doubly bottom tetraquark states (in MeV)  using heavy-heavy diquarks.}
\label{t16}
\begin{tabular}{|c|c|c|c|c|c|c|c|c|c|c|c|}\hline
\multirow{2}{*}{\textbf{States}} & \multirow{2}{*}{$I(J^{P})$} & \multirow{2}{*}{\textbf{Configuration}} & \multirow{2}{*}{\textbf{This work}} & \textbf{CMIM} & \textbf{Regge} & \textbf{HQET} & \textbf{BM} & \textbf{RQM} & \textbf{RQM} & \textbf{HQS} \\
&  &  & & \bf \cite{Luo:2017eub} & \bf \cite{Song:2023izj} & \bf \cite{Braaten:2020nwp} & \bf \cite{Zhang:2021yul} & \bf \cite{Ebert:2007rn} & \bf \cite{Lu:2020rog} & \bf \cite{Eichten:2017ffp} \\ \hline
		
$T(bb\bar u\bar d)$ & $0(1^+)$ & $\big|\{bb\}_1^{\bar 3}[\bar u\bar d]_0^3 \big\rangle$ & $10355.07$ & $10483$ & $10530$ & $10471(25)$ & $10654$ &  $10502$ & $10550$ & $10482$ \\
& & $\big|\{bb\}_0^{6}[\bar u\bar d]_1^{\bar 6} \big\rangle$ & $10850.06$ & $10617$ & - & - & $10982$ &  - & $10951$ & - \\ \cline{2-11}
& $1(0^+)$ & $\big|\{bb\}_1^{\bar 3}\{\bar u\bar d\}_1^3 \big\rangle$ & $10436.89$ & $10734$ & $10726$ & $10664(25)$ & $11092$ &  $10648$ & $10765$ & $10674$ \\
& & $\big|\{bb\}_0^{6}\{\bar u\bar d\}_0^{\bar 6} \big\rangle$ & $10903.73$ & $10637$ & - & - & $10834$ &  - & $11019$ & - \\ \cline{2-11}
& $1(1^+)$ & $\big|\{bb\}_1^{\bar 3}\{\bar u\bar d\}_1^3 \big\rangle$ & $10449.65$ & $10671$ & $10733$ & $10671(25)$ & $10854$ &  $10657$ & $10779$ & $10681$ \\ \cline{2-11}
& $1(2^+)$ & $\big|\{bb\}_1^{\bar 3}\{\bar u\bar d\}_1^3 \big\rangle$ & $10475.16$ & $10694$ & $10747$ & $10685(25)$ & $10878$ &  $10673$ & $10799$ & $10695$ \\ \hline
    
$T(bb\bar u\bar u)/$ & $1(0^+)$ & $\big|\{bb\}_1^{\bar 3}\{\bar u\bar u\}_1^3 \big\rangle$ & $10434.97/10437.68$ & - & - & - & - & - & - & - \\
$T(bb\bar d\bar d)$ & & $\big|\{bb\}_0^{6}\{\bar u\bar u\}_0^{\bar 6} \big\rangle$ & $10901.51/10904.30$ & - & - & - & - & - & - & - \\ \cline{2-11}
& $1(1^+)$ & $\big|\{bb\}_1^{\bar 3}\{\bar u\bar u\}_1^3 \big\rangle$ & $10447.76/10450.42$ & - & - & - & - & - & - & -  \\ \cline{2-11}
& $1(2^+)$ & $\big|\{bb\}_1^{\bar 3}\{\bar u\bar u\}_1^3 \big\rangle$ & $10473.33/10475.90$ & - & - & - & - & - & - & -  \\ \hline
    
$T(bb\bar u\bar s)$/ & $\frac{1}{2}(1^+)$ & $\big|\{bb\}_1^{\bar 3}[\bar u\bar s]_0^3 \big\rangle$ & $10488.00/10469.44$ & $10619$ & $10713$ & $10644(26)$ & $10811$ &  $10706$ & $10734$ & $10643$ \\
$T(bb\bar d\bar s)$ & & $\big|\{bb\}_0^{6}[\bar u\bar s]_1^{\bar 6} \big\rangle$ & $11023.29/11021.15$ & $10745$ & - & - & $11068$ &  - & $11046$ & - \\ \cline{2-11}
& $\frac{1}{2}(0^+)$ & $\big|\{bb\}_1^{\bar 3}\{\bar u\bar s\}_1^3 \big\rangle$ & $10639.71/10647.58$ & $10804$ & $10855$ & $10781(25)$ & $11160$ &  $10802$ & $10883$ & - \\
& & $\big|\{bb\}_0^{6}\{\bar u\bar s\}_0^{\bar 6} \big\rangle$ & $11109.19/11120.19$ & $10707$ & - & - & $10955$ &  - & $11098$ & - \\ \cline{2-11}
& $\frac{1}{2}(1^+)$ & $\big|\{bb\}_1^{\bar 3}\{\bar u\bar s\}_1^3 \big\rangle$ & $10649.76/10657.55$ & $10718$ & $10861$ & $10788(25)$ & $10974$ &  $10809$ & $10897$ & - \\ \cline{2-11}
& $\frac{1}{2}(2^+)$ & $\big|\{bb\}_1^{\bar 3}\{\bar u\bar s\}_1^3 \big\rangle$ & $10669.85/10677.48$ & $10769$ & $10873$ & $10802(25)$ & $10997$ &  $10823$ & $10915$ & - \\ \hline
    
$T(bb\bar s\bar s)$ & $0(0^+)$ & $\big|\{bb\}_1^{\bar 3}\{\bar s\bar s\}_1^3 \big\rangle$ & $10788.45$ & $10875$ & $10976$ & $10898(26)$ & $11232$ &  $10932$ & $10972$ & - \\
& & $\big|\{bb\}_0^{6}\{\bar s\bar s\}_0^{\bar 6} \big\rangle$ & $11237.00$ & $10777$ & - & - & $11078$ &  - & $11155$ & - \\ \cline{2-11}
& $0(1^+)$ & $\big|\{bb\}_1^{\bar 3}\{\bar s\bar s\}_1^3 \big\rangle$ & $10797.13$ & $10820$ & $10981$ & $10905(26)$ & $11099$ &  $10939$ & $10986$ & - \\ \cline{2-11}
& $0(2^+)$ & $\big|\{bb\}_1^{\bar 3}\{\bar s\bar s\}_1^3 \big\rangle$ & $10814.51$ & $10844$ & $10991$ & $10919(26)$ & $11119$ &  $10950$ & $11004$ & - \\ \hline
\end{tabular}
\end{table}
\begin{figure}[t]
    \centering
    \includegraphics[width=1.0\linewidth]{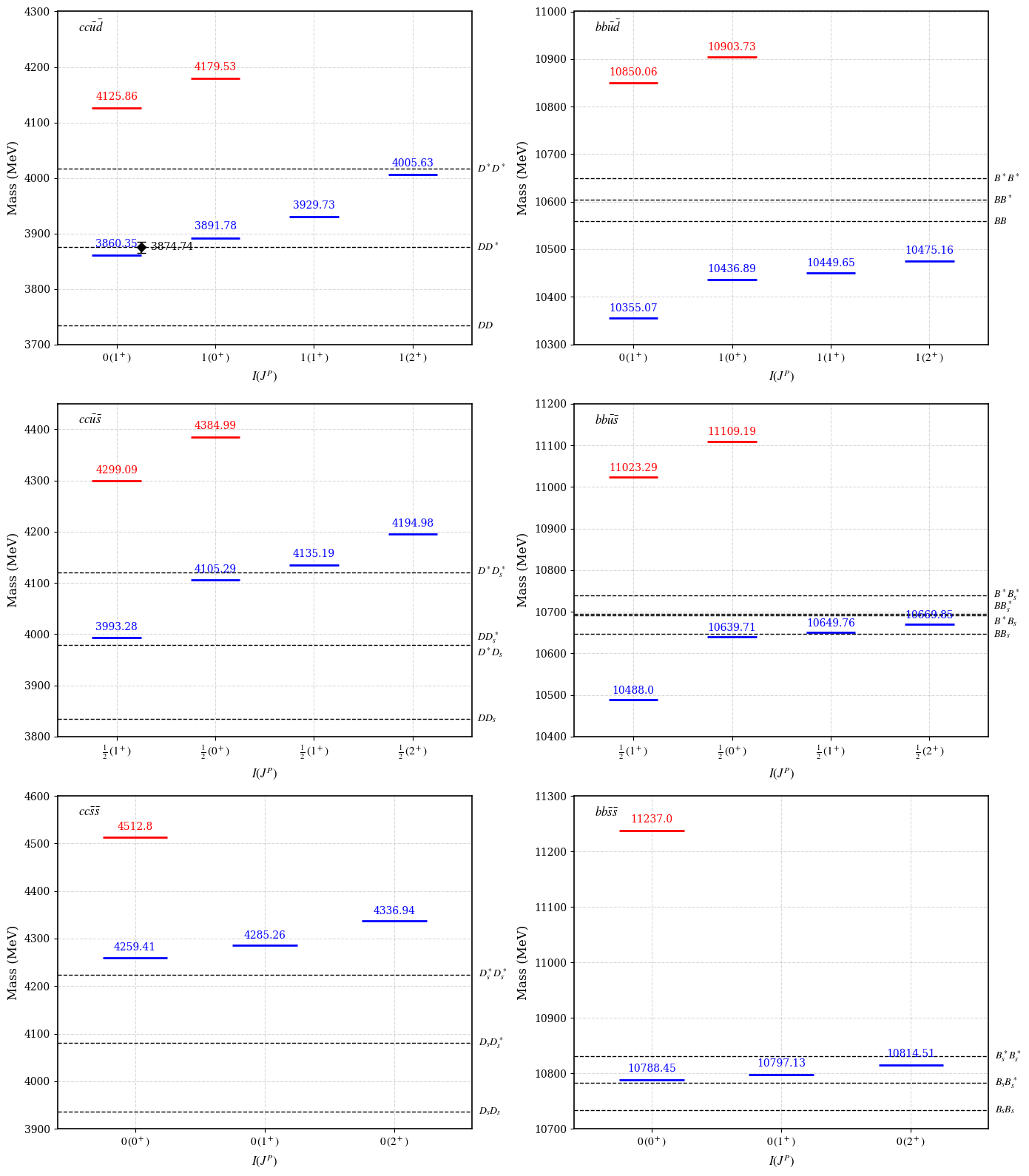}
    \caption{Mass spectra of doubly charm and doubly bottom tetraquark states: $cc\bar u\bar d, cc\bar u\bar s, cc\bar s\bar s$ (left panel) and $bb\bar u\bar d, bb\bar u\bar s, bb\bar s\bar s$ (right panel).}
    \label{Plot_cc_bb}
\end{figure}
\begin{figure}[t]
    \centering
    \includegraphics[width=0.5\linewidth]{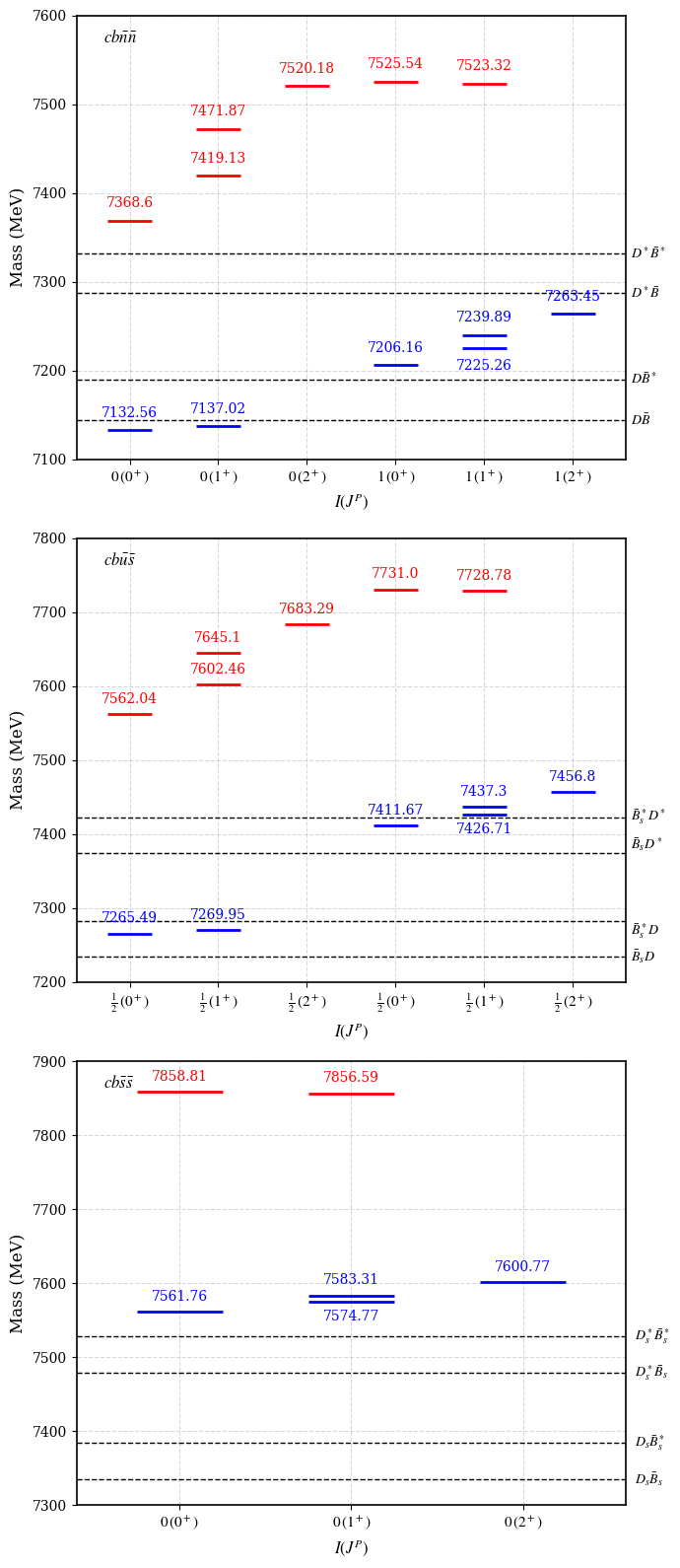}
    \caption{Mass spectra of charmed bottom tetraquark states: $cb\bar u\bar d, cb\bar u\bar s, cb\bar s\bar s$.}
    \label{Plot_cb}
\end{figure}
\begin{table}[ht]
\centering
\captionof{table}{Masses of $(\overline{\mb3}_c\mb3_c)$ and $(\mb6_c\overline{\mb6}_c)$ hidden charm $cn\bar c\bar n$ tetraquark states using heavy-light diquarks.}
\label{t17}
\scalebox{0.9}{
\begin{tabular}{|c|c|c|c|c|c|c|}\hline
\multirow{2}{*}{\textbf{States}} & \multirow{2}{*}{$I(J^P)$} & \multirow{2}{*}{\textbf{Configuration}} & \multirow{2}{*}{\textbf{This work}} & \textbf{RQM} & \textbf{BM} & \textbf{PDG} \\ 
& & & & \bf \cite{Yu:2024ljg} & \bf \cite{Yan:2023lvm} & \bf \cite{ParticleDataGroup:2024cfk} \\ \hline
$T(cu\bar c\bar u)/$ & $0/1(0^+)$ & $\big|(cu)_0^{\bar 3}(\bar c\bar u)_0^3 \big\rangle$ & $3864.96/3867.96$ & $3970$ & $3954$ & - \\
$T(cd\bar c\bar d)$ & & $\big|(cu)_1^{\bar 3}(\bar c\bar u)_1^3 \big\rangle$ & $4114.60/4116.70$ & $4038$ & $3715$ & - \\ \cline{3-7}
& & $\big|(cu)_0^{6}(\bar c\bar u)_0^{\bar 6} \big\rangle$ & $4194.54/4196.48$ & $4102$ & $4259$ & - \\
& & $\big|(cu)_1^{6}(\bar c\bar u)_1^{\bar 6} \big\rangle$ & $3933.79/3936.32$ & $3834$ & $4127$ & - \\ \cline{2-7}
& $0/1(1^+)$ & $\big|(cu)_0^{\bar 3}(\bar c\bar u)_1^3 \big\rangle$ & $4011.44/4013.97$ & $4039$ & $4210(4172)$ & - \\
& & $\big|(cu)_1^{\bar 3}(\bar c\bar u)_1^3 \big\rangle$ & $4136.26/4138.34$ & $4076$ & $4079$ & - \\ \cline{3-7}
& & $\big|(cu)_0^{6}(\bar c\bar u)_1^{\bar 6} \big\rangle$ & $4121.30/4123.47$ & $4077$ & $4233(4626)$ & - \\
& & $\big|(cu)_1^{6}(\bar c\bar u)_1^{\bar 6} \big\rangle$ & $3990.92/3993.39$ & $4066$ & $4263$ & - \\ \cline{2-7}
& $0/1(2^+)$ & $\big|(cu)_1^{\bar 3}(\bar c\bar u)_1^3 \big\rangle$ & $4179.58/4181.62$ & $4142$ & $4152$ & - \\
& & $\big|(cu)_1^{6}(\bar c\bar u)_1^{\bar 6} \big\rangle$ & $4105.20/4107.53$ & $4147$ & $4219$ & - \\ \hline

$T(cu\bar c\bar d)/$ & $0/1(0^+)$ & $\big|(cu)_0^{\bar 3}(\bar c\bar d)_0^3 \big\rangle$ & $3866.46$ & - & - & - \\
$T(cd\bar c\bar u)$ & & $\big|(cu)_1^{\bar 3}(\bar c\bar d)_1^3 \big\rangle$ & $4115.65$ & - & - & - \\ \cline{3-7}
& & $\big|(cu)_0^{6}(\bar c\bar d)_0^{\bar 6} \big\rangle$ & $4195.51$ & - & - & - \\
& & $\big|(cu)_1^{6}(\bar c\bar d)_1^{\bar 6} \big\rangle$ & $3935.06$ & - & - & - \\ \cline{2-7}
& $0/1(1^+)$ & $\big|(cu)_0^{\bar 3}(\bar c\bar d)_1^3 \big\rangle$ & $4012.47$ & - & - & $3887.1\pm 2.6/4196^{+35}_{-32}$ \\
& & $\big|(cu)_1^{\bar 3}(\bar c\bar d)_0^3 \big\rangle$ & $4012.94$ & - & - & - \\
& & $\big|(cu)_1^{\bar 3}(\bar c\bar d)_1^3 \big\rangle$ & $4137.30$ & - & - & - \\ \cline{3-7}
& & $\big|(cu)_0^{6}(\bar c\bar d)_1^{\bar 6} \big\rangle$ & $4122.50$ & - & - & - \\
& & $\big|(cu)_1^{6}(\bar c\bar d)_0^{\bar 6} \big\rangle$ & $4122.27$ & - & - & - \\
& & $\big|(cu)_1^{6}(\bar c\bar d)_1^{\bar 6} \big\rangle$ & $3992.16$ & - & - & - \\ \cline{2-7}
& $0/1(2^+)$ & $\big|(cu)_1^{\bar 3}(\bar c\bar d)_1^3 \big\rangle$ & $4180.60$ & - & - & - \\
& & $\big|(cu)_1^{6}(\bar c\bar d)_1^{\bar 6} \big\rangle$ & $4106.36$ & - & - & - \\\hline
\end{tabular} }
\end{table}
\begin{table}[ht]
\centering
\captionof{table}{Masses of $(\overline{\mb3}_c\mb3_c)$ and $(\mb6_c\overline{\mb6}_c)$ hidden charm $(cn\bar c\bar s, cs\bar c\bar s)$ tetraquark states using heavy-light diquarks.}
\label{t18}
\scalebox{0.9}{
\begin{tabular}{|c|c|c|c|c|c|c|c|}\hline
\multirow{2}{*}{\textbf{States}} & \multirow{2}{*}{$I(J^P)$} & \multirow{2}{*}{\textbf{Configuration}} & \multirow{2}{*}{\textbf{This work}} & \textbf{RQM} & \textbf{NRPM} & \textbf{BM} & \textbf{PDG} \\ 
& & & & \bf \cite{Yu:2024ljg} & \bf \cite{Liu:2024fnh} & \bf \cite{Yan:2023lvm} & \bf \cite{ParticleDataGroup:2024cfk} \\ \hline
$T(cu\bar c\bar s)/$ & $\frac{1}{2}(0^+)$ & $\big|(cu)_0^{\bar 3}(\bar c\bar s)_0^3 \big\rangle$ & $4010.01$ & $4059$ & - & - & - \\
$T(cs\bar c\bar u)$ & & $\big|(cu)_1^{\bar 3}(\bar c\bar s)_1^3 \big\rangle$ & $4260.59$ & $4129$ & - & - & - \\ \cline{3-8}
& & $\big|(cu)_0^{6}(\bar c\bar s)_0^{\bar 6} \big\rangle$ & $4395.66$ & $4174$ & - & - & - \\
& & $\big|(cu)_1^{6}(\bar c\bar s)_1^{\bar 6} \big\rangle$ & $4145.64$ & $3946$ & - & - & - \\ \cline{2-8}
& $\frac{1}{2}(1^+)$ & $\big|(cu)_0^{\bar 3}(\bar c\bar s)_1^3 \big\rangle$ & $4153.95$ & $4118$ & - & - & $3988\pm 5/4220^{+50}_{-40}$ \\
& & $\big|(cu)_1^{\bar 3}(\bar c\bar s)_0^3 \big\rangle$ & $4156.49$ & $4130$ & - & - & - \\
& & $\big|(cu)_1^{\bar 3}(\bar c\bar s)_1^3 \big\rangle$ & $4280.51$ & $4160$ & - & - & - \\ \cline{3-8}
& & $\big|(cu)_0^{6}(\bar c\bar s)_1^{\bar 6} \big\rangle$ & $4323.69$ & $4156$ & - & - & - \\
& & $\big|(cu)_1^{6}(\bar c\bar s)_0^{\bar 6} \big\rangle$ & $4322.42$ & $4149$ & - & - & - \\
& & $\big|(cu)_1^{6}(\bar c\bar s)_1^{\bar 6} \big\rangle$ & $4198.05$ & $4142$ & - & - & - \\ \cline{2-8}
& $\frac{1}{2}(2^+)$ & $\big|(cu)_1^{\bar 3}(\bar c\bar s)_1^3 \big\rangle$ & $4320.35$ & $4217$ & - & - & - \\
& & $\big|(cu)_1^{6}(\bar c\bar s)_1^{\bar 6} \big\rangle$ & $4302.85$ & $4210$ & - & - & - \\ \hline

$T(cd\bar c\bar s)/$ & $\frac{1}{2}(0^+)$ & $\big|(cd)_0^{\bar 3}(\bar c\bar s)_0^3 \big\rangle$ & $4011.51$ & - & - & - & - \\
$T(cs\bar c\bar d)$ & & $\big|(cd)_1^{\bar 3}(\bar c\bar s)_1^3 \big\rangle$ & $4261.64$ & - & - & - & - \\ \cline{3-8}
& & $\big|(cd)_0^{6}(\bar c\bar s)_0^{\bar 6} \big\rangle$ & $4396.63$ & - & - & - & - \\
& & $\big|(cd)_1^{6}(\bar c\bar s)_1^{\bar 6} \big\rangle$ & $4146.90$ & - & - & - & - \\ \cline{2-8}
& $\frac{1}{2}(1^+)$ & $\big|(cd)_0^{\bar 3}(\bar c\bar s)_1^3 \big\rangle$ & $4155.45$ & - & - & - & $3988\pm 5/4220^{+50}_{-40}$ \\
& & $\big|(cd)_1^{\bar 3}(\bar c\bar s)_0^3 \big\rangle$ & $4157.52$ & - & - & - & - \\
& & $\big|(cd)_1^{\bar 3}(\bar c\bar s)_1^3 \big\rangle$ & $4281.55$ & - & - & - & - \\ \cline{3-8}
& & $\big|(cd)_0^{6}(\bar c\bar s)_1^{\bar 6} \big\rangle$ & $4324.66$ & - & - & - & - \\
& & $\big|(cd)_1^{6}(\bar c\bar s)_0^{\bar 6} \big\rangle$ & $4323.62$ & - & - & - & - \\
& & $\big|(cd)_1^{6}(\bar c\bar s)_1^{\bar 6} \big\rangle$ & $4199.28$ & - & - & - & - \\ \cline{2-8}
& $\frac{1}{2}(2^+)$ & $\big|(cd)_1^{\bar 3}(\bar c\bar s)_1^3 \big\rangle$ & $4321.37$ & - & - & - & - \\
& & $\big|(cd)_1^{6}(\bar c\bar s)_1^{\bar 6} \big\rangle$ & $4304.02$ & - & - & - & - \\ \hline

$T(cs\bar c\bar s)$ & $0(0^+)$ & $\big|(cs)_0^{\bar 3}(\bar c\bar s)_0^3 \big\rangle$ & $4155.06$ & $4144$ & $4356$ & $4254$ & - \\
& & $\big|(cs)_1^{\bar 3}(\bar c\bar s)_1^3 \big\rangle$ & $4406.30$ & $4210$ & $4526$ & $4091$ & - \\ \cline{3-8}
& & $\big|(cs)_0^{6}(\bar c\bar s)_0^{\bar 6} \big\rangle$ & $4596.78$ & $4237$ & $4174$ & $4492$ & - \\
& & $\big|(cs)_1^{6}(\bar c\bar s)_1^{\bar 6} \big\rangle$ & $4356.71$ & $4031$ & $4424$ & $4378$ & - \\ \cline{2-8}
& $0(1^+)$ & $\big|(cs)_0^{\bar 3}(\bar c\bar s)_1^3 \big\rangle$ & $4299.00$ & $4204$ & $4529(4448)$ & $4572(4596)$ & - \\
& & $\big|(cs)_1^{\bar 3}(\bar c\bar s)_1^3 \big\rangle$ & $4424.62$ & $4236$ & $4322$ & $4529$ & - \\ \cline{3-8}
& & $\big|(cs)_0^{6}(\bar c\bar s)_1^{\bar 6} \big\rangle$ & $4524.81$ & $4217$ & $4492(4382)$ & $4655(4843)$ & - \\
& & $\big|(cs)_1^{6}(\bar c\bar s)_1^{\bar 6} \big\rangle$ & $4404.78$ & $4208$ & $4239$ & $4638$ & - \\ \cline{2-8}
& $0(2^+)$ & $\big|(cs)_1^{\bar 3}(\bar c\bar s)_1^3 \big\rangle$ & $4461.26$ & $4290$ & $4528$ & $4421$ & - \\
& & $\big|(cs)_1^{6}(\bar c\bar s)_1^{\bar 6} \big\rangle$ & $4500.90$ & $4270$ & $4389$ & $4450$ & - \\ \hline
\end{tabular} }
\end{table}
\begin{table}[ht]
\centering
\captionof{table}{Masses of $(\overline{\mb3}_c\mb3_c)$ and $(\mb6_c\overline{\mb6}_c)$ hidden bottom $bn\bar b\bar n$ tetraquark states using heavy-light diquarks.}
\label{t19}
\scalebox{0.9}{
\begin{tabular}{|c|c|c|c|c|}\hline
\multirow{2}{*}{\textbf{States}} & \multirow{2}{*}{$I(J^P)$} & \multirow{2}{*}{\textbf{Configuration}} & \multirow{2}{*}{\textbf{This work}} & \textbf{PDG} \\ 
& & & & \bf \cite{ParticleDataGroup:2024cfk} \\ \hline
$T(bu\bar b\bar u)/$ & $0/1(0^+)$ & $\big|(bu)_0^{\bar 3}(\bar b\bar u)_0^3 \big\rangle$ & $10531.9/10534.8$ & - \\
$T(bd\bar b\bar d)$ & & $\big|(bu)_1^{\bar 3}(\bar b\bar u)_1^3 \big\rangle$ & $ 10797.5/10799.5$  & - \\ \cline{3-5}
& & $\big|(bu)_0^{6}(\bar b\bar u)_0^{\bar 6} \big\rangle$ & $10837.9/10839.8$  & - \\
& & $\big|(bu)_1^{6}(\bar b\bar u)_1^{\bar 6} \big\rangle$ & $10685.5/10687.9$  & - \\ \cline{2-5}
& $0/1(1^+)$ & $\big|(bu)_0^{\bar 3}(\bar b\bar u)_1^3 \big\rangle$ & $10667.9/10670.4$  & $ 10609\pm 6$ \\
& & $\big|(bu)_1^{\bar 3}(\bar b\bar u)_1^3 \big\rangle$ & $10800.7/10802.8$  & - \\ \cline{3-5}
& & $\big|(bu)_0^{6}(\bar b\bar u)_1^{\bar 6} \big\rangle$ & $10769.9/10772.1$  & - \\ 
& & $\big|(bu)_1^{6}(\bar b\bar u)_1^{\bar 6} \big\rangle$ & $10693.7/10696.1$  & - \\ \cline{2-5}
& $0/1(2^+)$ & $\big|(bu)_1^{\bar 3}(\bar b\bar u)_1^3 \big\rangle$ & $10807.1/10809.2$  & - \\
& & $\big|(bu)_1^{6}(\bar b\bar u)_1^{\bar 6} \big\rangle$ & $10710.1/10712.5$  & - \\ \hline

$T(bu\bar b\bar d)/$ & $0/1(0^+)$ & $\big|(bu)_0^{\bar 3}(\bar b\bar d)_0^3 \big\rangle$ & $10533.4$  & - \\
$T(bd\bar b\bar u)$ & & $\big|(bu)_1^{\bar 3}(\bar b\bar d)_1^3 \big\rangle$ & $10798.5$  & - \\ \cline{3-5}
& & $\big|(bu)_0^{6}(\bar b\bar d)_0^{\bar 6} \big\rangle$ & $10838.9$  & - \\
& & $\big|(bu)_1^{6}(\bar b\bar d)_1^{\bar 6} \big\rangle$ & $10686.7$  & - \\ \cline{2-5}
& $0/1(1^+)$ & $\big|(bu)_0^{\bar 3}(\bar b\bar d)_1^3 \big\rangle$ & $10668.9$  & $10607.2\pm 2.0/10652.2\pm 1.5$ \\
& & $\big|(bu)_1^{\bar 3}(\bar b\bar d)_0^3 \big\rangle$ & $10669.4$  & - \\ 
& & $\big|(bu)_1^{\bar 3}(\bar b\bar d)_1^3 \big\rangle$ & $10801.7$  & - \\ \cline{3-5}
& & $\big|(bu)_0^{6}(\bar b\bar d)_1^{\bar 6} \big\rangle$ & $10771.1$  & - \\
& & $\big|(bu)_1^{6}(\bar b\bar d)_0^{\bar 6} \big\rangle$ & $10770.9$  & - \\
& & $\big|(bu)_1^{6}(\bar b\bar d)_1^{\bar 6} \big\rangle$ & $10694.9$  & - \\ \cline{2-5}
& $0/1(2^+)$ & $\big|(bu)_1^{\bar 3}(\bar b\bar d)_1^3 \big\rangle$ & $10808.1$  & - \\
& & $\big|(bu)_1^{6}(\bar b\bar d)_1^{\bar 6} \big\rangle$ & $10711.3$  & - \\\hline
\end{tabular} }
\end{table}
\begin{table}[ht]
\centering
\captionof{table}{Masses of $(\overline{\mb3}_c\mb3_c)$ and $(\mb6_c\overline{\mb6}_c)$ hidden bottom $(bn\bar b\bar s, bs\bar b\bar s)$ tetraquark states using heavy-light diquarks.}
\label{t20}
\scalebox{0.9}{
\begin{tabular}{|c|c|c|c|}\hline
\multirow{2}{*}{\textbf{States}} & \multirow{2}{*}{$I(J^P)$} & \multirow{2}{*}{\textbf{Configuration}} & \multirow{2}{*}{\textbf{This work}} \\ 
& & & \\ \hline
$T(bu\bar b\bar s)/$ & $\frac{1}{2}(0^+)$ & $\big|(bu)_0^{\bar 3}(\bar b\bar s)_0^3 \big\rangle$ & $10658.4$ \\
$T(bs\bar b\bar u)$ & & $\big|(bu)_1^{\bar 3}(\bar b\bar s)_1^3 \big\rangle$ & $10936.4$ \\ \cline{3-4}
& & $\big|(bu)_0^{6}(\bar b\bar s)_0^{\bar 6} \big\rangle$ & $11048.3$  \\
& & $\big|(bu)_1^{6}(\bar b\bar s)_1^{\bar 6} \big\rangle$ & $10890.4$  \\ \cline{2-4}
& $\frac{1}{2}(1^+)$ & $\big|(bu)_0^{\bar 3}(\bar b\bar s)_1^3 \big\rangle$ & $10806.6$ \\
& & $\big|(bu)_1^{\bar 3}(\bar b\bar s)_0^3 \big\rangle$ & $10794.4$   \\
& & $\big|(bu)_1^{\bar 3}(\bar b\bar s)_1^3 \big\rangle$ & $10939.5$  \\ \cline{3-4}
& & $\big|(bu)_0^{6}(\bar b\bar s)_1^{\bar 6} \big\rangle$ & $10974.2$ \\
& & $\big|(bu)_1^{6}(\bar b\bar s)_0^{\bar 6} \big\rangle$ & $10980.3$ \\
& & $\big|(bu)_1^{6}(\bar b\bar s)_1^{\bar 6} \big\rangle$ & $10898.3$ \\ \cline{2-4}
& $\frac{1}{2}(2^+)$ & $\big|(bu)_1^{\bar 3}(\bar b\bar s)_1^3 \big\rangle$ & $10945.7$ \\
& & $\big|(bu)_1^{6}(\bar b\bar s)_1^{\bar 6} \big\rangle$ & $10914.1$  \\ \hline

$T(bd\bar b\bar s)/$ & $\frac{1}{2}(0^+)$ & $\big|(bd)_0^{\bar 3}(\bar b\bar s)_0^3 \big\rangle$ & $10659.9$ \\
$T(bs\bar b\bar d)$ & & $\big|(bd)_1^{\bar 3}(\bar b\bar s)_1^3 \big\rangle$ & $10937.4$ \\ \cline{3-4}
& & $\big|(bd)_0^{6}(\bar b\bar s)_0^{\bar 6} \big\rangle$ & $11049.3$  \\
& & $\big|(bd)_1^{6}(\bar b\bar s)_1^{\bar 6} \big\rangle$ & $10891.6$  \\ \cline{2-4}
& $\frac{1}{2}(1^+)$ & $\big|(bd)_0^{\bar 3}(\bar b\bar s)_1^3 \big\rangle$ & $10808.1$ \\
& & $\big|(bd)_1^{\bar 3}(\bar b\bar s)_0^3 \big\rangle$ & $10795.4$  \\
& & $\big|(bd)_1^{\bar 3}(\bar b\bar s)_1^3 \big\rangle$ & $10940.5$  \\ \cline{3-4}
& & $\big|(bd)_0^{6}(\bar b\bar s)_1^{\bar 6} \big\rangle$ & $10975.2$   \\
& & $\big|(bd)_1^{6}(\bar b\bar s)_0^{\bar 6} \big\rangle$ & $10981.5$   \\
& & $\big|(bd)_1^{6}(\bar b\bar s)_1^{\bar 6} \big\rangle$ & $10899.5$  \\ \cline{2-4}
& $\frac{1}{2}(2^+)$ & $\big|(bd)_1^{\bar 3}(\bar b\bar s)_1^3 \big\rangle$ & $10946.7$  \\
& & $\big|(bd)_1^{6}(\bar b\bar s)_1^{\bar 6} \big\rangle$ & $10915.3$  \\ \hline

$T(bs\bar b\bar s)$ & $0(0^+)$ & $\big|(bs)_0^{\bar 3}(\bar b\bar s)_0^3 \big\rangle$ & $10784.9$ \\
& & $\big|(bs)_1^{\bar 3}(\bar b\bar s)_1^3 \big\rangle$ & $11075.3$  \\ \cline{3-4}
& & $\big|(bs)_0^{6}(\bar b\bar s)_0^{\bar 6} \big\rangle$ & $11258.7$  \\
& & $\big|(bs)_1^{6}(\bar b\bar s)_1^{\bar 6} \big\rangle$ & $11095.2$  \\ \cline{2-4}
& $0(1^+)$ & $\big|(bs)_0^{\bar 3}(\bar b\bar s)_1^3 \big\rangle$ & $10933.1$  \\
& & $\big|(bs)_1^{\bar 3}(\bar b\bar s)_1^3 \big\rangle$ & $11078.3$  \\ \cline{3-4}
& & $\big|(bs)_0^{6}(\bar b\bar s)_1^{\bar 6} \big\rangle$ & $11184.6$  \\
& & $\big|(bs)_1^{6}(\bar b\bar s)_1^{\bar 6} \big\rangle$ & $11102.9$  \\ \cline{2-4}
& $0(2^+)$ & $\big|(bs)_1^{\bar 3}(\bar b\bar s)_1^3 \big\rangle$ & $11084.3$ \\
& & $\big|(bs)_1^{6}(\bar b\bar s)_1^{\bar 6} \big\rangle$ & $11118.1$  \\ \hline
\end{tabular} }
\end{table}
\begin{figure}[t]
    \centering
    \includegraphics[width=1.0\linewidth]{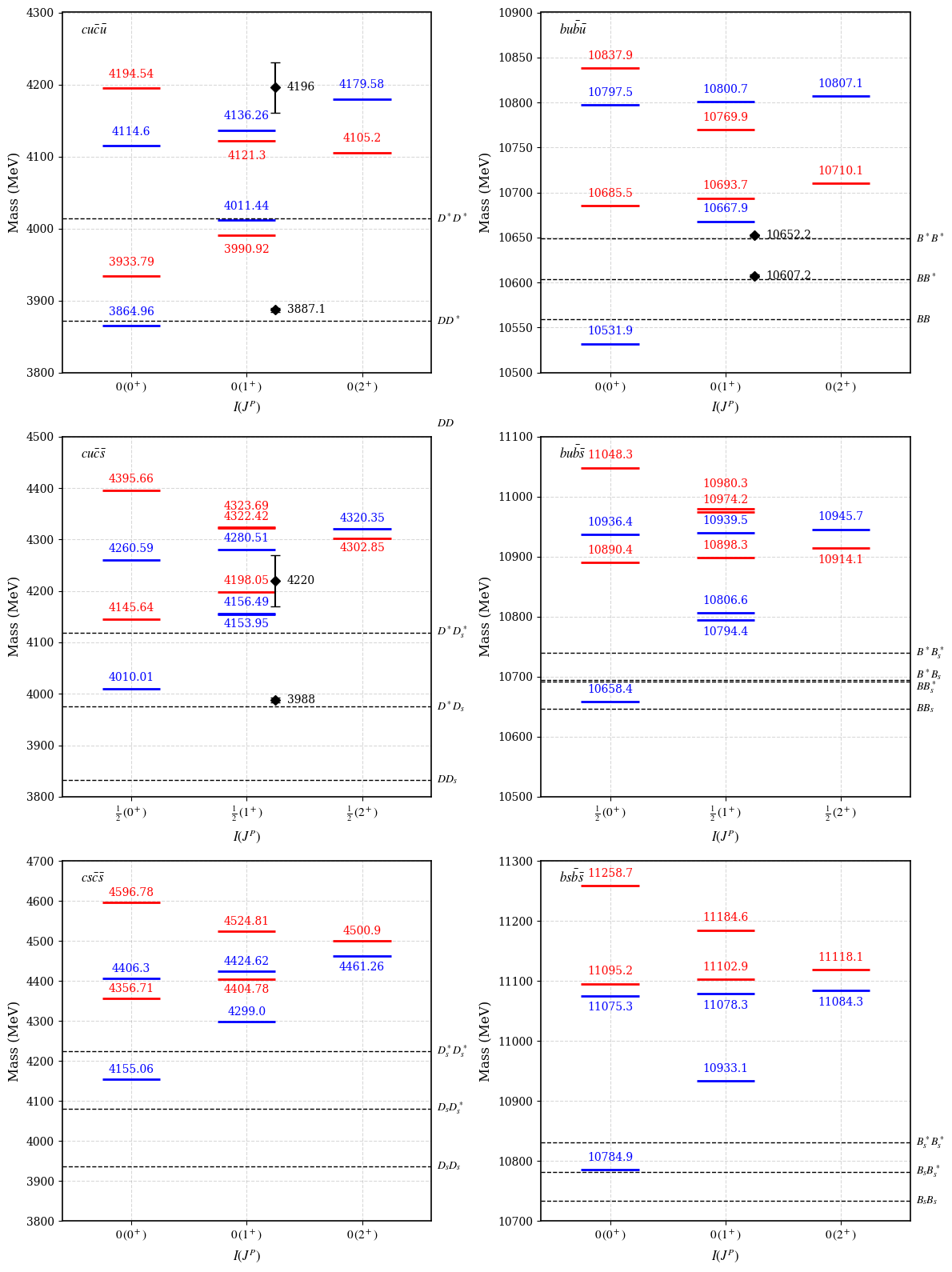}
    \caption{Mass spectra of hidden heavy tetraquark states: $cu\bar c\bar u, cu\bar c\bar s, cs\bar c\bar s$ (left panel) and $bu\bar b\bar u, bu\bar b\bar s, bs\bar b\bar s$ (right panel).}
    \label{Plot_HiddenHeavy}
\end{figure}
\begin{table*}[ht]
\centering
\caption{Masses of fully heavy $(\overline{\mb3}_c\mb3_c)$ and $(\mb6_c\overline{\mb6}_c)$ tetraquark states (in MeV).}
\label{t14_modified}
\begin{tabular}{|c|c|c|c|c|c|c|}
\hline
\multirow{2}{*}{\textbf{States}} & \multirow{2}{*}{$I(J^{P})$} & \multirow{2}{*}{\textbf{Configuration}} & \multirow{2}{*}{\textbf{This work}} & \textbf{PM} & \textbf{CMIM} & \textbf{DMC} \\
& & & & \bf \cite{Liu:2019zuc} & \bf \cite{Weng:2020jao} & \bf \cite{Gordillo:2020sgc} \\
\hline

$T(cc\bar c\bar c)$ 
& $0(0^+)$ & $\big|\{cc\}_1^{\bar 3}\{\bar c\bar c\}_1^{3} \big\rangle$ & 6452.95 & 6487 & 6271.3 & 6351 \\
& & $\big|\{cc\}_0^{6}\{\bar c\bar c\}_0^{\bar 6} \big\rangle$ & 6866.34 & 6518 & 6044.9 & -- \\ \cline{2-7}
& $0(1^+)$ & $\big|\{cc\}_1^{\bar 3}\{\bar c\bar c\}_1^{3} \big\rangle$ & 6461.22 & 6500 & 6230.6 & 6441 \\ \cline{2-7}
& $0(2^+)$ & $\big|\{cc\}_1^{\bar 3}\{\bar c\bar c\}_1^{3} \big\rangle$ & 6477.74 & 6524 & 6287.3 & 6471 \\\hline

$T(cc\bar c\bar b)$ 
& $0(0^+)$ & $\big|\{cc\}_1^{\bar 3}\{\bar c\bar b\}_1^{3} \big\rangle$ & 9737.83 & 9740 & 9505.9 & 9615 \\
& & $\big|\{cc\}_0^{6}\{\bar c\bar b\}_0^{\bar 6} \big\rangle$ & 10212.35 & 9763 & 9317.5 & -- \\ \cline{2-7}
& $0(1^+)$ & $\big|\{cc\}_1^{\bar 3}\{\bar c\bar b\}_0^{3} \big\rangle$ & 9741.69 & 9746 & 9498.5 & 9610 \\
& & $\big|\{cc\}_1^{\bar 3}\{\bar c\bar b\}_1^{3} \big\rangle$ & 9741.99 & 9749 & 9484.3 & -- \\
& & $\big|\{cc\}_0^{6}\{\bar c\bar b\}_1^{\bar 6} \big\rangle$ & 10210.13 & 9757 & 9335.1 & -- \\ \cline{2-7}
& $0(2^+)$ & $\big|\{cc\}_1^{\bar 3}\{\bar c\bar b\}_1^{3} \big\rangle$ & 9750.31 & 9768 & 9525.9 & 9719 \\\hline

$T(cc\bar b\bar b)$ 
& $0(0^+)$ & $\big|\{cc\}_1^{\bar 3}\{\bar b\bar b\}_1^{3} \big\rangle$ & 12958.65 & 12953 & 12711.9 & -- \\
& & $\big|\{cc\}_0^{6}\{\bar b\bar b\}_0^{\bar 6} \big\rangle$ & 13590.54 & 13032 & 12596.3 & -- \\ \cline{2-7}
& $0(1^+)$ & $\big|\{cc\}_1^{\bar 3}\{\bar b\bar b\}_1^{3} \big\rangle$ & 12961.42 & 12960 & 12671.7 & -- \\ \cline{2-7}
& $0(2^+)$ & $\big|\{cc\}_1^{\bar 3}\{\bar b\bar b\}_1^{3} \big\rangle$ & 12966.98 & 12972 & 12703.1 & -- \\\hline

$T(cb\bar c\bar b)$ 
& $0(0^+)$ & $\big|\{cb\}_0^{\bar 3}\{\bar c\bar b\}_0^{3} \big\rangle$ & 13013.90 & 13050 & 12746.9 & 12534 \\
& & $\big|\{cb\}_1^{\bar 3}\{\bar c\bar b\}_1^{3} \big\rangle$ & 13018.64 & 13035 & 12681.6 & -- \\
& & $\big|\{cb\}_0^{6}\{\bar c\bar b\}_0^{\bar 6} \big\rangle$ & 13558.36 & 12864 & 12509.3 & -- \\
& & $\big|\{cb\}_1^{6}\{\bar c\bar b\}_1^{\bar 6} \big\rangle$ & 13543.45 & 12835 & 12362.8 & -- \\ \cline{2-7}

& $0(1^+)$ & $\big|\{cb\}_0^{\bar 3}\{\bar c\bar b\}_1^{3} \big\rangle$ & 13018.36 & 13052(13056) & 12744.1(12703.2) & 12510 \\
& & $\big|\{cb\}_1^{\bar 3}\{\bar c\bar b\}_1^{3} \big\rangle$ & 13020.73 & 13047 & 12720.0 & 12569 \\
& & $\big|\{cb\}_0^{6}\{\bar c\bar b\}_1^{\bar 6} \big\rangle$ & 13556.14 & 12864(12870) & 12477.2(12523.6) & -- \\
& & $\big|\{cb\}_1^{6}\{\bar c\bar b\}_1^{\bar 6} \big\rangle$ & 13548.69 & 12852 & 12424.9 & -- \\ \cline{2-7}

& $0(2^+)$ & $\big|\{cb\}_1^{\bar 3}\{\bar c\bar b\}_1^{3} \big\rangle$ & 13024.91 & 13070 & 12754.9 & 12582 \\
& & $\big|\{cb\}_1^{6}\{\bar c\bar b\}_1^{\bar 6} \big\rangle$ & 13559.15 & 12884 & 12537.4 & -- \\ \hline

$T(bb\bar c\bar b)$ 
& $0(0^+)$ & $\big|\{bb\}_1^{\bar 3}\{\bar c\bar b\}_1^{3} \big\rangle$ & 16238.08 & 16158 & 15862.0 & 16040 \\
& & $\big|\{bb\}_0^{6}\{\bar c\bar b\}_0^{\bar 6} \big\rangle$ & 16936.55 & 16173 & 15711.9 & -- \\ \cline{2-7}
& $0(1^+)$ & $\big|\{bb\}_1^{\bar 3}\{\bar c\bar b\}_0^{3} \big\rangle$ & 16236.41 & 16157 & 15854.4 & 16013 \\
& & $\big|\{bb\}_1^{\bar 3}\{\bar c\bar b\}_1^{3} \big\rangle$ & 16239.47 & 16164 & 15851.3 & -- \\
& & $\big|\{bb\}_0^{6}\{\bar c\bar b\}_1^{\bar 6} \big\rangle$ & 16934.33 & 16167 & 15719.1 & -- \\ \cline{2-7}
& $0(2^+)$ & $\big|\{bb\}_1^{\bar 3}\{\bar c\bar b\}_1^{3} \big\rangle$ & 16242.27 & 16176 & 15882.3 & 16129 \\\hline

$T(bb\bar b\bar b)$ 
& $0(0^+)$ & $\big|\{bb\}_1^{\bar 3}\{\bar b\bar b\}_1^{3} \big\rangle$ & 19457.05 & 19322 & 18981.0 & 19199 \\
& & $\big|\{bb\}_0^{6}\{\bar b\bar b\}_0^{\bar 6} \big\rangle$ & 20314.74 & 19338 & 18836.1 & -- \\ \cline{2-7}
& $0(1^+)$ & $\big|\{bb\}_1^{\bar 3}\{\bar b\bar b\}_1^{3} \big\rangle$ & 19457.99 & 19329 & 18969.4 & 19276 \\ \cline{2-7}
& $0(2^+)$ & $\big|\{bb\}_1^{\bar 3}\{\bar b\bar b\}_1^{3} \big\rangle$ & 19459.85 & 19341 & 19000.1 & 19289 \\\hline
\end{tabular}
\end{table*}
\FloatBarrier
\begin{figure}
    \centering
    \includegraphics[width=1.0\linewidth]{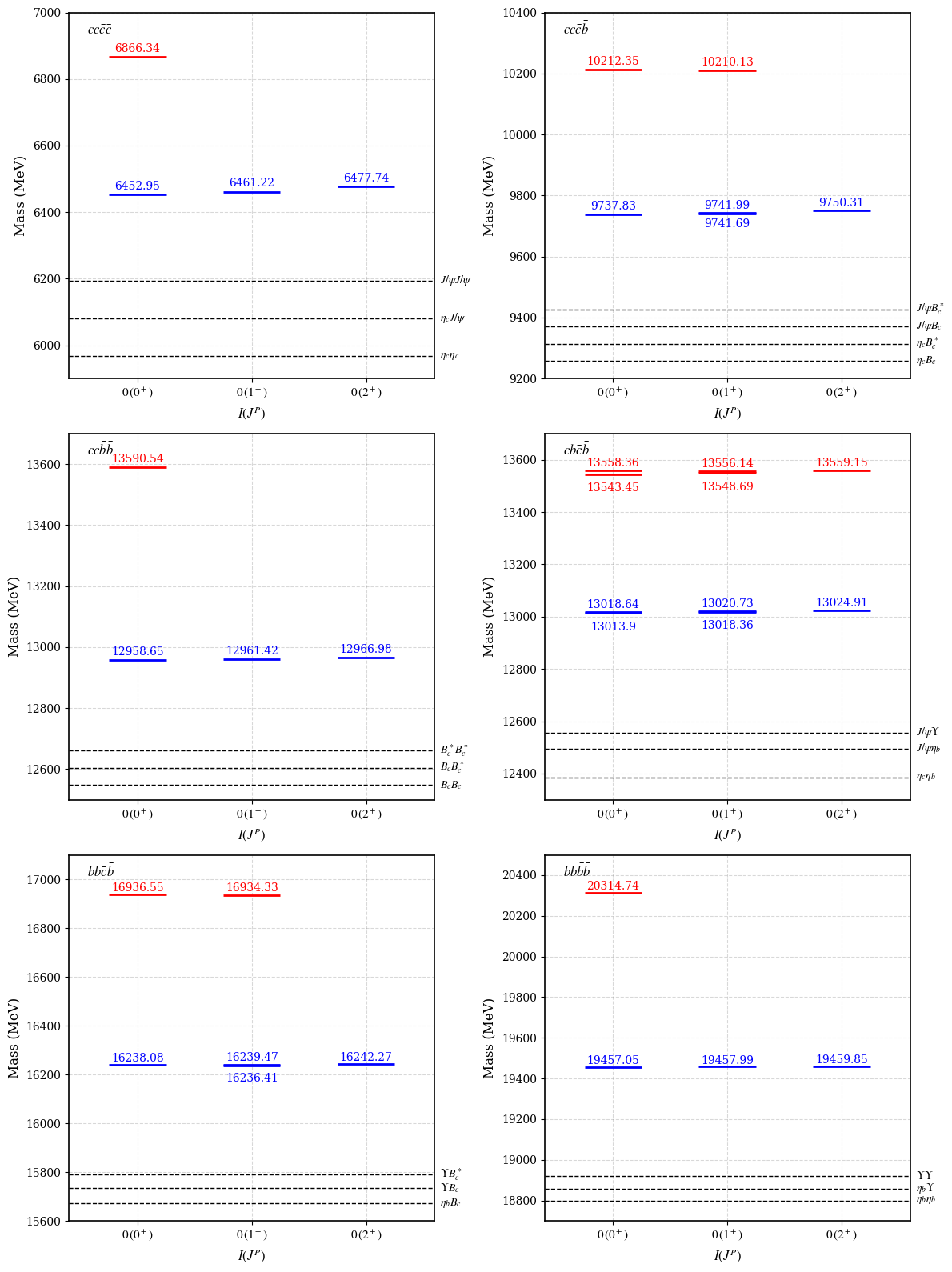}
    \caption{Mass spectra of fully heavy tetraquark states: $cc\bar c\bar c, cc\bar c\bar b, cc\bar b\bar b, cb\bar c\bar b, bb\bar c\bar b$, and $bb\bar b\bar b$.}
    \label{Plot_FullyHeavy}
\end{figure}
\begin{table*}[t]
\centering
\caption{Masses of compact hidden charm non-strange pentaquark states $P_{c\bar c}=(cu)(ud)\bar c$ (in MeV).}
\label{Penta1}
\renewcommand{\arraystretch}{1.25}
\scalebox{0.9}{\begin{tabular}{ccccccc}
\hline\hline
$(S_{D_1},S_{D_2},S_{\bar q})$ & $S_{D_1}+S_{D_2}$ & $J^P$ 
& Mass expressions\footnote{$m_{[ij]}$ and $m_{\{ij\}}$ denote scalar ($0^+$) and axial-vector ($1^+$) diquark masses, respectively.} 
& This work\footnote{The effective couplings used in our work are:
$b_{\{cu\}\{ud\}}=30.72$~MeV, $b_{\{cu\}\bar c}=24.84$~MeV, $b_{\{ud\}\bar c}=16.18$~MeV.} 
& PDG \cite{ParticleDataGroup:2024cfk} \\
\hline
$(0,0,\frac{1}{2})$ & $0$ & $\tfrac{1}{2}^-$ & $m_{[cu]} + m_{[ud]} + m_{\bar c}$ & $4247.94$ & - \\ \hline
\multirow{2}{*}{$(0,1,\frac{1}{2})$} & \multirow{2}{*}{$1$} & $\tfrac{1}{2}^-$ & $m_{[cu]} + m_{\{ud\}} + m_{\bar c} - b_{\{ud\}\bar c}$ & $4339.10$ & $P_{c\bar c}(4312)$ \\
& & $\tfrac{3}{2}^-$ & $m_{[cu]} + m_{\{ud\}} + m_{\bar c} + \tfrac{1}{2} b_{\{ud\}\bar c}$ & $4363.37$ & $P_{c\bar c}(4380)$ \\ \hline
\multirow{2}{*}{$(1,0,\frac{1}{2})$} & \multirow{2}{*}{$1$} & $\tfrac{1}{2}^-$ & $m_{\{cu\}} + m_{[ud]} + m_{\bar c} - b_{\{cu\}\bar c}$ & $4369.59$ & $P_{c\bar c}(4337)$ \\
& & $\tfrac{3}{2}^-$ & $m_{\{cu\}} + m_{[ud]} + m_{\bar c} + \tfrac{1}{2} b_{\{cu\}\bar c}$ & $4406.84$ & - \\ \hline
\multirow{5}{*}{$(1,1,\frac{1}{2})$} & $0$ & $\tfrac{1}{2}^-$ & $m_{\{cu\}} + m_{\{ud\}} + m_{\bar c} - 2 b_{\{cu\}\{ud\}}$ & $4440.32$ & $P_{c\bar c}(4440)$ \\ \cline{2-6}
& \multirow{2}{*}{$1$} & $\tfrac{1}{2}^-$ & $m_{\{cu\}} + m_{\{ud\}} + m_{\bar c} - b_{\{cu\}\{ud\}} - \tfrac{1}{2}(b_{\{cu\}\bar c}+b_{\{ud\}\bar c})$ & $4450.53$ & $P_{c\bar c}(4457)$ \\
& & $\tfrac{3}{2}^-$ & $m_{\{cu\}} + m_{\{ud\}} + m_{\bar c} - b_{\{cu\}\{ud\}} + \tfrac{1}{4}(b_{\{cu\}\bar c}+b_{\{ud\}\bar c})$ & $4481.29$ & - \\ \cline{2-6}
& \multirow{2}{*}{$2$} & $\tfrac{3}{2}^-$ & $m_{\{cu\}} + m_{\{ud\}} + m_{\bar c} + b_{\{cu\}\{ud\}} - \tfrac{3}{4}(b_{\{cu\}\bar c}+b_{\{ud\}\bar c})$ & $4501.72$ & - \\
& & $\tfrac{5}{2}^-$ & $m_{\{cu\}} + m_{\{ud\}} + m_{\bar c} + b_{\{cu\}\{ud\}} + \tfrac{1}{2}(b_{\{cu\}\bar c}+b_{\{ud\}\bar c})$ & $4552.99$ & - \\\hline\hline
\end{tabular} }
\end{table*}

\begin{table*}[t]
\centering
\caption{Masses of compact hidden charm strange pentaquark states $P_{c\bar cs}=(cs)(ud)\bar c$ (in MeV).}
\label{Penta2}
\renewcommand{\arraystretch}{1.25}
\scalebox{0.9}{\begin{tabular}{ccccccc}
\hline\hline
$(S_{D_1},S_{D_2},S_{\bar q})$ & $S_{D_1}+S_{D_2}$ & $J^P$ 
& Mass expressions\footnote{$m_{[ij]}$ and $m_{\{ij\}}$ denote scalar ($0^+$) and axial-vector ($1^+$) diquark masses, respectively.} 
& This work\footnote{The effective couplings used in our work are: $b_{\{cs\}\{ud\}}=28.25$~MeV, $b_{\{cs\}\bar c}=34.06$~MeV, $b_{\{ud\}\bar c}=16.18$~MeV.}
& PDG \cite{ParticleDataGroup:2024cfk} \\
\hline
$(0,0,\frac{1}{2})$ & $0$ & $\tfrac{1}{2}^-$ & $m_{[cs]} + m_{[ud]} + m_{\bar c}$ & $4392.99$ & $P_{c\bar cs}(4338)$ \\ \hline
\multirow{2}{*}{$(0,1,\frac{1}{2})$} & \multirow{2}{*}{$1$} & $\tfrac{1}{2}^-$ & $m_{[cs]} + m_{\{ud\}} + m_{\bar c} - b_{\{ud\}\bar c}$ & $4484.15$ & $P_{c\bar cs}(4459)$ \\
& & $\tfrac{3}{2}^-$ & $m_{[cs]} + m_{\{ud\}} + m_{\bar c} + \tfrac{1}{2} b_{\{ud\}\bar c}$ & $4508.42$ & - \\ \hline
\multirow{2}{*}{$(1,0,\frac{1}{2})$} & \multirow{2}{*}{$1$} & $\tfrac{1}{2}^-$ & $m_{\{cs\}} + m_{[ud]} + m_{\bar c} - b_{\{cs\}\bar c}$ & $4502.88$ & - \\
& & $\tfrac{3}{2}^-$ & $m_{\{cs\}} + m_{[ud]} + m_{\bar c} + \tfrac{1}{2} b_{\{cs\}\bar c}$ & $4553.96$ & - \\ \hline
\multirow{5}{*}{$(1,1,\frac{1}{2})$} & $0$ & $\tfrac{1}{2}^-$ & $m_{\{cs\}} + m_{\{ud\}} + m_{\bar c} - 2 b_{\{cs\}\{ud\}}$ & $4587.77$ & - \\ \cline{2-6}
& \multirow{2}{*}{$1$} & $\tfrac{1}{2}^-$ & $m_{\{cs\}} + m_{\{ud\}} + m_{\bar c} - b_{\{cs\}\{ud\}} - \tfrac{1}{2}(b_{\{cs\}\bar c}+b_{\{ud\}\bar c})$ & $4590.90$ & - \\
& & $\tfrac{3}{2}^-$ & $m_{\{cs\}} + m_{\{ud\}} + m_{\bar c} - b_{\{cs\}\{ud\}} + \tfrac{1}{4}(b_{\{cs\}\bar c}+b_{\{ud\}\bar c})$ & $4628.58$ & - \\ \cline{2-6}
& \multirow{2}{*}{$2$} & $\tfrac{3}{2}^-$ & $m_{\{cs\}} + m_{\{ud\}} + m_{\bar c} + b_{\{cs\}\{ud\}} - \tfrac{3}{4}(b_{\{cs\}\bar c}+b_{\{ud\}\bar c})$ & $4634.84$ & - \\
& & $\tfrac{5}{2}^-$ & $m_{\{cs\}} + m_{\{ud\}} + m_{\bar c} + b_{\{cs\}\{ud\}} + \tfrac{1}{2}(b_{\{cs\}\bar c}+b_{\{ud\}\bar c})$ & $4697.64$ & - \\\hline\hline
\end{tabular} }
\end{table*}
\FloatBarrier
\begin{figure}
    \centering
    \includegraphics[width=1.0\linewidth]{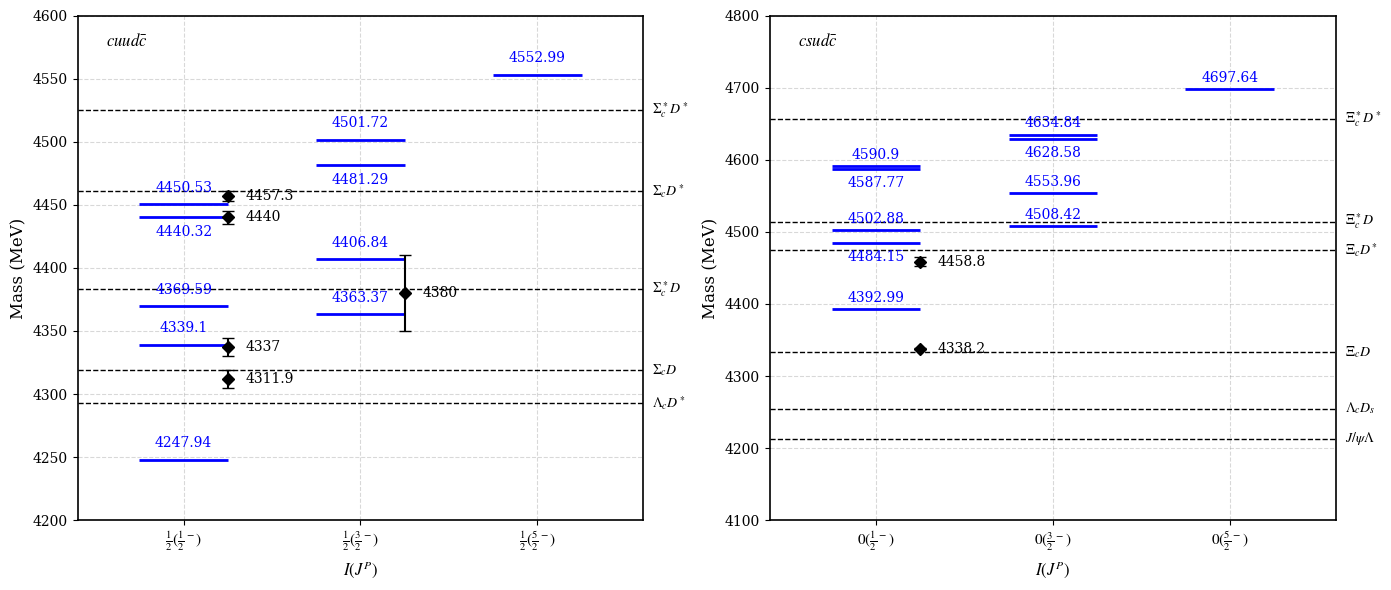}
    \caption{Mass spectra of hidden charm pentaquark states: $P_{c\bar c}=(cu)(ud)\bar c$ (left panel) and $P_{c\bar cs}=(cs)(ud)\bar c$ (right panel).}
    \label{Plot_Penta}
\end{figure}
\begin{table}
\caption{Comparison of our mass predictions with experimental data.}\label{table_last}
\renewcommand{\arraystretch}{1.15}
\begin{tabular}{llcc} 
\hline
Sector & State & PDG \cite{ParticleDataGroup:2024cfk} (MeV) & $|\Delta m|$ (MeV) \\
\hline
Singly charm
  & $T(cd\bar{u}\bar{s}),\;0^+$
    & $2892\pm 21$ & $21$ \\
  & $T(cs\bar{n}\bar{n}),\;0^+$
    & $2872\pm 16$ & $31$ \\
Doubly charm
  & $T_{cc}^+,\;0(1^+)$
    & $3874.74\pm 0.10$ & $14.8$ \\
Hidden charm
  & $T_{c\bar{c}1}(3900),\;1^+$
    & $3887\pm 3$ & $105^{*}$ \\
  & $T_{c\bar{c}1}(4200),\;1^+$
    & $4196^{+35}_{-32}$ & $60$ \\
  & $T_{c\bar{c}\bar{s}1}(4220),\;1^+$
    & $4220^{+50}_{-40}$ & $22$ \\
Hidden bottom
  & $T_{b\bar{b}1}(10610),\;1^+$
    & $10607\pm 2$ & $60$ \\
  & $T_{b\bar{b}1}(10650),\;1^+$
    & $10652\pm 2$ & $42$ \\
Pentaquark
  & $P_c(4440),\;\tfrac{1}{2}^-$
    & $4440.3\pm 1.3$ & $\phantom{0}0.3$ \\
  & $P_c(4457),\;\tfrac{1}{2}^-$
    & $4457.3\pm 0.6$ & $\phantom{0}6.5$ \\
\hline\hline
\multicolumn{4}{l}{\footnotesize
  ${}^{*}$Near $D\bar{D}^*$ threshold; compact-diquark value.}
\end{tabular}
\end{table}

\end{document}